\pdfoutput=1
\documentclass[11pt,letterpaper]{article}
\usepackage{jcappub}
\usepackage{caption}
\usepackage{floatrow}
\usepackage{hyperref}

\usepackage{graphicx}
\usepackage{amsmath,array}
\newcolumntype{P}[1]{>{\raggedright\arraybackslash}p{#1}}

\usepackage{mathtools}

\usepackage{bm}

\def\be{\begin{eqnarray}}
\def\ee{\end{eqnarray}}

\def\f{\mathrm{\textsl{f}}\,}
\def\k{\bm{k}}
\def\o{\bm{w}}
\def\d{\delta_D}

\def\la{\langle}
\def\ra{\rangle}

\renewcommand{\L}[2]{L_{#1 #2}}

\title{Dominated-Convergence Failure in Cosmological Perturbation Theory and a Numerical Foundation for BBGKY+ZA}
\author{Svetlin V. Tassev}
\emailAdd{stassev@alum.mit.edu}
\affiliation{Braintree High School\\Braintree, MA 02184, USA}

\abstract{A common ingredient in cosmological perturbation theory (PT) is the expansion of the dark matter overdensity $\delta$ in the Lagrangian displacement $\bm s$, which amounts to enforcing mass conservation perturbatively. In Eulerian PT (EPT), that expansion occurs already at the level of the continuity equation; in Lagrangian PT (LPT) it is done in the Poisson equation. We show that the resulting perturbative solutions for $\delta$ can diverge not because of the expansion in $\bm s$ per se, but because of an exchange of an infinite sum with a Fourier integral that violates the conditions of Lebesgue's dominated-convergence (DC) theorem. We show that this DC obstruction (DCO) is one clear reason why the convergence of EPT is controlled by advection terms beyond the linear $\delta$. The same DCO underlies LPT: LPT's region of validity is the resummation region of a DC-violating series, bounded by shell crossing on one side and severely underdense regions on the other. Effective field theories (EFT) of large-scale structure need to smooth at short scales just to recover from that DCO, independent of whether non-linearities beyond mass conservation are important or not. An alternative is to never expand $\delta$ in $\bm s$: instead evolve phase-space cumulants using the BBGKY hierarchy, initialized with the Zel'dovich approximation (ZA). The DCO is then absent by construction, so an EFT of BBGKY can focus on physics beyond mass conservation, which may allow pushing PT beyond shell crossing. The trade-off is the need for a closure relation, for which one can again use the ZA. We provide the building blocks for such a BBGKY+ZA recipe. A bottleneck for implementing it has been the ZA phase-space two-point function $\mathcal{P}$, which we successfully integrate numerically; we then write the higher ZA phase-space correlators needed for closure as products and convolutions of $\mathcal{P}$.
}
\begin{document}
\maketitle

\section{Introduction}\label{sec:introd}

Understanding the evolution of large-scale structure (LSS) in the universe is an important problem in cosmology. Neglecting the complications introduced by baryons, that evolution is governed by cold dark matter (CDM) and, at scales much smaller than the horizon, by the Vlasov--Poisson system. Even in that simplified setting, the analytical study of the statistics of LSS remains an open problem because of non-linearities.

Progress in understanding the mildly nonlinear and nonlinear regimes has relied heavily on $N$-body simulations and on semi-analytical schemes (e.g., COLA \cite{Tassev_Zaldarriaga_Eisenstein_2013,Tassev_Eisenstein_Wandelt_Zaldarriaga_2015}). Analytical approaches remain essential for physical understanding and for fast exploration of cosmological parameter space. Standard Eulerian perturbation theory (EPT\footnote{Often standard perturbation theory in Eulerian space is abbreviated as SPT. Here we choose to highlight its Eulerian nature, and call it EPT.}), Lagrangian perturbation theory (LPT), and their effective-field-theory extensions (EFT) have all led to important insights, and much has been written about what controls their convergence, regime of validity, and eventual breakdown (for example, \cite{Valageas_2007,Bernardeau_2002,Catelan_1995,Matsubara_2008,Tassev_Zaldarriaga_2012,Tassev_2014,Porto_Senatore_Zaldarriaga_2014,Porto_2014,Baumann:2010tm,Carrasco_Hertzberg_Senatore_2012}).

\subsection{Dominated convergence of PT}

In the first part of this paper, rather than asking only which parameter becomes large in perturbative studies of LSS, we ask what precise step in the perturbative calculation actually fails. We restrict our analysis solely to enforcing mass conservation as applied to CDM and disregard any non-linearities from particle dynamics beyond mass conservation. That allows us to focus on one ingredient common to any perturbative  treatment of large-scale structure: the expression of the fractional CDM overdensity $\delta$ in terms of the CDM particle displacement $\bm s$. 

One can understand the functional $\delta[\bm s]$ and its relationship to mass conservation in the language of Particle Mesh $N$-body simulations. In those, particles are displaced from a uniform grid (their Lagrangian positions, $\bm q$) by an amount $\bm s(\bm q)$ (with the time dependence not explicitly shown) to their final positions (their Eulerian coordinates, $\bm x=\bm q+\bm s(\bm q)$). Then one can assign each particle's mass to the Eulerian voxel in 3D it ends up in, in order to recover the Eulerian overdensity $\delta(\bm x)$. This is clearly a mass preserving assignment.  Any expansion of $\delta[\bm s]$ in $\bm s$, \textit{independent} of whether $\bm s$ is the linear or non-linear displacement, is therefore equivalent to making an approximation to mass conservation. It is precisely this expansion that we are going to focus our analysis on.

\paragraph{Eulerian Perturbation Theory.}
The Eulerian overdensity $\delta(\bm x)$ plays the role of the inertial overdensity in the continuity and Euler equations, as well as the active gravitational overdensity in Poisson's equation \cite{Will_2014}. In EPT, all of these equations are expanded in powers of the linear displacement, $\bm s_L(\bm q=\bm x)$. Therefore, mass conservation (both inertial and active gravitational) is enforced only perturbatively. We study when the expansion of $\delta(\bm x)$ in $\bm s(\bm q)$ fails to converge, which results in the failure to satisfy the continuity equation and failure to obtain the correct gravitational potential. To make the analysis transparent, we will do it for a single realization of $\bm s(\bm q)$ as opposed to taking expectation values. Moreover, we will begin by studying that problem in 1D, where each step can be easily traced, and only then move on to 3D.

\begin{figure}[h]
	\centering
	\includegraphics[width=0.7\textwidth]{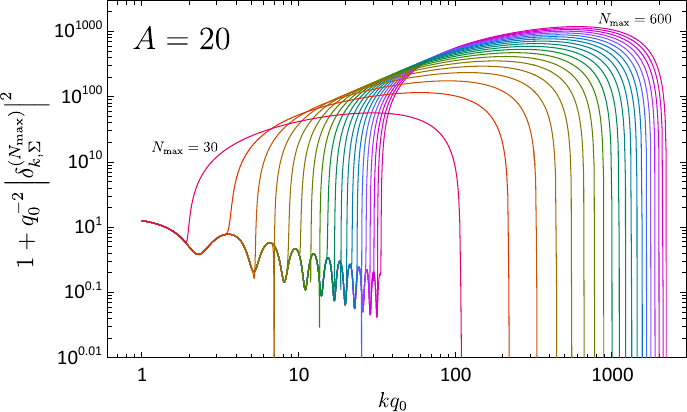}
	\caption{A ``sliding bump'' demonstrating the DC obstruction for a 1D toy model. Finite-order partial sums $\delta_{k,\Sigma}^{(N_{\mathrm{max}})}$ of the Fourier-space expansion of the overdensity $\delta(k)$ in powers of $s$ develop a peak that slides to higher $k$ as the maximum expansion order $N_{\mathrm{max}}$ grows. At any fixed $k$ the pointwise limit is finite: clearly all curves eventually converge to the same answer (oscillating $\delta(k)$ at lower left), but the integral under the bump does not vanish. The order-by-order real-space expansion of $\delta(x)$ exchanges the limit $N_{\mathrm{max}}\to\infty$ with the Fourier integral over $k$, which is precisely the step that Lebesgue's dominated convergence theorem does not permit because of the presence of the sliding bump.  Note that the vertical scale is \textit{logarithmic in the exponent}. Here $q_0$ is a length-scale for this particular toy model and $A$ plays the role of a growth factor. The figure will be revisited in full detail in Section~\ref{sec:approx}.\label{fig:escape}}
\end{figure}

In this paper, we show that in 1D (and later in 3D), the \textit{Fourier-space}\footnote{A wavevector $\bm k$ argument indicates that a quantity is evaluated in Fourier space.} expansion of $\delta(\bm k)$ in powers of $\bm s$ \textit{converges absolutely} for all Eulerian wavevectors $\bm k$ even after shell crossing. A breakdown appears only when one writes down the Eulerian overdensity order by order \textit{in real space}: the order-by-order construction requires exchanging an infinite sum (or a limit) with a Fourier integral over the Eulerian wavevector $\bm k$, and for finite-order partial sums of the expansion for $\delta(\bm x)$ that exchange is not justified. The reason is that the truncated partial sums of $\delta(\bm k)$ (up to some order $N_{\mathrm{max}}$) develop a ``sliding bump'' at high $k$ whose integral does not vanish even as the bump slides off to infinity with increasing $N_{\mathrm{max}}$. That is the textbook example demonstrating the violation of Lebesgue's dominated convergence theorem (DC). Figure~\ref{fig:escape} shows this sliding bump for a 1D toy model (the full discussion is in Section~\ref{sec:approx}).

When considering EPT, the expansion of the real-space $\delta(\bm{x})$ in powers of $s(\bm q)$ is a relevant step to look at since EPT attempts to solve the nonlinear continuity and Euler equations, which contain products of fields evaluated at the same Eulerian point (a composite operator in Quantum Field Theory parlance), say $a(\bm x)b(\bm x)$ (for some $a$ and $b$, for instance the density and the velocity fields in the continuity equation). A product at the same point is a convolution in Fourier space. Expanding each field in powers of the linear overdensity (or equivalently the linear $s$) and keeping terms of total order $N$ (order in $s$ indicated by subscript) in 1D schematically gives (we drop the vector notation in 1D):
\be\label{abN_intro_skel}
(ab)_N(k) \;=\; \sum_{n=0}^{N} \int dk'\, a_n(k')\, b_{N-n}(k-k')\,,
\ee
so any sliding-bump contribution in $a_n$ or $b_{N-n}$, no matter how high in $k$, is convolved into every mode of $(ab)_N$. Therefore, order-by-order cancellations of the bump between different terms cannot occur. Thus, even when working entirely in Fourier space, Eulerian perturbation theory cannot avoid the DC obstruction (DCO) due to the sliding bump.

We demonstrate that the DC failure of EPT in 1D (and similarly in 3D) can be entirely recast in the usual language of a perturbation parameter becoming large, leading to failure of convergence. The relevant parameter is not the linear overdensity $|\delta_L|=|s'|$ alone (with a prime indicating a derivative with respect to $q$) but a family of derivative-displacement combinations of $s$: $|s\,s''|$ (familiar from e.g. \cite{Tassev_Zaldarriaga_2012,Tassev_2014,Porto_Senatore_Zaldarriaga_2014}) is the leading entry, with $|s^2 s'''|$, $|s\,s'\,s''|, \ldots$ entering at higher orders. We trace the origin of all these parameters to the advection of particles from Lagrangian to Eulerian space, and therefore we call them advection parameters. In 3D, these advection terms take on indices, but otherwise the same family controls convergence. As the displacement amplitude is scaled up, the first divergence of $\delta_{x,\Sigma}$ occurs at the $x$ where either $|s'|$ or any one of these advection parameters first reaches $\mathcal{O}(1)$; which parameter and where depends on the shape of $s(q)$. For our 1D toy models in Section~\ref{sec:toy}, that parameter is $|s\,s''|$ at extrema of $s$ (where $|\delta_L|=0$); for other realizations, the first divergence can occur elsewhere and can be determined by other advection terms (see Sections~\ref{sec:saddle}--\ref{sec:saddle1} as well as Appendix~\ref{app:advection-form} for details). What this paper does is identify these previously known advection terms with a DCO appearing when enforcing mass conservation perturbatively by expanding $\delta[\bm s]$.

\paragraph{Lagrangian  Perturbation Theory.}

In Lagrangian PT, often one \textit{does not} expand $\delta(\bm x)$ in $\bm s$ when calculating the density power spectrum. Therefore, mass conservation at the level of the continuity equation is exact (or nearly exact, as most ``large'' terms are kept unexpanded, e.g. \cite{Carlson_2012}). However, that is no longer true at the level of the Poisson equation. In calculating $\bm s$ using the Euler equation, the gravitational acceleration is obtained through Poisson's equation, where the overdensity  (and more precisely the active gravitational overdensity) sources the gravitational potential, $\phi$. Taking the divergence of the Euler equation and combining with the Poisson equation, up to a choice of units we have: $$\left.{\bm \nabla}_{\bm x}\cdot\ddot{\bm s}(\bm q)\right|_{\bm x(\bm q)=\bm q + \bm s(\bm q)}=\left.-\nabla^2_{\bm x}\phi\right|_{\bm x(\bm q)=\bm q + \bm s(\bm q)}=-\delta(\bm x(\bm q))\equiv- \delta_q(\bm q)\ ,$$ where dots indicate derivative with respect to time at fixed $\bm q$. It is on the right-hand side\footnote{The left-hand side of the above equation, can be written as $(\partial q_j/\partial x_i)\partial_{q_j}\ddot {s_i}$, with the first term being the inverse of the deformation tensor. In LPT, the deformation tensor is moved to the right-hand side and expanded together with $\delta_q(\bm q)$. Here we focus only on the expansion of the source $\delta_q(\bm q)$, however.\label{ftnt:lhsLPT}} of the above equation that LPT expands the overdensity $\delta_q(\bm q)$ evaluated in Lagrangian space in orders of $\bm s$. Thus, a failure to converge of that expansion in LPT violates the conservation of the Lagrangian active gravitational mass sourcing the gravitational potential.

Similar to EPT, we demonstrate a DC obstruction underlying the LPT expansion of $\delta_q(\bm q)$. We show that the expansion of $\delta_q(\bm q)$ in powers of $\bm s(\bm q)$ can again be done with an infinite radius of convergence, similar to EPT in Fourier space. However, that expansion does not allow one to separate the individual orders of the overdensity as they are embedded inside an integral. To extract the overdensity order by order, one has to perform the same dominated-convergence-violating exchange as done in EPT. The resulting series can then be ``resummed'' to recover the familiar single-stream (ss) formula $\delta_{q,\mathrm{ss}}(q)=1/|1+s'(q)|-1$. Its validity region $|s'|<1$ is the resummation region of the DC-violating exchanged series; its boundary $|s'|=1$ is the usual shell crossing\footnote{Note that the radius of convergence $|s'(q)|<1$ excludes severely underdense regions ($s'\ge1$, $\delta_L\le-1$), where the LPT series fails to converge. See, for example, footnote 5 of \cite{Porto_Senatore_Zaldarriaga_2014} as well as the toy examples later in our paper.\label{ftnt:single}}, and this time around the advection terms identified in EPT do not make an appearance. 

\paragraph{The DC obstruction and the EFT of LSS.} What we isolate in the first part of the paper is \textit{one} clear failure mode of PT. Further failure modes from solving the full nonlinear dynamics are not ruled out. Indeed, our convergence analysis  makes no attempt to solve for the actual particle trajectories given by $\bm s(\bm q)$ but assumes them as given when investigating the convergence of $\delta[\bm s]$. To reiterate, the DC violation we identify comes purely from mass conservation at different entry points in PT: in all equations in the case of EPT, and in the Poisson equation in LPT. The parameters (the advection terms and/or the linear overdensity) we identify controlling this failure mode of different PTs are not new \cite{Tassev_Zaldarriaga_2012,Tassev_2014,Porto_Senatore_Zaldarriaga_2014}. What we show is their inevitable appearance from a DC violation when enforcing mass conservation perturbatively.

Similar to EPT and LPT, EFTs of LSS (e.g. \cite{Baumann:2010tm,Carrasco_Hertzberg_Senatore_2012,Porto_2014}), whether in Eulerian or Lagrangian space, have to tackle the DC obstruction we identified once structure has grown enough for the sliding bump to develop. Therefore short-scale smoothing in EFTs is already needed just to suppress the sliding bump in $k$ that produces that DCO, independent of whether non-linearities beyond mass conservation are important or not.

An alternative is to never expand $\delta$ in $\bm s$. The second part of the paper turns to such a recipe involving the BBGKY (Bogoliubov--Born--Green--Kirkwood--Yvon) hierarchy \cite{peebles}. The eventual hope is that any EFT of the BBGKY hierarchy can solely focus on curing divergences arising from non-linear physics beyond mass conservation, which may allow pushing the limits of PT beyond shell crossing.

\subsection{Towards BBGKY+ZA}

In the second part of this paper, we turn to a complementary route that never expands $\delta$ in $\bm s$ in the first place. Instead, one evolves the cumulants of the CDM phase-space distribution function $f(\bm x,\bm v,D)$ (with $\bm v$ the Eulerian velocity and $D$ the linear growth factor, used here as time variable), whose equal-time evolution is governed by the BBGKY hierarchy \cite{peebles}. Schematically, the hierarchy relates the $n$-th connected phase-space cumulant both to the $(n{+}1)$-th cumulant and to products of lower cumulants through operators inherited from the Vlasov--Poisson equation. The explicit forms are given in Section~\ref{sec:BBGKY}. The upward coupling to the $(n{+}1)$-th cumulant is the familiar closure problem: the hierarchy requires both a closure prescription and initial conditions.
Following \cite{Tassev_2011} (hereafter HH, for ``Helmholtz Hierarchy''), we take the Zel'dovich approximation (ZA) to supply both. We refer to this combination as BBGKY+ZA: a BBGKY hierarchy truncated with a ZA $(n{+}1)$-point cumulant and initialized with the ZA cumulants up to order $n$.

The dominated-convergence obstruction identified in the first part of this paper is absent from BBGKY+ZA by construction\footnote{Note that BBGKY+ZA does not expand the phase-space distribution directly in powers of the gravitational potential as done in the Introduction of HH -- a perturbative scheme which was eventually implemented as an $N$-body-LPT hybrid simulation in work that led up to the COLA method (see Footnote~1 of \cite{Tassev_Zaldarriaga_Eisenstein_2013}). Such an expansion was also explored in \cite{PhRvD91j3507A} and \cite{Nascimento:2024hle}. Instead of expanding in the potential, the explicit underlying approximation behind BBGKY+ZA is the ZA $n$-point phase-space distribution truncating the BBGKY hierarchy, which could in principle be \textit{further} truncated at a fixed order in the potential (see Section~\textit{Particle interpretation} of HH for further discussion).}, since $\delta$ is never expanded in $\bm s$ (as long as one uses the exact unexpanded ZA cumulants). That is the primary motivation for the approach. It does not imply that BBGKY+ZA is guaranteed to converge, or to converge to the correct answer: other modes of failure from the full nonlinear dynamics may still arise, and we expect EFT-like short-scale regulation to be required even if the hierarchy is truncated at high order.

What this paper adds to BBGKY+ZA is a stable numerical program. A central bottleneck has been the ZA phase-space two-point function $\mathcal{P}$, which, to our knowledge, has not previously been integrated numerically. We reduce $\mathcal{P}$ to a stable numerical summation, and we derive the ZA three-point and higher $n$-point phase-space correlators required for closure beyond the two-point level as products and convolutions of $\mathcal{P}$. The numerical approach we use is to trade cancellations between large values of an oscillating integrand (which requires carefully engineered quadrature) for cancellations between discrete summand terms, which arbitrary-precision arithmetic can control to an arbitrary tolerance.

There is a broader motivation for BBGKY+ZA as well. $N$-body simulations, which similarly do not expand $\delta(\bm x)$ (or the gradient of the gravitational potential) in powers of $\bm s$, evolve CDM beyond shell crossing, but per realization. The statistics of the phase-space distribution are recovered only by averaging over many realizations. BBGKY, by contrast, acts directly on cumulants -- already ensemble-averaged, and therefore smooth functions of their arguments. If BBGKY+ZA can eventually be pushed through shell crossing at the cumulant level, it may give access to the statistics of the post-shell-crossing regime without the per-realization averaging burden of $N$-body simulations. We do not demonstrate that here; what this paper provides is one ingredient toward it -- a stable numerical footing for BBGKY+ZA at the two-point level.

The paper is organized as follows. In Section~\ref{sec:approx} we treat the Lagrangian displacement $\bm s(\bm q)$ as a fixed predetermined realization and study the convergence of the Eulerian overdensity $\delta(\bm x)$ expanded in $\bm s$, working first in 1D and then in 3D. The aim is to identify the precise step in the perturbative calculation at which convergence breaks down. In Section~\ref{sec:prelims} we review the Zel'dovich approximation and rewrite the Vlasov--Poisson system in the phase-space variables used later. In Section~\ref{sec:BBGKY} we derive the BBGKY hierarchy in these variables and make explicit what ZA input is needed for BBGKY+ZA. In Section~\ref{sec:corrZA} we derive the ZA one-, two-, three-, and $n$-point phase-space correlators and obtain a stable numerical representation for $\mathcal{P}$. Section~\ref{sec:summary} summarizes the main lessons and the route forward. The appendices contain a derivative-form expansion of $\delta_{x,\Sigma}$ in 1D, in which each advection parameter appears as the controlling parameter of a separate partial sum, together with a per-partial-sum convergence-radius analysis (Appendix~\ref{app:advection-form}); the derivation of the angular-radial decomposition of $\mathcal{P}$ (Appendix~\ref{app:P}); the calculation of the Zel'dovich overdensity power spectrum ($P_Z$) (Appendix~\ref{app:PZ}); the  asymptotic regimes of $P_Z$ (Appendix~\ref{app:asympt}); the details of the numerical $\mathcal{P}$ pipeline (Appendix~\ref{app:numerics}); and the interpolation-based checks of the FFTLog implementation (Appendix~\ref{app:interp}).

\section{Approximations for the overdensity, and their convergence properties}\label{sec:approx}

In this section, we will study the convergence properties of $\delta$ when expanded in $s(\bm q)$. To simplify our analysis, we will present everything in 1D first. Then we will generalize to 3D, where all 1D arguments transparently follow through (despite the fact that the equations get a bit more cumbersome because of vector indices and contractions).
Here we will not focus on evolving the density as a function of time. Moreover, we will focus on $\delta$ as calculated for a single realization of the displacement field $\bm s(\bm q)$ of particles, instead of ensemble averaging over the stochastic initial conditions.

Along the way we will meet five expressions for the fractional overdensity evaluated in different spaces -- Lagrangian and Eulerian, real and Fourier: $\delta_{k,\Sigma}$, $\delta_{x,\Sigma}$, $\delta_{\mathrm{rec}}$, $\delta_{q,\Sigma}$, and $\delta_{q,\mathrm{ss}}$. They are \textit{formally equivalent} in the sense that they all describe the same overdensity field -- the Eulerian and Lagrangian forms related by the map $x=q+s(q)$, and the three Eulerian (respectively, two Lagrangian) forms matching order by order in $s$ -- even though their convergence properties need not coincide. We will trace how they are related, where each is valid, and where each fails (summarized later in eq.~(\ref{deltaSummary})). The central finding is that the Eulerian real-space expansion $\delta_{x,\Sigma}(x)$ diverges well before shell crossing, even though the underlying Fourier-space expansion $\delta_{k,\Sigma}(k)$ is unconditionally convergent: the obstruction is a \textit{sliding bump} in  $|\delta_{k,\Sigma}^{(N_{\mathrm{max}})}(k)|$ (which is  $\delta_{k,\Sigma}(k)$ expanded to order $N_{\mathrm{max}}$ in $s$) that escapes to higher $k$ as the truncation order $N_{\mathrm{max}}$ is raised, violating the conditions for Lebesgue's dominated convergence. We find an analogous DC obstruction in Lagrangian space as well.

\subsection{Exact solutions for the density in 1D}

Let us consider a 1D universe, which starts out homogeneous (with infinitesimal perturbations). Over time, particles get displaced from an initial (Lagrangian) position $q$  ($q$ can be any number on the real axis) to a final Eulerian position $x$ given by:
\be\label{s1d}
x=q+s(q)\ ,
\ee
where $s$ is also a function of time. Note that we drop the bolding and indexing of vectors in this section, as in 1D vectors can be represented simply by their single real components.

Since the initial density ($\bar \rho$) of the particles is uniform, the final density, $\rho(x)$, is  $\bar\rho$ scaled by the Jacobian of the transformation from $q$ to $x$. 
In the beginning, one $q$ is mapped to one $x$, corresponding to the single-stream solution for CDM. Eventually (at shell crossing), particles from different $q$'s can end up at the same $x$ -- a situation usually referred to as multistreaming. Generally, the fractional overdensity $\delta(x)=(\rho(x)-\bar\rho)/\bar\rho$ can then be written as:
\be\label{deltaX}
1+\delta(x)=\sum\limits_{\substack{{\tilde q} \mathrm{\ such\ that}\\ 
		 x=\tilde{ q}+{s}(\tilde{ q})}}\left|\frac{1}{dx(\tilde q)/d\tilde q}\right|=\sum\limits_{\substack{{\tilde q} \mathrm{\ such\ that}\\ 
		 x=\tilde{ q}+{s}(\tilde{ q})}}\frac{1}{|1+s'(\tilde{ q})|}\ ,
\ee
where the sum is over all streams that started at initial/Lagrangian position $\tilde q$ and ended up at Eulerian position $x$. A prime in $s'(q)$ denotes a derivative with respect to $q$.

In the single-stream regime (one term in the sum), the above equation for the fractional overdensity is the exact solution to the continuity equation written in Eulerian space. Indeed, if we restore the time  ($t$) dependence of $s$, then the velocity in Eulerian space is given by $v(x,t)=\partial s(q,t)/\partial t|_{q=q(x)}$ evaluated at $q$ which solves eq.~(\ref{s1d}) for given $x$. With that identification in mind, direct substitution of $v(x,t)$ and $\delta(x)$ from eq.~(\ref{deltaX}) in the continuity equation leads to an identity. That shows that $\delta(x)$ as given in eq.~(\ref{deltaX})
is the exact solution for the fractional overdensity for which Eulerian perturbation theory is trying to find a good approximation by expanding in orders of  $s$.

Moreover, as discussed in the Introduction, the displacement $s$ is the solution to the Poisson equation (schematically, $\ddot s\propto -\partial_x \phi$), where the acceleration $\ddot s$ (a dot denotes the partial derivative in time at fixed $q$) of a particle is proportional to the negative of the gradient of the gravitational potential, $\phi$. In Lagrangian perturbation theory, one usually takes the divergence (in $x$) of both sides of the Poisson equation, to find (in 1D): $\partial_x \ddot s(q,t)|_{q=q(x)}\propto -\delta(x(q))$. Thus, LPT in the end is trying to find an approximation to $\delta_q(q)\equiv \delta(x(q))$ or is relying on (a perturbative expansion of) that quantity to find an approximation for $s$. We can write $\delta_q(q)$ explicitly as:
\be\label{deltaQinit}
1+\delta_q(q)=\sum\limits_{\substack{{\tilde q} \mathrm{\ such\ that}\\ 
		\tilde{ q}+{s}(\tilde{ q})= q+ s( q)}}\frac{1}{|1+s'(\tilde{ q})|}\ .
\ee

Note that stream crossing occurs when caustics first form in $\delta(x)$, which corresponds to vanishing denominator in eq.~(\ref{deltaX}) (or equivalently, eq.~(\ref{deltaQinit})), which occurs when $s'(q)=-1$. Neither EPT nor LPT can be extended beyond that point. The expressions above seem to imply a  radius of convergence of any expansion  to be $|s'(q)|<1$.  In the following section, we will explore what a perturbative expansion in $s$ looks like for $\delta(x)$ and $\delta_q(q)$, which would correspond to solutions in EPT and LPT respectively.  We will see that in certain cases, the radius of convergence is even smaller than $|s'(q)|<1$, and in other cases, it is in fact infinity.

\subsection{Perturbative solutions for the density in 1D}

Even though it is tempting to directly expand $\delta(x)$ and $\delta_q(q)$ in powers of $s$, such expansions in EPT and LPT always come with an  additional restriction to the single-stream regime. From the narrow perspective of eqs.~(\ref{deltaX}, \ref{deltaQinit}) such a restriction is \textit{independent}\footnote{Indeed, one can picture a realization, where multiple streams overlap,  yet where each has a very small overdensity. For each stream, $|s'|$ can be chosen arbitrarily small, and its respective overdensity can be expanded safely in series in $s'$. The presence of multiple streams would make the total overdensity large and equal to the number of streams minus 1 plus small corrections. Thus, the single stream approximation and an expansion in $s$ can at least in principle be treated as separate approximations.} from an expansion in $s$ and is imposed by truncating the sums to one term by hand which seems ad hoc. To understand that ad hoc restriction better, we will first take a detour through Fourier space, where we will see that expanding in $s$ does \textit{not} imply we are in the single stream regime (see eqs.~\ref{deltaK0} and \ref{deltaQ3}). We will show that the single stream regime restriction arises as a consequence of violating the Lebesgue's dominated convergence conditions. In other words, by going through Fourier space, we will gain a better understanding of the assumptions that go into the  approximations in EPT and LPT.

The fractional overdensity in Eulerian space can be expressed identically as:
\be\label{deltaX1}
1+\delta(x)=\int\limits_{-\infty}^\infty dq \delta_{D}(x-q-s(q))=\int\limits_{-\infty}^\infty dq\int\limits_{-\infty}^\infty \frac{dk}{2\pi}e^{ik(x-q-s(q))}\ ,
\ee
which is valid both in the single- and multi-streaming regimes as the Dirac delta function $\delta_D$ picks up contributions for each $q$ for which its argument is zero. In Lagrangian space, to write $\delta_q(q)$ one can just rename the integration variable $q$ to $q'$, and replace $x$ with $q+s(q)$, see eq.~(\ref{deltaQ2}).

\subsubsection{Expanding the fractional overdensity in (Eulerian) Fourier space\label{sectionF}}

In Fourier space, eq.~(\ref{deltaX1}) gives\footnote{This is consistent with our choice for Fourier transform in Footnote~\ref{ft:fourier}.}:
\be\label{deltaK1}
\delta(k)=\int\limits_{-\infty}^\infty dx e^{-ikx}\delta(x)=-(2\pi)\delta_D(k)+\int\limits_{-\infty}^\infty dx e^{-ikx}\int\limits_{-\infty}^\infty dq \delta_D\big(x-q-s(q)\big)\ ,
\ee
where $k$ is the Fourier wavevector.  When it will not lead to confusion, we distinguish between $\delta$ in real and Fourier space by its argument. When needed, we add $x$ or $k$ as a subscript to $\delta$ to make that difference explicit.

The integrals over $x$ and $q$ in eq.~(\ref{deltaK1}), although written as improper integrals, are in fact over finite (here in 1D; later on in 3D) volumes. First, our Newtonian treatment breaks down as we approach scales comparable to the Hubble horizon, where the relation between Newtonian dynamics and observables requires relativistic care \cite{Chisari_Zaldarriaga_2011}. Second, very large-scale modes have very low amplitude in the real universe and can be treated linearly, or absorbed into a separate-universe description \cite{Wagner_2015}. Over such finite volumes, the second term of the expression above is absolutely converging: the double integral over the magnitude of the integrand reduces to the finite integration volume. Therefore we can exchange the order of the $x$ and $q$ integrations and find:
\be
\delta(k)=\int\limits_{-\infty}^\infty dq e^{-ikq}\left(e^{-iks(q)}-1\right)\ .
\ee

Let us now follow the steps of PT and expand in $s$:
\be
\delta(k)=\int\limits_{-\infty}^\infty dq e^{-ikq}\sum\limits_{n=1}^\infty\frac{(-ik)^n}{n!}s(q)^n\ .
\ee
For a fixed value of $k$, the sum in $n$ is \textit{absolutely} converging as well as \textit{uniformly} converging for any value of $q$ since for physically relevant displacements, $s(q)$ is bounded. Therefore, we are allowed to exchange the sum with the integral in $q$ to find:
\be\label{deltaK0}
\delta(k)=\sum\limits_{n=1}^\infty \int\limits_{-\infty}^\infty  dq e^{-ikq}\frac{(-ik)^n}{n!}s(q)^n\equiv\sum\limits_{n=1}^\infty\delta_{k,n}(k)\equiv \delta_{k,\Sigma}(k)\ .
\ee
The last two equalities define $\delta_{k,n}(k)$ and $\delta_{k,\Sigma}(k)$. The former  is the $n$-th order term of $\delta(k)$ in an expansion in $s$. And the sum $\delta_{k,\Sigma}(k)$ is the sum over $n$ over all $\delta_{k,n}(k)$. Given the steps we took in obtaining eq.~(\ref{deltaK0}), we see that the sum over $\delta_{k,n}$ must be absolutely converging with an \textit{infinite radius of convergence}, i.e. $\delta_{k,\Sigma}$ should match the correct result for $\delta(k)$ even after shell-crossing (we show that explicitly in toy models later on). That may come as a surprise given that in real space the radius of convergence of an expansion of $\delta(x)$ (eq.~(\ref{deltaX})) in $s$ is clearly at most $|s'(q)|<1$. The $s$-expansion itself is not the source of trouble; the trouble must enter when we move back to real space, and we discuss next where exactly that happens.

\subsubsection{Expanding the fractional overdensity in Eulerian space\label{sec:euleriandelta}}

The expansion of the fractional overdensity in Eulerian space  in $s$ is simply the inverse Fourier transform (IFT) of $\delta(k)$ given by eq.~(\ref{deltaK0}):
\be\label{deltax1st}
\delta(x)=\mathrm{IFT}[\delta(k)]=
\mathrm{IFT}\left[\sum\limits_{n=1}^\infty\delta_{k,n}\right]
=\int\limits_{-\infty}^\infty \frac{dk}{2\pi}e^{ikx}\sum\limits_{n=1}^\infty\left( \frac{(-ik)^n}{n!}\int\limits_{-\infty}^\infty  dq e^{-ikq}s(q)^n\right)\ .
\ee

Note that the sum for given $q$ and $x$ is \textit{absolutely} converging but is \textit{non-uniformly} converging in $k$ (for a given large $ks(q)$ value, the series converges only after we pick a correspondingly large $n$). Thus, we \textit{cannot} exchange the order of the summation in $n$ and the integral in $k$ without risking getting wrong results.

But let us pretend the rules of mathematical analysis do not apply to us, and do the exchange of the sum and the $k$-integral.
To flag the unjustified step, we write $A\,\mathclap{\hspace{1.35em}\times}=\,B$ instead of $A=B$. In other words $A\,\mathclap{\hspace{1.35em}\times}=\,B$ is only a formal equation: the two quantities $A$ and $B$ match order by order, but they may have different convergence properties. We find:
\be\label{deltaXn-1}
\delta(x)\mathclap{\hspace{1.35em}\times}=   \sum\limits_{n=1}^\infty \mathrm{IFT}[\delta_{k,n}]= \sum\limits_{n=1}^\infty\frac{1}{n!}\left(
\int\limits_{-\infty}^\infty \frac{dk}{2\pi}e^{ikx}(-ik)^n\int\limits_{-\infty}^\infty  dq e^{-ikq}s(q)^n\right)\ .
\ee
The quantity in parenthesis is nothing but the integral over $k$ of the Fourier transform of the $n$-th derivative of the Dirac delta function multiplied by the Fourier transform of $s(q)^n$. Using the theory of distributions, and the definition of the Fourier transform of a distribution, we obtain:
\be\label{deltaXn1}
\delta(x)\mathclap{\hspace{1.35em}\times}= \sum\limits_{n=1}^\infty \mathrm{IFT}[\delta_{k,n}]= \sum\limits_{n=1}^\infty\frac{\partial_x^n}{n!} \left(-s(x)\right)^n\equiv \sum\limits_{n=1}^\infty \delta_{x,n}(x)\equiv \delta_{x,\Sigma}(x)\ .
\ee
Here we assumed that $s(q)$ is infinitely differentiable.

The last two equalities in eq.~(\ref{deltaXn1}) define the $n$-th order fractional overdensities in Eulerian space, $\delta_{x,n}(x)$,  as well as their sum (if it converges), which we denote as $\delta_{x,\Sigma}(x)$. To avoid any confusion, let us state explicitly that in eq.~(\ref{deltaXn1}) by $\partial_x^n (-s(x))^n $ we mean $\partial_q^n (-s(q))^n|_{q=x} $, as opposed to derivatives in $x$ of powers of $s(q(x))$.

It may seem reasonable that we could have arrived at the expression above by just taking the shortcut of inverse Fourier transforming each individual $\delta_{k,n}(k)$ in the sum given by eq.~(\ref{deltaK0}). However, we took the steps above to better understand the underlying assumptions behind such a shortcut. Namely, the exchange of the inverse-Fourier integral in $k$ with the infinite sum in $n$ that produced $\delta_{x,\Sigma}$ is not justified, because the partial sums converge non-uniformly in $k$. The resulting $\delta_{x,\Sigma}(x)$ therefore need not equal $\delta(x)$; we will see in toy models that it can indeed diverge well before shell crossing.

\subsubsection{The recurrence form $\delta_{\mathrm{rec}}$ and the single-stream limit $\delta_{x,\mathrm{ss}}$ in Eulerian space\label{sec:recdelta}}

To understand better the above result, one can explicitly check\footnote{\label{ftnt:rec}By expansion in $s$, we imply the following procedure, which is standard in perturbation theory. Every $s$, $s'$ and its higher derivatives are multiplied by an ordering parameter $\epsilon$, and the expansion is done in $\epsilon$, which is then set to $\epsilon=1$. We checked eq.~(\ref{deltaXn2}) explicitly to 14-th order in $s$ (or rather, $\epsilon$). We checked the 3D counterpart, eq.~(\ref{rec3d}), to 7-th order. As the coefficients of these expansions are non-trivial, we expect these equalities to hold to all orders, and will not be surprised if one can prove that by induction. However, at this point, the author's interest in proving those equalities to all orders has faded.} that order-by-order the  expansion in eq.~(\ref{deltaXn1})  (plus 1) equals:
\be\label{deltaXn2}
1+\sum\limits_{n=1}^\infty\frac{\partial_x^n}{n!} \left(-s(x)\right)^n \mathclap{\hspace{1.35em}\times}= \frac{1}{1+s'(x-s(x-s(x-s(\cdots))))}\equiv 1+\delta_{\textrm{rec}}(x)\ ,
\ee
where again we use $\mathclap{\hspace{1.35em}\times}=$ in the first equality, as the two sides of that equation can clearly end up producing different results because they have different convergence properties. 
The last equality defines $\delta_{\textrm{rec}}(x)$, where ``rec'' stands for ``recurrence'' (see below). Note that $'$ in the equation above denotes derivative in $q$, i.e. the denominator equals $1+s'(q)|_{q=x-s(x-s(\cdots))}$.

Even at this stage, before any specific toy model is fixed, eq.~(\ref{deltaXn2}) already shows that the expansion parameter is not captured by $|s'|$ alone. Each term in the sum on the left is built from $\partial_x^n\, s(x)^n$, which generates combinations of $s$ and its higher derivatives up to order $n$. Controlling the sum therefore requires the joint smallness of products such as $|s\,s''|$, $|s^2 s'''|$, and their higher-order analogues, not only of $|s'|$. The same structure is behind the $|s(x)\delta_L'(x)|$ criterion that will appear in the Lorentzian toy model below (see eq.~(\ref{limitToy1}) and the discussion that follows). We will explore these parameters in detail in Section~\ref{sec:saddle1} and Appendix~\ref{app:advection-form}.

The convergence requirements of both sides of the first equality in eq.~(\ref{deltaXn2}) need not be the same. Even though $\delta_{\mathrm{rec}}$ may seem more ``resummed'', it could in fact be more restricted in validity than $\delta_{x,\Sigma} $, eq.~(\ref{deltaXn1}), as it assumes that $s(q)$ is analytic (and therefore it matches its Taylor expansion) at least over the domain it is evaluated at in the equation above\footnote{In all three toy models below we find that $\delta_{\mathrm{rec}}$ does actually converge over a larger domain than $\delta_{x,\Sigma}$. See the next section for details.}. Moreover, the convergence of $\delta_{\mathrm{rec}}$ depends on the convergence to a fixed point of the infinitely-many nested functions entering in its definition. We discuss those below.

The expression $x-s(x-s(x-s(\cdots)))$ in eq.~(\ref{deltaXn2}) attempts to solve  eq.~(\ref{s1d}) recursively for $q(x)$. To see that, one has to realize that the expression equals the expanded form of $Q_\infty(x)$ which is the limit $n\to\infty$ of the recurrence relation
\be\label{its1}
Q_{n+1}(x)=x-s(Q_n(x)), \textrm{with } Q_0(x)=x\ ,
\ee
with $x$ being treated as a parameter. Indeed, the above recurrence converges to the fixed point $\lim_{n\to\infty}Q_n(x=q+s(q))=q$ for all $x$ as long as\footnote{
	It is manifest that the value of $q$ which solves $q+s(q)=x $ is a fixed point  for the recurrence relation, eq.~(\ref{its1}) .  Now, let us check if it is a stable point. We can perturb around $q$ by adding $\Delta q$ to it. We find $Q_{n+1}(q+\Delta q)=x-s(Q_n(q+\Delta q))\approx x-s(Q_n(q))-\Delta q s'(Q_n(q))= x-s(q)-\Delta q s'(q)=q-\Delta q s'(q)$, where we expanded to linear order in $\Delta x$. Iterating again, we get: $Q_{n+2}(q+\Delta q)= x-s(Q_{n+1}(q+\Delta q))\approx x-s(q-\Delta q s'(q))\approx x-s(q)+\Delta q s'(q)^2$. Repeating these iterations $m$ times, we find that $Q_{n+m}$ receives a correction of $(-1)^m \Delta x s'(q)^m$, which converges to zero with increasing $m$ only if $|s'(q)|<1$. If that inequality holds, then there is a basin of attraction around $q$, which will converge to $q$ for a given initial condition. Notice, that this inequality also implies that we are working in the single stream approximation. Therefore, $x=q+s(q)$ has only one solution $q$ for each $x$, and thus only one fixed point for each $x$. } 
$|s'(q)|<1$ for all $q$, and as long as the basin of attraction of each fixed point $q$ is large enough to include the corresponding initial condition $ Q_0(x)=x$.
%(which is relegated to infinite order in $s$ if the Taylor expansion of the right-hand side of eq.~(\ref{deltaXn2}) exists). 
 Note that $|s'(q)|<1$ implies that we are in the single-stream regime. 
So, the recurrence solution for $q(x)$ above converges to the correct solution for single-stream solutions as long as $|s'(q)|<1$, and as long as the fixed points have sufficiently large basins of attraction. If the basins of attraction are not large enough or $|s'(q)|\geq 1$, then the recurrence relation above can converge to the wrong $q$, or to a limit cycle, or go into the chaotic regime (similar to the Logistic map \cite{1964Tell...16....1L}).

Combining eqs.~(\ref{deltaXn1}), (\ref{deltaXn2}) and (\ref{its1}) we can see that   $\delta_{\mathrm{rec}}$ is valid only in the single-stream regime. Therefore, at least under certain conditions (see below) it should equal the exact single-stream solution $\delta_{x,\mathrm{ss}}(x)$:
\be\label{deltaXnF}
\delta(x)\mathclap{\hspace{1.35em}\times}= \delta_{\mathrm{rec}}\mathclap{\hspace{1.35em}\times}= -1+\frac{1}{1+s'(q(x))}\ \equiv \delta_{x,\textrm{ss}}(x)\ ,
\ee
where $\delta_{x,\mathrm{ss}}$ makes sense only in the single-stream regime (hence no need for absolute values in the denominator), when it reproduces the exact result. Note that $\delta_{\mathrm{rec}}$ does not need to have the same convergence properties as $\delta_{x,\textrm{ss}}$, because the former relies on large enough basin of attraction of the fixed points of eq.~(\ref{its1}). Thus, we use $\mathclap{\hspace{1.35em}\times}=$ to relate them.

\subsubsection{Forward and backward Eulerian relations.}

The chain of approximations above can be packaged into two formal sequences of equalities. The forward direction, $s$-expansion of $\delta(x)$,
\be\label{forwardDeltaX}
\delta(x)\mathclap{\hspace{1.35em}\times}=\delta_{x,\Sigma}(x)\ ,
\ee
holds when: the integral in $k$ commutes with the sum in $n$, $s(q)$ is infinitely differentiable, and the sum converges. The first and third conditions both fail well before shell crossing in the toy models of Section~\ref{sec:toy}. 

The backward direction, working from the closed-form single-stream answer through the recurrence (eqs.~\ref{deltaXnF}, \ref{deltaXn2}, \ref{deltaXn1}), is
\be\label{backwardDeltaX}
\textrm{In the single-stream regime:  }\delta(x)=\delta_{x,\mathrm{ss}}(x)\mathclap{\hspace{1.35em}\times}=\delta_{\textrm{rec}}(x)\mathclap{\hspace{1.35em}\times}=\delta_{x,\Sigma}(x)\ ,
\ee
where the first $\mathclap{\hspace{1.35em}\times}=$ assumes the recurrence (\ref{its1}) converges to the correct root, and the second additionally assumes that $s(q)$ is analytic over the relevant domain and that the sum in $n$ converges. Next we explore the convergence of the fractional overdensity in Lagrangian space.

\subsubsection{Expanding the fractional overdensity in Lagrangian space}

Next, we focus on the fractional overdensity, $\delta_q(q)\equiv \delta_x(x=q+s(q))$, evaluated in Lagrangian space, see eq.~(\ref{deltaQinit}). We will follow a procedure parallel to the previous subsection to expand $\delta_q(q)$ in a series in $s$.

The density in Lagrangian space can be written as:
\be\label{deltaQ2}
1+\delta_q(q)=\sum\limits_{\substack{{\tilde q} \mathrm{\ such\ that}\\ 
		\tilde{ q}+{s}(\tilde{ q})= q+ s( q)}}\frac{1}{|1+s'(\tilde{ q})|}=\int\limits_{-\infty}^\infty dq'\int\limits_{-\infty}^\infty \frac{dk}{2\pi}e^{ik(q+s(q)-q'-s(q'))}\ ,
\ee
where the sum is over multiple streams in the multistreaming regime. Now let us check what an expansion in $s$ would give us:
\be\label{deltaQ3}
\delta_q(q)=\int\limits_{-\infty}^\infty dq'{\int\limits_{-\infty}^\infty\frac{dk}{2\pi}}e^{ik(q-q')}{\sum\limits_{n=1}^\infty} \frac{(ik)^n}{n!}\big(s(q)-s(q')\big)^n\ .
\ee
The expansion of the exponential function has an infinite radius of convergence, so the above equalities are exact and are valid in the multi-stream regime as well. 

Similar to the previous section, for fixed $q$ and $q'$, the sum is \textit{absolutely} but \textit{non-uniformly} converging as the maximum $n$ we need for convergence is proportional to $k$. Thus, we are \textit{not allowed} to exchange the sum and the integral in $k$ without risking getting a nonsensical result. Yet, in what follows, we will do that anyway to see where that would lead us. 

After the exchange of the sum and the integral over $k$, performing the integrals in $k$ for each $n$ is equivalent to the following procedure: We promote the variable $q$ entering $e^{ikq}$ to be a new variable $\tilde q$, which should be evaluated at $\tilde q=q$. We can then write $(ik)^n e^{ik\tilde q}=\partial_{\tilde q}^ne^{ik\tilde q}$ to find:
\be
\delta_q(q)\mathclap{\hspace{1.35em}\times}= \int\limits_{-\infty}^\infty dq' {\sum\limits_{n=1}^\infty} \frac{\partial_{\tilde q}^n}{n!}{\int\limits_{-\infty}^\infty\frac{dk}{2\pi}}e^{ik(\tilde q-q')}\big(s(q)-s(q')\big)^n\Big|_{\tilde q=q}\nonumber\\
\mathclap{\hspace{1.35em}\times}= \sum\limits_{n=1}^\infty \int\limits_{-\infty}^\infty dq' \left(\frac{\partial_{\tilde q}^n}{n!}\delta_D(\tilde q-q')\right)\big(s( q)-s(q')\big)^n\Big|_{\tilde q=q}\nonumber\\
=\sum\limits_{n=1}^\infty \frac{\partial_{\tilde q}^n}{n!} \big(s(q)-s(\tilde q)\big)^n\Big|_{\tilde q=q}\label{deltaQ1}\ .
\ee
To write down the second line, we assumed infinitely differentiable $s(q)$, and an absolutely converging integrand, so that exchanging the sum and the integral in $q'$ is allowed, which is again a questionable step. But that exchange is necessary if we are to make sense of the derivative of the delta function.

Order by order, the last line of eq.~(\ref{deltaQ1}) corresponds to the Taylor expansion in $s$ of $1/(1+s'(q))$, i.e. all derivatives of $s(q)$ with respect to $q$ beyond the first derivative cancel in eq.~(\ref{deltaQ1}) at each order as long as $s(q)$ is an analytic function in a domain around $q$. To see that explicitly, let us expand  $s(\tilde q)$ around $\tilde q=q$, so that $s(q)-s(\tilde q)= (q-\tilde q)s'(q)-(q-\tilde q)^2s''(q)/2+(q-\tilde q)^3s'''(q)/6-\dots$. Raising that last quantity to power $n$ in eq.~(\ref{deltaQ1}), and then taking the $n$-th derivative in $\tilde q$, and then setting $\tilde q=q$, we see that only the term proportional to $s'(q)$ in the Taylor expansion has a non-zero contribution. Thus:
\be\label{DELTAQF}
\delta_q(q)\mathclap{\hspace{1.35em}\times}= \sum\limits_{n=1}^\infty \frac{\partial_{\tilde q}^n}{n!} \big(s(q)-s(\tilde q)\big)^n\Big|_{\tilde q=q}=\sum\limits_{n=1}^\infty \frac{\partial_{\tilde q}^n}{n!} \big((q-\tilde q)s'(q)\big)^n\Big|_{\tilde q=q}=\\
=\sum\limits_{n=1}^\infty \big(-s'(q)\big)^n\equiv \sum\limits_{n=1}^\infty \delta_{q,n}(q)\equiv \delta_{q,\Sigma}(q)\ ,
\ee
where the last two equalities define the $n-$th order fractional overdensities, $\delta_{q,n}$, in Lagrangian space, as well as their sum (if it converges) as $\delta_{q,\Sigma}(q)$.

The sum above has a radius of convergence given by $|s'(q)|<1$, and therefore, it can only be applied in the single stream regime (see Footnote~\ref{ftnt:single}). One, however, could be tempted to perform the sum and extend it beyond its region of validity and write\footnote{Note that we introduced by hand the expected absolute value in the denominator in eq.~(\ref{LPTstart}), even though a ``pure'' resummation of eq.~(\ref{DELTAQF}) does not imply it. That absolute value matters only outside the radius of convergence, and therefore does not affect the single stream regime itself.}:
\be\label{LPTstart}
\delta_q(q)\mathclap{\hspace{1.35em}\times}=\frac{1}{|1+s'(q)|}-1\equiv\delta_{q,\mathrm{ss}}(q)\ ,
\ee
where the last equality defines the fractional overdensity in the single-stream regime.  Clearly, the above equality is exact in the single stream regime and breaks down after shell crossing. 
This compact single-stream expression is usually taken as the starting point in standard LPT calculations \cite{Bernardeau_2002,Catelan_1995,Matsubara_2008} as sourcing the gravitational potential in the Poisson equation.

One may wonder why, in the above calculations, we took the route through the Fourier-space representation of the Dirac delta function to obtain $\delta_{q,\Sigma}(q)$  when we could have just expanded $\delta_{q,\mathrm{ss}}$. Indeed, by expanding $\delta_{q,\mathrm{ss}}$ directly in $s$, it is immediately clear that $\delta_{q,\Sigma}(q)$ has a finite convergence radius. The reason we do not do that is that starting from $\delta_{q,\mathrm{ss}}$ assumes an ad hoc truncation of the sum over all streams in the expression for the overdensity. 

Moreover, when we go through Fourier space, we start with  eq.~(\ref{deltaQ3}), which is an absolutely converging expansion in $s$ of infinite radius of convergence. That stark difference in the convergence radii may seem like a contradiction. That is, until we realize that the reason we do end up with a finite radius of convergence when going through Fourier space is \textit{not} the expansion in $s$, but rather, the steps we took in eq.~(\ref{deltaQ1}) and those leading up to it, which involve exchanging improper integrals and infinite sums when those exchanges are not allowed. If confusing, the toy models of the next section will cast more light on these statements.

To summarize, the compact expression in eq.~(\ref{LPTstart}) is therefore somewhat deceptive: the final formula is simple, but getting to it from the exact Fourier representation requires a sequence of manipulations that are allowed only when the conditions for applying Lebesgue's dominated convergence theorem are satisfied. The finite radius of convergence $|s'|<1$ of $\delta_{q,\Sigma}=\sum(-s')^n$ is thus inherited not from the $s$-expansion of eq.~(\ref{deltaQ3}) (which has infinite radius of convergence) but from the same DC-violating exchange of sum and Fourier integral; the familiar single-stream formula $1/|1+s'|-1$ is the resummation of a DC-violating series, with shell crossing at its boundary. Clearly, one must take extreme care when ``resumming'' blindly in PT.

\subsubsection{Summary\label{sec:summDelta}}

Let us summarize the different approximations   for the fractional overdensity presented until now, namely $\delta_{k,\Sigma}$ (eq.~\ref{deltaK0}), $\delta_{x,\Sigma}$ (eq.~\ref{deltaXn1}), $\delta_{\mathrm{rec}}$ (eq.~\ref{deltaXn2}), $\delta_{q,\Sigma}(q)$ (eq.~\ref{DELTAQF}), $\delta_{q,\mathrm{ss}}$ (eq.~\ref{LPTstart}):

\begin{equation}
	\label{deltaSummary}
	\renewcommand{\arraystretch}{1.15}
	\begin{array}{@{}>{\displaystyle}r@{}>{\displaystyle}c@{}>{\displaystyle}l@{\qquad}P{0.34\linewidth}@{}}
		\delta_{k,\Sigma}(k) &=&
		\sum\limits_{n=1}^\infty \delta_{k,n}(k)
		= \sum\limits_{n=1}^\infty \int\limits_{-\infty}^\infty dq\, e^{-ikq}\frac{(-ik)^n}{n!}s(q)^n
		&
		\textrm{Exact}
		\\[0.7ex]
		\delta_{x,\Sigma}(x) &=&
		\sum\limits_{n=1}^\infty \delta_{x,n}(x)
		= \sum\limits_{n=1}^\infty \frac{\partial_x^n}{n!}\left(-s(x)\right)^n
		&
		\textrm{EPT expansion}\newline
		\textrm{(Integrable dominating function, i.e.\ no DC violation; $C^\infty$; convergent sum)}
		\\[0.7ex]
		\delta_{\mathrm{rec}}(x) &=&
		-1+\frac{1}{1+s'(x-s(x-s(x-s(\cdots))))}
		&
		\textrm{Recursion expansion of $\delta_{x,\mathrm{ss}}$}\newline
		\textrm{(Single stream; $|s'(q)|<1$, which excludes severely underdense regions; large enough basin of attraction)}
		\\[0.7ex]
		\delta_{q,\Sigma}(q) &=&
		\sum\limits_{n=1}^\infty \bigl(-s'(q)\bigr)^n
		&
		\textrm{LPT expansion\newline(Single stream; $|s'(q)|<1$, which excludes severely underdense regions; no DC violation)}
		\\[0.7ex]
		\delta_{q,\mathrm{ss}}(q) &=&
		\frac{1}{|1+s'(q)|}-1
		&
		\textrm{Single stream LPT}
	\end{array}
\end{equation}
In parenthesis we list the assumptions made to get to the approximation for the fractional overdensity on the corresponding line (see the previous section for details). ``Single stream'' implies $s'(q)>-1$ as well as no overlapping streams at a given $q(x)$.  For the full list of assumptions required for $\delta_{x,\Sigma}$ see ca. eqs.~(\ref{forwardDeltaX}) and (\ref{backwardDeltaX}). Among the quantities\footnote{The reason we include $\delta_{\mathrm{rec}}(x)$ is curiosity. We would like to see how its radius of convergence compares with that of $\delta_{x,\Sigma}(x)$ and to see whether $|s'(q)|<1$ is a sufficient condition for its convergence in the toy models we list. } listed above, we expect only $\delta_{k,\Sigma}(k)$ to converge to the correct answer, while we expect all the rest to fail at least sometimes for the reasons outlined in the previous sections and summarized above. 

For our discussion in the next section it would be helpful to write down the relationship between $\delta_{x,\Sigma}$, $\delta(x)$ and $\delta_{k,\Sigma}$ in the form below. 
We start by writing eq.~(\ref{deltax1st}) as a limit:
\be\label{ksigmaIssue2}
\delta(x)=\int\limits_{-\infty}^\infty \frac{dk}{2\pi}e^{ikx}\left( \lim\limits_{N_{\mathrm{max}}\to\infty}\delta_{k,\Sigma}^{(N_{\mathrm{max}})}(k)\right)=\mathrm{IFT}\left[\lim\limits_{N_{\mathrm{max}}\to\infty}\delta_{k,\Sigma}^{(N_{\mathrm{max}})}\right]\ ,
\ee
where we used eq.~(\ref{deltaK0}). Above,  we used the partial sum:
\be\label{defSigmaNmax}
\delta_{k,\Sigma}^{(N_{\mathrm{max}})}(k) \equiv \sum\limits_{n=1}^{N_{\mathrm{max}}}\delta_{k,n}(k)\ .
\ee

Similarly, we can write eq.~(\ref{deltaXn1}) as the limit of a partial sum:
\be\label{tmpA}
\delta_{x,\Sigma}=\lim\limits_{N_{\mathrm{max}}\to\infty}\left(\sum\limits_{n=1}^{N_{\mathrm{max}}}\mathrm{IFT}[\delta_{k,n}]\right)\ .
\ee
In making the transition from eq.~(\ref{deltax1st}) to eq.~(\ref{deltaXn-1}), we argued that swapping the integral in $k$ and the \textit{infinite} sum in $n$ is not allowed as the sum is non-uniformly convergent. For the \textit{finite} sum inside the limit in eq.~(\ref{tmpA}), that swap is allowed assuming both $\delta_{k,n}$ and its IFT, $\delta_{x,n}$, exist (which they should for physically relevant $s(q)$; see for example the toy models in the next section). We obtain:
\be\label{ksigmaIssue1}
\delta_{x,\Sigma}(x)=\lim\limits_{N_{\mathrm{max}}\to\infty}\left( \int\limits_{-\infty}^\infty \frac{dk}{2\pi}e^{ikx}\delta_{k,\Sigma}^{(N_{\mathrm{max}})}(k)\right)=\lim\limits_{N_{\mathrm{max}}\to\infty}\left(\mathrm{IFT}\left[\delta_{k,\Sigma}^{(N_{\mathrm{max}})}\right]\right)\ .
\ee
Comparing equations~(\ref{ksigmaIssue1}) and (\ref{ksigmaIssue2}), we see that the exchange of the order of the limit and the integral corresponding to the IFT must be the culprit\footnote{This is just a restatement of our previous result, that the failure of $\delta_{x,\Sigma}$ stems from the non-uniform convergence of the sum in eq.~(\ref{deltax1st}) which prevented us from exchanging the integral in $k$ and the infinite sum in $n$ to obtain eq.~(\ref{deltaXn-1}).} behind any failure of $\delta_{x,\Sigma}$ to converge to $\delta(x)$. 

A similar failure occurs when doing a DC-violating exchange in the expansion of  $\delta_q(q)$ between eq.~(\ref{deltaQ3}) and eq.~(\ref{deltaQ1}). In a way, that exchange is less transparent in LPT than in EPT, since in Lagrangian Fourier space, the presence of $s(q)$ in eq.~(\ref{deltaQ3}) makes it much more complicated to separate the orders of $\delta_q$ into absolutely converging Fourier order-by-order terms, the way we did it for the Eulerian Fourier-space $\delta(k)$ as in eq.~(\ref{defSigmaNmax}). To make the DC violations more transparent, we therefore explore those EPT and LPT failures in the context of three toy models in the next section.

\subsection{1D Toy Models}\label{sec:toy}

In this section we use toy models to explore the convergence properties of the different approximations for the fractional overdensity presented in eq.~(\ref{deltaSummary}). We consider three choices of $s(q)$ -- a Lorentzian, a Gaussian, and a Cosine -- chosen so that the relevant order-by-order sums can be tracked mostly analytically to high order, and we use them to check explicitly the central claim of the previous section: that $\delta_{k,\Sigma}$ converges while the real-space $\delta_{x,\Sigma}$ breaks down well before shell crossing.

\subsubsection{Toy model 1: Lorentzian displacement}

Here we are going to explore the quantities in eq.~(\ref{deltaSummary}) in the context of a 1-dimensional toy model, with a displacement given by the following Lorentzian:
\be\label{1dmodel}
 s( q)=-\frac{A\, q_0}{3+\left(\frac{q}{q_0}\right)^2}\ ,
\ee
with $A>0$ being a dimensionless parameter scaling the overall displacement, and $q_0>0$ -- a parameter having units of length. Note that before shell crossing, in 1D, the ZA solution is an exact solution to the equations of motion, and thus $A$ can be identified with the growth factor. However, here we will explore the convergence of the different approximations of the overdensity both before and after shell crossing. In the latter regime, the identification of $A$ with the growth factor is no longer valid. So, we keep $A$ as just a parameter.

\begin{figure}[t!]
	\centering
	\includegraphics{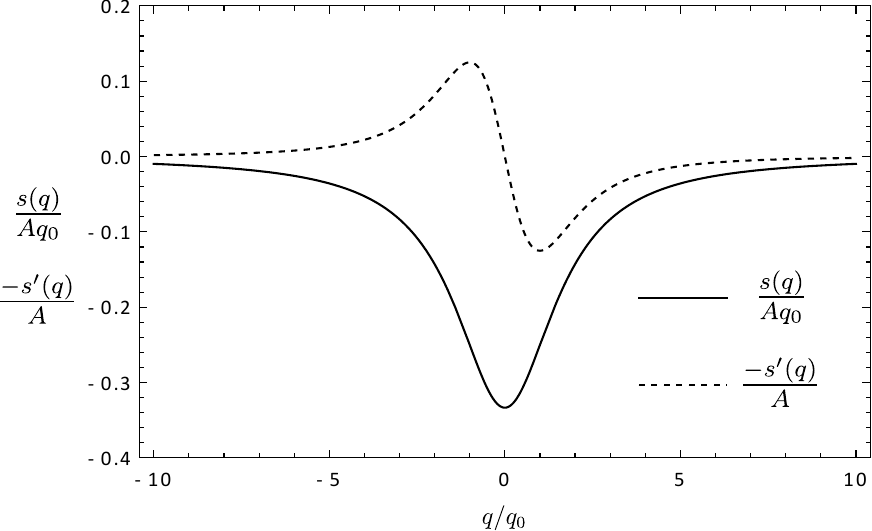}
	\caption{Here we show the Lorentzian displacement toy model $s(q)$ as well as the resulting linear fractional overdensity $\delta_L=-s'(q)$ scaled with the linear growth factor $A$; and in the case of $s(q)$, divided by $q_0$ to render the displacement dimensionless. Note that as the displacement is negative, mass is being advected from right to left, causing an underdensity at positive $q$, and an overdensity for $q<0$ (at first order). The minimum  of $s(q)$ at $q=0$ is where Eulerian perturbation theory (in real space) first breaks down. }
	\label{fig:S}
\end{figure}

The reason we start with the model above is that it allows for closed-form solutions for many quantities of interest. Later on, we will explore a Gaussian model, as well as a Cosine model of $s(q)$. 

\begin{figure}[t!]
	\centering
	\includegraphics{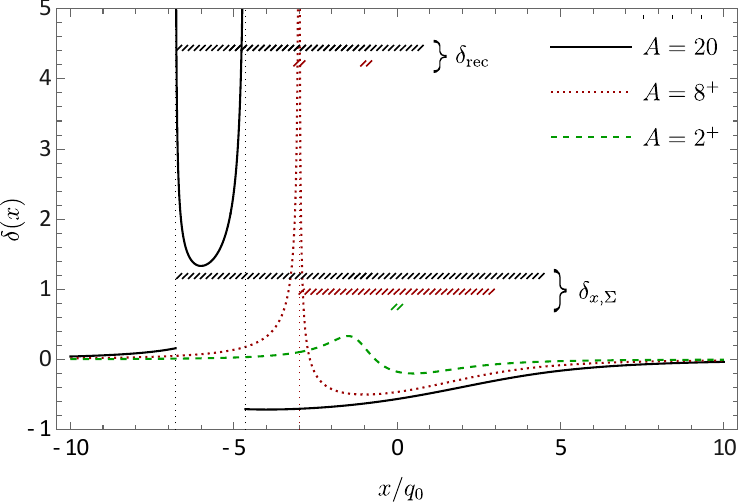}
	\caption{The figure shows the exact solution to the non-linear fractional overdensity in Eulerian space corresponding to the Lorentzian displacement toy model. The overdensity is shown for three different parameter values $A=2^+$ (right after Eulerian perturbation theory breaks down), $A=8^+$ (right after first stream crossing) and $A=20$ (well after first stream-crossing). One can see the appearance of a caustic for $A=8$ at $x=-3q_0$ (red dotted vertical asymptote), which is where stream crossing first occurs. For $A=20$, we have three streams overlapping in the domain between the two caustics  (black dotted vertical asymptotes) around $x/q_0\approx -6$. Outside of that interval, the model produces single streams. The hatched intervals in $x/q_0$  correspond to locations where different approximations for $\delta(x)$, namely $\delta_{\mathrm{rec}}(x)$ and $\delta_{x,\Sigma}(x)$ break down either because they fail to converge, or because they converge to the wrong value. The overdensity $\delta_{\mathrm{rec}}(x)$ fails to converge to the correct value for $A=20$ (top hatched interval) and $A=8^+$ (bottom hatched interval) in intervals where either $|s'(q(x))|\geq1$ or when we have multiple overlapping streams. And indeed, note that $s'\geq 1$ corresponds to $\delta\leq-0.5$ which corresponds to locations  covered by the hatched intervals; and $s'=-1$ gives rise to caustics -- another clear set of breakdown locations. The overdensity $\delta_{\Sigma}(x)$ fails to converge to the correct value for all three $A$'s in the hatched intervals indicated (arranged top-to-bottom in $A$ in the same order as the legend). Those are intervals where the assumptions behind Lebesgue's Dominated Convergence Theorem are violated. See the text for further discussion.}
	\label{fig:deltaX}
\end{figure}

For the Lorentzian model in eq.~(\ref{1dmodel}), the displacement and associated linear density $\delta_L\equiv\delta_{x,n=1}=-s'(q=x)$ are shown in Fig.~\ref{fig:S} as a reference. One can see that the negative displacement leads to mass being advected from positive $q$ to negative $q$. Thus, in the linear regime,  the model develops an underdensity for $q>0$, and an overdensity for $q<0$. The non-linear (exact) density, eq.~(\ref{deltaX}), is shown in Fig.~\ref{fig:deltaX} for different values of $A$. For $A=2$, which is a value of $A$ for which there is no shell crossing, one can see that the exact density behaves similarly to the linear density -- with an overdensity on the left of an underdense region. For $A=8$ and $A=20$, the model leads to shell crossing. We turn to that regime next.

The Eulerian position $x$ is given by eq.~(\ref{s1d}). Inverting that equation to solve for $q$ using eq.~(\ref{1dmodel}), we obtain a cubic polynomial for $q$, which has closed-form solutions. When there is one real root for $q$, the model is in the single-stream regime. When for a given $x$, all three roots for $q$ are real, the displacement model above leads to shell crossing and then three overlapping streams of particles coexist at that $x$.

At small $A$, the displacement leads to a single stream for all $x$. As $A$ increases, the displacement field eventually leads to stream crossing. It occurs when the exact fractional overdensity becomes infinite (a caustic forms), which in turn happens when the linear fractional overdensity, $\delta_L$, reaches $\delta_{L,\mathrm{sc}}= -s'(q_{\mathrm{sc}})=1$, see eq.~(\ref{deltaX}). Here ``sc'' denotes values at first shell-crossing -- ``first'' meaning at smallest value of $A$ resulting in shell-crossing (remembering that $A$ until then can be treated as the growth factor).

For our toy model, shell crossing ($\delta_{L}=1$) first occurs when $A$ reaches a value of $A_{\mathrm{sc}}=8$ (see Fig.~\ref{fig:deltaX}). The location of stream crossing for that $A$ is at $q_{\mathrm{sc}}=-q_0$, which corresponds to $x_{\mathrm{sc}}\equiv-3q_0$. For larger $A$, shell crossing occurs at two Eulerian positions $x$ which correspond to the two real roots of the polynomial $p_{\mathrm{sc}}(x)\equiv 12 (x^2+3q_0^2)^2 + 4 A x q_0 ( x^2+27q_0^2)+27 A^2q_0^4$. Let us denote those\footnote{Since $p_{\mathrm{sc}}(x)$ is a quartic polynomial,  closed-form expressions exist for $X_1$ and $X_2$ but we do not find them particularly illuminating, and so do not include them here. Those expressions, however, were used in constructing the figures accompanying the text.} with $X_1$ and $X_2$, with $X_1<X_2$. Thus, for $x<X_1$ and $x>X_2$, the displacement given by eq.~(\ref{1dmodel}) leads to single streams  (a single real solution for $q$ of $x=q+s(q)$), whereas for $X_1<x<X_2$, three streams overlap (all three solutions for $q$ to $x=q+s(q)$ are real). In Fig.~\ref{fig:deltaX} that can be seen in the graph for $A=20$.

\begin{figure}[t!]
	\centering
	\includegraphics{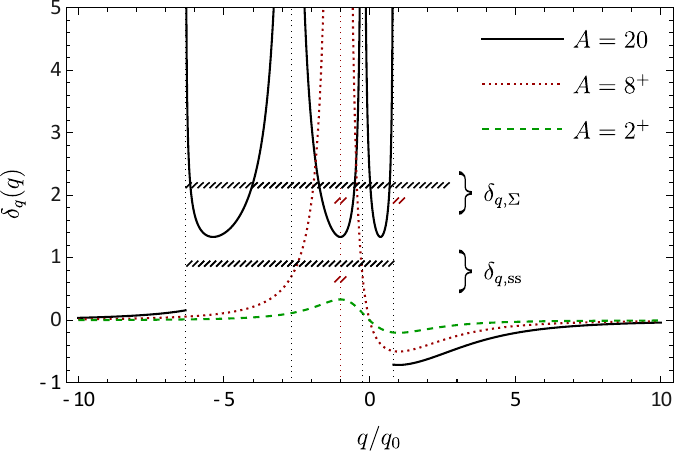}
	\caption{
		The figure shows the exact solution to the non-linear fractional overdensity in Lagrangian space,  $\delta_q(q)$, corresponding to the Lorentzian displacement toy model. The overdensity is shown for three different parameter values $A=2^+$ (right after Eulerian perturbation theory breaks down), $A=8^+$ (right after first stream crossing) and $A=20$ (well after first stream-crossing). One can see the appearance of a caustic for $A=8$ at $q=-q_0$ (red dotted vertical asymptote), which is where stream crossing first occurs. For $A=20$, we have three streams overlapping in Eulerian space with corresponding two caustics in said space (see Fig.~\ref{fig:deltaX}). Once mapped back to Lagrangian space, the number of caustics doubles as each $x$ at a caustics is mapped to two locations $q$ in Lagrangian space. That results in 4 caustics for $A=20$ marked with black dotted vertical asymptotes. For $A=20$, all $q$'s between the leftmost and rightmost caustics  correspond to positions $x(q)$, where multiple streams coexist. Outside of that range, the model produces single streams. Analogous to Fig.~\ref{fig:deltaX}, the ranges in $q/q_0$ marked by the hatched intervals correspond to locations where different approximations for $\delta_q(q)$, namely $\delta_{q,\mathrm{ss}}(q)$ and $\delta_{q,\Sigma}(q)$, break down either because they fail to converge, or because they converge to the wrong value. Both approximations converge for $A=2^+$ unlike in Eulerian space, where $\delta_{x,\Sigma}(x)$ diverges at $x=0$ (Fig.~\ref{fig:deltaX}).
		The overdensity $\delta_{q,\Sigma}(q)$ fails to converge to the correct value for $A=20$ (top hatched range) and $A=8^+$ (bottom hatched range) in intervals where either $|s'(q(x))|\geq1$ or when we have multiple overlapping streams. The overdensity $\delta_{q,\mathrm{ss}}(q)$ fails to converge to the correct value for the same values of $A$ (hatched ranges arranged the same way as for $\delta_{q,\Sigma}(q)$) but in a narrower range -- only when multi-streaming is occurring. }
	\label{fig:deltaQ}
\end{figure}

Note that in Lagrangian space, after first shell crossing, there are four caustics\footnote{The locations of the caustics correspond to the two real roots of the polynomial $p_{1,q}(q)=(q^2+3 q_0^2)^2 + 2A q_0^3   q $ and the two real roots of $p_{2,q}(q)=4 A q_0 q  (q^2+3 q_0^2) + 12 (q^2+3q_0^2)^2-A^2q_0^2$.}, not two, as then at each caustic location, $x=q+s(q)$ has two distinct real roots for $q$, so each shell-crossing location $x$ corresponds to two locations in $q$. That is illustrated by the $\delta_q(q)$ curve for $A=20$ in Fig.~\ref{fig:deltaQ}.

The hatched intervals in Fig.~\ref{fig:deltaX} and Fig.~\ref{fig:deltaQ}, which are labeled  (next to a curly bracket) with one of the approximations for $\delta$ from eq.~(\ref{deltaSummary}), show the intervals in $x$ and $q$ respectively, when the given approximations fail. The failure can  either be because the corresponding approximation diverges or because it converges to the wrong result. 

In Lagrangian space (Fig.~\ref{fig:deltaQ}), we see no surprises in the convergence properties of the approximations given by $\delta_{q,\mathrm{ss}}$ and $\delta_{q,\Sigma}$. By construction, $\delta_{q,\mathrm{ss}}$ converges correctly in the single stream approximation ($s'(q)>-1$ and when no streams overlap), and fails otherwise. Similarly, $\delta_{q,\Sigma}$ fails outside its radius of convergence ($|s'(q)|<1$),  as well as when multiple streams overlap (compare with eq.~(\ref{deltaSummary})). Both approximations converge for all $q$ before shell crossing (curve corresponding to $A=2$).

In Eulerian space (Fig.~\ref{fig:deltaX}), we find numerically that $\delta_{\mathrm{rec}}$ converges to the correct $\delta(x)$ when the following two conditions are satisfied: when $x$ corresponds to single-stream regions, and when $|s'(q)|<1$ (see discussion ca. eq.~(\ref{its1})). At least in this example, we  find that the third condition in eq.~(\ref{deltaSummary}), namely that all basins of attraction of the stable points of the recurrence relation, eq.~(\ref{its1}), are sufficiently large, does not impose further convergence constraints.

Next we turn our attention to the convergence of $\delta_{x,\Sigma}$ (Fig.~\ref{fig:deltaX}). To the precision of our numerical investigations, we found that in the presence of caustics (curves corresponding to $A=20$ and $A=8^+$), the radius of convergence is determined by the locations of those caustics and their distances to the origin -- something which is completely expected given that caustics correspond to singularities in $\delta(x)$. What we further find  is the failure of $\delta_{x,\Sigma}$ to converge well within the single-stream regime, namely for $8>A>2$. The failure starts right after $A=2$ (curve at $A=2^+$ is shown) occurring first at $x=0$. That failure requires an explanation.

In real space, $\delta_{x,\Sigma}(x)$ is a sum over $n$ of $\delta_{x,n}$, which are terms proportional to the $n$-th derivative of $s(q)^n$ (see eq.~\ref{deltaSummary}). For our model, $s(q)$ peaks at $q=0$, and therefore as we take $n\to\infty$,  $(-3s(q)/(Aq_0))^n|_{q=x}$ converges pointwise to 1 at $x=0$ and to 0 otherwise. So, in a sense $s(q)^n$ is being ``sharpened'' to become a more and more spiky function of $q$ with increasing $n$. One can see qualitatively how taking the $n$-th derivative of such a spike can lead to divergences unrelated to shell crossing, and occurring at a location where $\delta_L(x)=-s'(q)_{q=x}=0$. We proceed to show that explicitly.

For our Lorentzian model, it is straightforward to evaluate $\delta_{x,n}$ for even $n$ at the origin (all odd-$n$ $\delta_{x,n}(x=0)$'s vanish):
\be\label{disc}
\delta_{x,2n}(x=0)=2\times 3^{-3n}(iA)^{2n}\binom{3n-1}{2n}, \ \ n=1,2,3,\dots\ ,
\ee
where $\binom{n}{m}$ denotes the binomial coefficient. We then apply the ratio test to find that $\delta_{x,\Sigma}$ is guaranteed to converge when:
\be\label{limitToy1}
\lim\limits_{n\to\infty}\left|\frac{\delta_{x,2n+2}(x=0)}{\delta_{x,2n}(x=0)}\right|=\lim\limits_{n\to\infty}\frac{A^2(3n+1)(3n+2)}{18(n+1)(2n+1)}= \left(\frac{A}{2}\right)^2<1\ .
\ee
Thus, we verify our numerical result: For $A\geq A_{x,\Sigma}=2$, Eulerian perturbation theory breaks down in real space  ($A_{x,\Sigma}$ defined as the threshold beyond which that happens) and fails to produce the correct fractional overdensity at $x=0$. That is a location where the linear fractional overdensity is zero. That may seem surprising until we realize that the Eulerian expansion parameter need not be $\delta_L$ itself: here the familiar advection term $|s(x)\delta_L'(x)|$ is already large \cite{Tassev_Zaldarriaga_2012,Tassev_2014,Porto_Senatore_Zaldarriaga_2014}, and later (see Section~\ref{sec:saddle}) we will confirm that term is indeed  one of many controlling the convergence of Eulerian PT in general.

For larger $A$, the region of failure of convergence expands as can be seen from Fig.~\ref{fig:deltaX}. We would like to highlight that for $A=A_{x,\Sigma}$ the linear overdensity reaches a maximum of $\delta_L=1/4$ and the non-linear overdensity reaches a maximum of $|\delta|=1/3$. So, although the Lorentzian model is  clearly no longer in the linear regime, it is still well in the single-stream regime, and yet, Eulerian perturbation theory fails in real space. 

Next we would like to understand this failure through the lens of the previous sections. First, we would like to confirm that no such failure happens in Fourier space. It is in the transition from Fourier to real space that Eulerian perturbation theory fails (ca. eq.~\ref{ksigmaIssue1}) and we will find that the concrete reason is indeed that the conditions for applying Lebesgue's dominated convergence theorem (e.g.  \cite{tao2011introduction}) are violated (see below). So, to proceed, we now go to Fourier space.

For our toy model, we are able to find a closed-form solution for $\delta_{k,n}$, eq.~(\ref{deltaSummary}):
\be\label{deltakn}
\delta_{k,n}(k)= \frac{4\sqrt{3}q_0\left(i\frac{\, \operatorname{sgn}(k) (kq_0)^2 A}{2\sqrt{3}}\right)^n\mathrm{k}_{n-1}(\sqrt{3}q_0|k|)}{n!(n-1)!}\ ,
\ee
where $\operatorname{sgn}(k)$ is the signum function (equal to +1 when $k>0$, -1 for $k<0$ and 0 for $k=0$). We use $\mathrm{k}_n(x)\equiv \sqrt{\pi/(2 x)}K_{n+1/2}(x)$ to denote the modified spherical Bessel function of the second kind (eq.~10.47.9 of \cite{NIST:DLMF}). Here, $K_n(x)$ is the modified Bessel function of the second kind.

\begin{figure}[t!]
	\centering
	\includegraphics{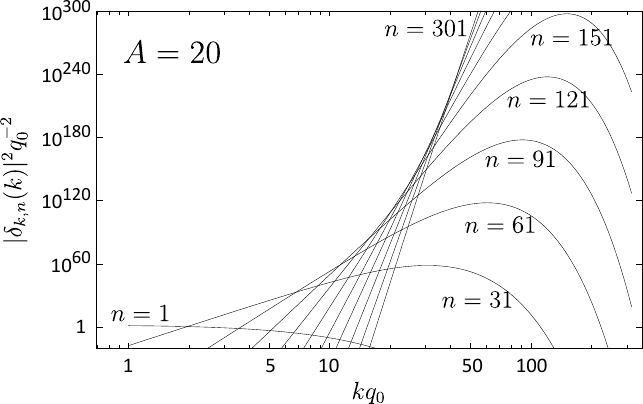}
	\caption{This figure shows $\delta_{k,n}(k)$ for different orders ($n$) for the Lorentzian displacement toy model. The graphs cover the range $n=1$ to $301$ in steps of $\Delta n=30$, with $n=181$ to $271$ not labeled. The graph is for large $A=20$ well after first shell crossing (see Fig.~\ref{fig:deltaX} for the corresponding $\delta(x)$). The range in $kq_0$ goes deep into the non-linear regime ($k\gtrsim 1/q_0$ for this $A$). Note that $\delta_{k,n}(k)$ grows rapidly with $n$, peaking at $n\approx kq_0$ -- a clear indication that the sum over $n$ is non-uniformly converging for different values of $k$. As a result, the sum $\delta_{k,\Sigma}(k)=\sum_{n\geq1} \delta_{k,n}(k)$ converges for a given $k$ to better than percent accuracy as long as the sum is done up to $n\sim 20 (kq_0)$ (for this $A$). The converged result for $\delta_{k,\Sigma}(k)$ is shown in Fig.~\ref{fig:P}. Comparing that figure with the individual terms in the sum shown here, we can see that the sum exhibits massive cancellations between different order terms, and one needs to keep increasingly more significant figures for $\delta_{k,n}(k)$  with increasing $k$ in order to obtain the final $\delta(k)$.} 
	\label{fig:deltaKN}
\end{figure}

We show $|\delta_{k,n}(k)|^2$ in Fig.~\ref{fig:deltaKN} for $A=20$ for orders $n$ spanning a range of $1$ to 301. Clearly, the amplitude of the separate orders  $\delta_{k,n}(k)$ blows up exponentially. Looking at this figure, one may naively conclude that summing all of those orders cannot lead to a converging $\delta_{k,\Sigma}(k)$ (see eq.~(\ref{deltaSummary})). Yet, our analysis in Section~\ref{sectionF} showed that $\delta_{k,\Sigma}(k)$ must converge to the correct result, and therefore very precise cancellations between the different orders $\delta_{k,n}(k)$ must be taking place. We confirm that next.

\begin{figure}[t!]
	\centering
	\includegraphics{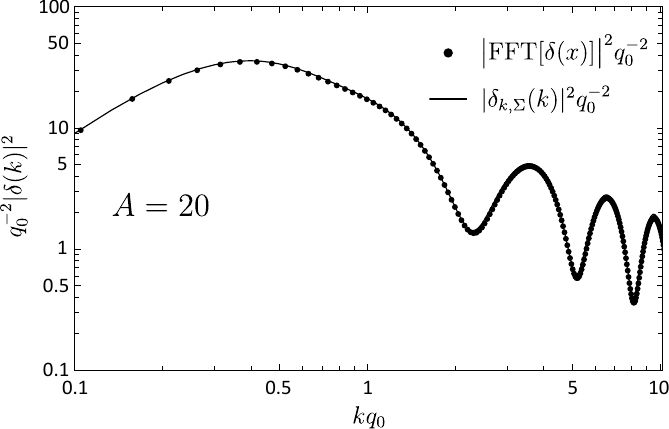}
	\caption{This figure shows $\delta_{k,\Sigma}(k)=\sum_{n\geq1} \delta_{k,n}(k)$ (solid line) along with the result of numerically Fourier transforming (using FFT) the exact non-linear $\delta(x)$ (dots), shown in Fig.~\ref{fig:deltaX}. The displayed quantities are for $A=20$ for the Lorentzian displacement toy model. We can see the two results match in amplitude well into the non-linear regime, which for this value of $A$ is $kq_0\gtrsim 1$. We found that the phase (not shown) of $\delta(k)$ is also faithfully reproduced by $\delta_{k,\Sigma}(k)$. To make sure $\delta_{k,\Sigma}(k)$ converges, we had to sum $\delta_{k,n}$ (shown in Fig.~\ref{fig:deltaKN}) up to $n= 210$. And because of massive cancellations between the individual orders, $\delta_{k,n}(k)$ (shown in Fig.~\ref{fig:deltaKN}), we had to keep more than 25 significant figures while performing the sum to get sensible results for the $k$ range shown.}
	\label{fig:P}
\end{figure}

In Fig.~\ref{fig:P} we show the result (solid line) of summing  over  $\delta_{k,n}$, which gives $\delta_{k,\Sigma}$, for $A=20$. We had to do the sum up to $n=210$ and along the way keep more than 25 significant figures of precision while performing the sum in order to reach good enough convergence to construct the plot and to prevent round-off errors. First of all notice that we get a finite  well-behaved $\delta_{k,\Sigma}$ as a result  despite the exponentially large values of the individual terms, $\delta_{k,n}$. Even though $\delta_{k,\Sigma}$ converges, one may still wonder whether  $\delta_{k,\Sigma}$ matches the exact result, or whether it converges to a nonsensical answer. To check that, we numerically Fourier transformed (using Fast Fourier Transform, FFT) the exact $\delta(x)$ shown in Fig.~\ref{fig:deltaX} (curve corresponding to $A=20$). The result of the FFT is shown as dots in Fig.~\ref{fig:P}, which indeed match the curve for  $\delta_{k,\Sigma}$. Therefore, as we expected in Section~\ref{sectionF},  $\delta_{k,\Sigma}$ does recover the exact non-linear $\delta(k)$ even deep in the non-linear regime, well after shell crossing.

\begin{figure}[h!]
	\centering
	\includegraphics{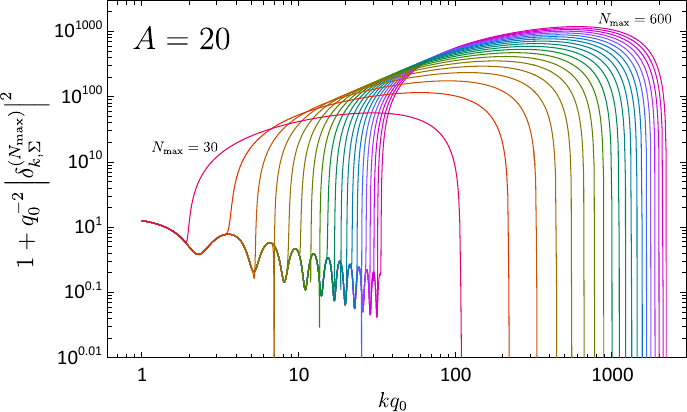}
	\caption{This figure shows the scaled $|\delta_{k,\Sigma}^{(N_{\mathrm{max}})}(k)|^2$ for $N_{\mathrm{max}}=$30 to 600 in steps of  $\Delta N_{\mathrm{max}}=30$ for the Lorentzian model with $A=20$. To construct the figure, we had to keep more than 120 significant figures of precision to ensure the massive cancellations between the different orders do not lead to round-off errors. Note that the vertical axis is \textit{logarithmic in the exponent}, with maximum values of $q_0^{-2}|\delta_{k,\Sigma}^{(N_{\mathrm{max}})}|^2> 10^{ 1000}$. 
	Clearly, the partial sums $|\delta_{k,\Sigma}^{(N_{\mathrm{max}})}|^2$ show convergence to higher and higher $k$ with increasing $N_{\mathrm{max}}$. As can be seen from the figure, we recover $\delta_{k,\Sigma}$ with a good precision for $k\lesssim 35/q_0$ -- that corresponds to the oscillating function on the bottom left, where all curves start merging. That same limit is shown in Fig.~\ref{fig:P} as well. 
	Note that as $N_{\mathrm{max}}$ increases, the exponentially large region, which we refer to as the sliding bump in the text, moves to proportionally higher $k$. Therefore, $\delta_{k,\Sigma}^{(N_{\mathrm{max}})}$ exhibits non-uniform convergence, which in the end is the culprit why Eulerian perturbation theory fails in real space for all $A\geq 2$ in this toy model (see the text for further discussion).
	  }
	\label{fig:esc}
\end{figure}

To summarize so far, we showed that $\delta_{x,\Sigma}$ starts diverging for some $x$ for $A\geq 2$ (eq.~\ref{limitToy1}), while  $\delta_{k,\Sigma}$ converges for all $k$ and all $A$. The reason for that as we noted in Section~\ref{sec:euleriandelta} should stem from a disallowed exchange of an improper integral and  an infinite sum (or equivalently a limit, see Section~\ref{sec:summDelta}). Let us see in further detail how that materializes  by exploring $\delta_{x,\Sigma}$ and $\delta_{k,\Sigma}$ for our particular toy model. 

%First, we confirm that the inverse Fourier transform (IFT) of $\delta_{k,n}$  found in eq.~(\ref{deltakn}) equals $\delta_{x,n}$. Indeed, we obtain a non-illuminating closed-form solution for $IFT[\delta_{k,n}] $expressed through Hypergeometric functions. Evaluating that solution order by order, we find that  $\mathrm{IFT}[\delta_{k,n}]$ matches $\delta_{x,n}=\partial_x^n(-s(x))^n/n!$ as anticipated in eq.~(\ref{deltaXn1}).

To that end, we use eqs.~(\ref{ksigmaIssue2}) and (\ref{ksigmaIssue1}), which depend on $\delta_{k,\Sigma}^{(N_{\mathrm{max}})}$. So,  in Figure~\ref{fig:esc} we graph that quantity for $N_{\mathrm{max}}=30$ to $600$. To construct the graph, we had to keep more than 120 significant figures of precision since the individual terms $\delta_{k,n}$ grow exponentially large, but their sum does not (see Figs.~\ref{fig:deltaKN} and \ref{fig:P}). 

Note that for $A=20$ each $\left|\delta_{k,\Sigma}^{(N_{\mathrm{max}})}\right|$ shown in Fig.~\ref{fig:esc} has a subdomain over which it is exponentially large. We refer to that region as the \textit{sliding bump} region as it moves\footnote{ The fact that the sliding bump region shifts to higher and higher values of $k\propto N_{\mathrm{max}}$  can also be explicitly seen from eq.~(\ref{deltakn}) by showing that the maximum of $|\delta_{k,n}|$ occurs at $k\propto n$ for large $n$. } to higher $k$ with increasing $N_{\mathrm{max}}$. For each $\delta_{k,\Sigma}^{(N_{\mathrm{max}})}$, the exponentially large values in the sliding bump region 
 are eventually canceled (at fixed $k$) by higher order $\delta_{k,n>N_{\mathrm{max}}}$ when $N_{\mathrm{max}}$ is increased. But by including those higher orders (which peak at higher $k$), we only end up moving the exponentially large bump to higher $k$.

In the sliding bump region\footnote{\label{ftntSpiky}Qualitatively, the reason for the presence of that bump is that in real space, $\delta_{k,\Sigma}^{(N_{\mathrm{max}})}$ is just a sum over $\delta_{x,n}$ (cf. eqs.~\ref{defSigmaNmax} and \ref{deltaXn1}) which are proportional to the derivatives of the ``spiky'' $(-s(q))^n$ (see the discussion preceding eq.~\ref{disc}). As $n$ increases, the width of $(-s(q))^n$ shrinks in $q$ with increasing $n$, and thus its Fourier transform can have amplified high-$k$ contributions which move to higher $k$ with $n$.}, $\delta_{k,\Sigma}^{(N_{\mathrm{max}})}$ is dominated by the last two terms in the sum (eq.~\ref{defSigmaNmax}): for even $N_{\mathrm{max}}$, the term $\delta_{k,N_{\mathrm{max}}}$ dominates the real part, and $\delta_{k,N_{\mathrm{max}}-1}$ dominates the imaginary part of $\delta_{k,\Sigma}^{(N_{\mathrm{max}})}$ (and vice versa for odd $N_{\mathrm{max}}$). Neither of those quantities oscillate around zero for $k>0$, since the phase of $\delta_{k,n}$ (eq.~\ref{deltakn}) is entirely determined by the factor $(i\operatorname{sgn}(k))^n$, with the signum function ensuring that $\delta_{k,n}(-k)$ equals the complex conjugate of $\delta_{k,n}(k)$, so that the inverse FT of  $\delta_{k,n}(k)$ is real. Thus, from eq.~(\ref{ksigmaIssue1}), we can see that $\delta_{x,\Sigma}$ will first diverge at positions $x$ for which $e^{ikx}=1$ as then no cancellations of the integral due to an oscillating phase with $k$ can occur. Thus, for our toy model we expect $\delta_{x,\Sigma}$ to first fail at $x=0$, corresponding to the peak of $s(q)|_{q=x}$ (see footnote \ref{ftntSpiky}), exactly as we observe numerically (see Fig.~\ref{fig:deltaX}). At that point, $\delta_{x,\Sigma}$ is simply proportional to the integral over $k$ of $\delta_{k,\Sigma}^{(N_{\mathrm{max}})}$, which should then be evaluated at $N_{\mathrm{max}}\to\infty$. That integral cannot converge because of the sliding bump region, which contributes an ever-increasing amount to $\delta_{x,\Sigma}(x=0)$ as $N_{\mathrm{max}}\to\infty$ -- an amount that would have canceled if we took the limit first \textit{before} evaluating the integral (as in eq.~(\ref{ksigmaIssue2})). And that is precisely the reason for the failure of $\delta_{x,\Sigma}$ to converge.

In the language of mathematical analysis, the issue is that the sequence of partial sums $\delta_{k,\Sigma}^{(N_{\mathrm{max}})}$ entering eq.~(\ref{ksigmaIssue1}) does not satisfy the conditions for applying Lebesgue's dominated convergence theorem: there is no $N_{\mathrm{max}}$-independent integrable function which dominates it. To understand this in a simpler context, consider the classical moving boxcar example of a sequence of functions $f_n(k)=1$ for $n<k<n+1$ and $f_n(k)=0$ otherwise, for which the boxcar escapes to ``horizontal infinity'' (i.e. its support moves to higher and higher $k$ with $n$) (see e.g. \cite{tao2011introduction}). The sequence of functions converges pointwise to $f_n(k)\to f(k)=0$ but the convergence is \textit{non-uniform} (for every $k$, one has to pick $n>k$ for the sequence to converge). The improper integral gives $\int_{-\infty}^\infty dk f_n(k)=1$ for all $n$ and therefore the limit as $n\to\infty$ of the integrals of $f_n$ converges to $1$. Yet, the integral of the limiting function gives $\int_{-\infty}^\infty dk f(k)=0\neq 1$, and thus exchanging the limit and the integral is not allowed. Another way of stating the problem is that the sequence $f_n(k)$ is dominated by a function $g(k)=1$ (such that $|f_n(k)|\leq g(k)$ for all $n$), which is not integrable ($\int_{-\infty}^\infty dk \,g(k)= \infty$), and therefore the limit of the integrals over $f_n(k)$ does not equal the integral over the limiting function, $f(k)$. 

Note that the above-described issue of non-uniform convergence leading to lack of convergence of the dominating function of $\delta_{k,\Sigma}^{(N_{\mathrm{max}})}$ is not reserved only for after shell crossing ($A\geq8$). For example, for $A=3$ (well before shell crossing), the peak of the sliding bump for $N_{\mathrm{max}}=600$ reaches $|\delta_{k,\Sigma}^{(N_{\mathrm{max}})}|^2q_0^{-2}>10^{200}$, sliding to higher and higher $k$ with $N_{\mathrm{max}}$. That again leads to the failure of $\delta_{x,\Sigma}$ for some $x$ around the origin.

Strictly speaking, however, an exponentially blowing-up $|\delta_{k,\Sigma}^{(N_{\mathrm{max}})}|^2$ at some $k$ is not by itself a sufficient condition for $\delta_{x,\Sigma}$ to diverge: the width of the sliding bump could in principle shrink fast enough that its integral remains finite, leaving $\delta_{x,\Sigma}$ well defined despite the local blow-up. Our numerical analysis does not show such behavior for any of the three toy models considered here. In every case where the sliding bump grows in magnitude with $N_{\mathrm{max}}$, $\delta_{x,\Sigma}$ is found to break down for some $x$. The Laplace analysis of Section~\ref{sec:saddle} explains why this is generic: at the location where $\delta_{x,\Sigma}$ first fails, the same parameter that drives the peak magnitude of $\delta_{k,n}$ to grow with $n$ is the parameter that drives the integral over $k$ to diverge.

Bringing $A$ closer to the threshold of  $A=A_{x,\Sigma}=2$ (ca. eq.~\ref{limitToy1}), for $A=2.01$ the sliding bump reaches a maximum of $|\delta_{k,\Sigma}^{(N_{\mathrm{max}})}|^2q_0^{-2}>10^{34}$ for $N_{\mathrm{max}}=10^4$, showing that large cancellations between the terms in the sum for $\delta_{k,\Sigma}$ occur even very close to that threshold. In contrast, the exponentially large sliding bump suddenly disappears for $A=1.99$: we get $|\delta_{k,\Sigma}^{(N_{\mathrm{max}})}|^2q_0^{-2}\approx 0.3$ for the same $k$ and $N_{\mathrm{max}}$ we used for $A=2.01$. These results for $A=2\pm 0.01$ serve as yet another confirmation that at $A=A_{x,\Sigma}=2$ the behavior of $\delta_{k,\Sigma}^{(N_{\mathrm{max}})}$ imposes a limit on Eulerian perturbation theory in real space, with the presence of the sliding bump of $\delta_{k,\Sigma}^{(N_{\mathrm{max}})}$ in eq.~(\ref{ksigmaIssue1})  being the reason for its failure.

We end our discussion of this toy model by noting that in the limit of  $n\to\infty$, we find that $|\delta_{k,n}|$ peaks at $k\to n/q_0$ -- an asymptotic behavior we confirmed up to $n=10^7$. Evaluating $\delta_{k,n}$ at the peak, we find the ratio:
\be \label{peakRat1}
\lim\limits_{n\to\infty}\left|\frac{\delta_{k,n+1}\big(k=(n+1)/q_0\big)}{\delta_{k,n}\big(k=n/q_0\big)}\right|=\frac{A}{2}= \frac{A}{A_{x,\Sigma}}\ .
\ee
Therefore, we see that for $A>A_{x,\Sigma}$, the magnitude of the peak of $|\delta_{k,n}|$ increases with $n$. Thus, it is not surprising that we find that $\delta_{k,\Sigma}$ exhibits an exponentially increasing sliding bump for those $A$'s. The same anatomy reappears in the Gaussian and Cosine models below.

\subsubsection{Toy model 2: Gaussian displacement}

In this and the next section, we will look at two different displacement models to see whether the reasons behind the failure of Eulerian perturbation theory in real space remain the same. In this section, we will focus on a 1D Gaussian toy model for the displacement:
\be
s(q)=A q_0 e^{-\frac{1}{2}(q/q_0)^2}\ ,
\ee
where again $A$ can be identified with the linear growth factor before stream crossing. For this model, stream crossing occurs first at $q=q_0$ (since $s''(q_0)=0$) when $A=A_{\mathrm{sc}}\equiv\exp(1/2)\approx1.65$. That value of $A$ is the solution to $s'(q_0)=-1$ at which point, a caustic forms for the exact $\delta(x)$, eq.~(\ref{deltaX}). Those  values of $q$ and $A$ allow  us to calculate the Eulerian location of that first caustic: $x=2q_0$.

We numerically investigated this model similar to what we did in the previous section, and found that similar to the Lorentzian model, Eulerian perturbation theory fails first at $x=0$ well before first shell crossing. So, similar to what we did in eq.~(\ref{disc}) for the Lorentzian model, we proceed to evaluate $\delta_{x,n}$ for even $n$ at the origin (all odd-$n$ $\delta_{x,n}(x=0)$'s vanish):
\be\label{disc1}
\delta_{x,2n}(x=0)=(iA)^{2n}\frac{n^n}{n!}, \ \ n=1,2,3,\dots\ .
\ee
We then apply the ratio test to find that $\delta_{x,\Sigma}$ converges when:
\be\label{limitToy2}
\lim\limits_{n\to\infty}\left|\frac{\delta_{x,2n+2}(x=0)}{\delta_{x,2n}(x=0)}\right|=\lim\limits_{n\to\infty}A^2\left(1+\frac{1}{n}\right)^n=A^2 e<1\ ,
\ee
where $e=\exp(1)$. Thus, the radius of convergence of Eulerian perturbation theory for the Gaussian model is:
\be\label{thresh2}
A<A_{x,\Sigma}=\exp(-1/2)\approx 0.607\ ,
\ee
which is clearly smaller than $A_{\mathrm{sc}}=\exp(1/2)$. We confirmed the above threshold, eq.~(\ref{thresh2}), numerically. At the threshold value of $A_{x,\Sigma}=\exp(-1/2)$, the linear fractional overdensity reaches a maximum of  $|\delta_L(x)|\approx 0.368$ and the exact overdensity reaches a maximum of $|\delta(x)|\approx0.582$. So, although this is clearly no longer the linear regime, Eulerian perturbation theory in real space breaks down well before shell crossing, similar to what we found for our previous toy model.

We now proceed to calculate the $n$-th order overdensity in Fourier space:
\be
\delta_{k,n}(k)=\sqrt{2\pi}q_0\frac{(-iA k q_0)^n}{n!\sqrt{n}}e^{-(kq_0)^2/(2n)}\ .
\ee
Clearly, $\delta_{k,\Sigma}=\sum_n \delta_{k,n}$ is non-uniformly convergent as in the previous toy model as the maximum $n$ to which we sum  needs to increase with $k$ to reach convergence. 

One can check that each $\delta_{k,n}(k)$ peaks at $k=n/q_0$. Thus, the peak of  $\delta_{k,n}(k)$ moves to higher and higher $k$ with $n$. At the location of the peak, similar to eq.~(\ref{peakRat1}) we find: 
\be\label{peakRat2}
\lim\limits_{n\to\infty}\left|\frac{\delta_{k,n+1}\left(k=\frac{n+1}{q_0}\right)}{\delta_{k,n}\left(k=\frac{n}{q_0}\right)}\right|= A\sqrt{e}=\frac{A}{A_{x,\Sigma}}\ .
\ee
 Thus, the peak magnitude of $\delta_{k,n}$ is not increasing with $n$ only when $A<A_{x,\Sigma}$. For $A>A_{x,\Sigma}$, we find that this translates to $|\delta_{k,\Sigma}^{(N_{\mathrm{max}})}|$ having an exponentially large sliding bump similar to the Lorentzian model -- something we confirmed numerically as well. Thus, our discussion in the previous section applies to this model as well without modifications.

\subsubsection{Toy model 3: Cosine displacement}

Now, we move on to our last toy model -- a periodic displacement given by:
\be\label{toy3}
s(q)=Ak_0^{-1}\cos(q k_0)\ ,
\ee
where again $A$ can be treated as the growth factor before shell crossing, which occurs at $A_{\mathrm{sc}}=1$.

By gradually increasing $A$, numerically we find that $\delta_{x,\Sigma}(x)$ first starts diverging at $x\approx 0$ (and similarly at $x\approx N\pi$, for $N$ -- an integer), i.e. around the extrema of $s(q)|_{q=x}$, similar to what we found for the other two toy models. To obtain $A_{x,\Sigma}$, beyond which $\delta_{x,\Sigma}$ starts breaking down, we calculate $\delta_{x,n}$ following these steps: We express the cosine in eq.~(\ref{toy3}) as $(e^{iqk_0}+e^{-iqk_0})/2$; then we apply the binomial formula  to find $(-s(q))^n$ in eq.~(\ref{deltaXn1}). After taking the $n$-th derivative of $(-s(x))^n$ in $x$, for even $n$ at the origin (all odd-$n$ $\delta_{x,n}(x=0)$'s vanish) we get:
\be\label{disc3}
\delta_{x,2n}(x=0)=(iA)^{2n}\sum\limits_{m=0}^{2n}\frac{(n-m)^{2n}}{m!(2n-m)!}, \ \ n=1,2,3,\dots\ .
\ee
This expression  has the same radius of convergence as the classical expansion of Kepler's equation in eccentricity as both sums have the same structure. Therefore,  $\delta_{x,\Sigma}(x)$ converges at $x=0$ as long as\footnote{This was confirmed numerically to about 5 significant figures by directly applying the ratio test to eq.~(\ref{disc3}). Additionally, by using the asymptotic behavior of sequence A209289 found in The Online Encyclopedia of Integer Sequences (\url{https://oeis.org/A209289}), we confirmed the result for the threshold in $A$ to precision of several tens of significant figures.} $A<A_{x,\Sigma}=C_{\mathrm{LL}}\approx 0.66274$, where $C_{\mathrm{LL}}$ equals Laplace's limit constant\footnote{See \url{https://oeis.org/A033259} and references therein.} (which, in turn, equals the maximum eccentricity for which the expansion of Kepler's equation converges).  Once again, we find that Eulerian perturbation theory fails in real space before shell crossing as $A_{x,\Sigma}<A_{\mathrm{sc}}$.

\begin{figure}[t!]
	\centering
	\includegraphics{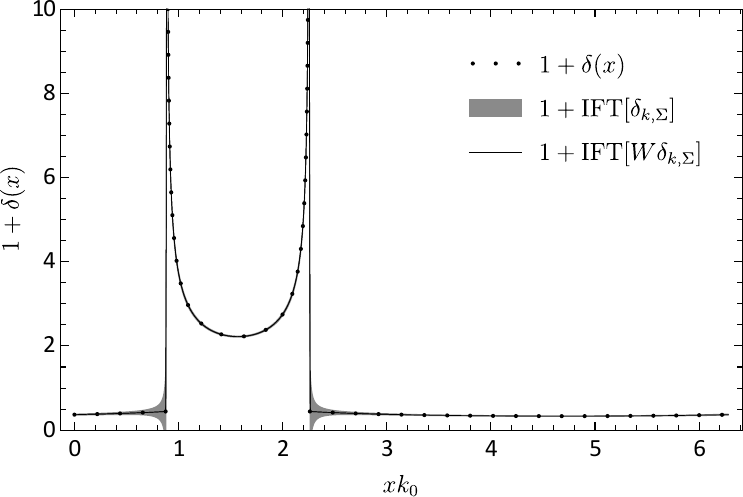}
	\caption{This figure shows the fractional density in Eulerian space for the
		Cosine displacement toy model for $A=2$ (well after first shell-crossing).
		We show $1+\mathrm{IFT}[\delta_{k,\Sigma}]$ (shaded region) using $\delta_{k,\Sigma}$ given by eq.~(\ref{toy3kS}) summed up to $\nu_{\mathrm{max}}=1000$. Because of the sharp cutoff in $\nu$ (corresponding to a sharp cutoff in $k$ at $k_{\mathrm{max}}=\nu_{\mathrm{max}} k_0$),   $1+\mathrm{IFT}[\delta_{k,\Sigma}]$ oscillates rapidly within the bounds of the shaded region with a wavelength of $2\pi/k_{\mathrm{max}}$. We also show the  exact non-linear $\delta(x)$ (dots) obtained by directly applying eq.~(\ref{deltaX}). Clearly, apart from some ringing near the caustics (indicated by the broadening of the shaded region) caused by the sharp cutoff in $k$, the approximate density recovers the exact density. We find that increasing $\nu_{\mathrm{max}}$ does lead to a reduction in the amplitude of the oscillation shown by the shaded region, showing that the resummed $\delta_{k,\Sigma}$ with a sharp cutoff in $k$ does converge to the correct result even after shell crossing. Additionally, we tried smoothing $1+\mathrm{IFT}[\delta_{k,\Sigma}]$ using  a  window function $W$ which matches the smoothing which one obtains from cloud-in-cell assignment done in Particle Mesh N-body codes in order to calculate the density field. After smoothing, we obtain a result (thin solid line) indistinguishable from the exact result shown. Note that the density for our Cosine toy model is periodic in $x$ with a wavelength of $2\pi/k_0$. In this figure we show the range of $xk_0$ between $(0,2\pi)$, since the positive displacement near the origin in the Cosine model shifts the caustics to the right of the origin.}
	\label{fig:cos}
\end{figure}

Now, let us move on to Fourier space. After rewriting the cosine in eq.~(\ref{toy3}) using exponents, and raising $s(q)$ to power $n$ using the binomial formula,  performing the integral in eq.~(\ref{deltaK0}) is straightforward. We find:
\be\label{toy3kn}
\delta_{k,n}(k)=\sum\limits_{m=0}^n 2\pi \delta_D\big(k-k_0(2m-n)\big) \left(\frac{-i A (k/k_0)}{2}\right)^n\frac{1}{m!(n-m)!}\ .
\ee
Given the sum in $m$ above, all $n$'s contribute to a given $k=\nu k_0$, where $\nu$ is an integer.

One can show that the peak location of $|\delta_{k,n}|$ approaches\footnote{ Note that $k$ in eq.~(\ref{toy3kn}) must be an integer multiple of $k_0$ as imposed by the Dirac delta function. But in the limit of large $n\to\infty$ that distinction is irrelevant.} $k\to nk_0/C_{\mathrm{CFP}}$ as $n\to\infty$. Here $C_{\mathrm{CFP}}$ is the fixed point\footnote{See \url{https://oeis.org/A085984}.} of the hyperbolic cotangent, such that $\coth(C_{\mathrm{CFP}})=C_{\mathrm{CFP}}\approx1.1997$. After some non-illuminating algebra we find that at the peak,   similar to eq.~(\ref{peakRat1}) and eq.~(\ref{peakRat2})  we have the ratio:
\be\label{peakRat3}
\lim\limits_{n\to\infty}\left|\frac{\delta_{k,n+1}\left(k=\frac{(n+1)k_0}{C_{\mathrm{CFP}}}\right)}{\delta_{k,n}\left(k=\frac{nk_0}{C_{\mathrm{CFP}}}\right)}\right|=\frac{A}{\sqrt{C_{\mathrm{CFP}}^2-1}}=\frac{A}{C_{\mathrm{LL}}}=\frac{A}{A_{x,\Sigma}}\ .
\ee
 Thus, the peak magnitude of $\delta_{k,n}$ is not increasing with $n$ only when $A<A_{x,\Sigma}$. For $A>A_{x,\Sigma}$,  $\delta_{k,n}$ increases in magnitude at its peak, which translates to $|\delta_{k,\Sigma}^{(N_{\mathrm{max}})}|$ having an exponentially large sliding bump similar to the Lorentzian and the Gaussian models -- something we confirmed numerically as well. Thus, our discussion in the previous two sections applies to this model as well.

Unlike the previous two toy models, here we can find a closed-form expression for $\delta_{k,\Sigma}$. So, we proceed to write that down, and then inverse Fourier transform it to obtain $\delta(x)$ using eq.~(\ref{ksigmaIssue2}). 
The summation over $n$ and $m$ of eq.~(\ref{toy3kn}) for a fixed $\nu=2m-n$ can be performed analytically to obtain:
\be\label{toy3kS}
\delta_{k,\Sigma}(k)+(2\pi)\delta_D(k)=\sum\limits_{\nu=-\infty}^{\infty}(2\pi)\delta_D(k-\nu k_0)J_{|\nu|}(A|\nu|)(-i)^\nu\ ,
\ee
with $J_n(x)$ being the Bessel function of the first kind. The above equation is the exact non-linear result\footnote{To confirm this, we use the fact that the Dirac delta function in eq.~(\ref{toy3kS}) can be easily inverse-Fourier transformed back to Eulerian space.}.

So, as a check of eq.~(\ref{toy3kS}), we numerically summed the resulting expression for $\delta(x)=\mathrm{IFT}[\delta_{k,\Sigma}]$ (eq.~\ref{ksigmaIssue2}) in the range $-\nu_{\mathrm{max}}\leq\nu\leq\nu_{\mathrm{max}}$ with $\nu_{\mathrm{max}}=1000$, in the case of $A=2$, i.e. well after first shell crossing, for which  $A_{\mathrm{sc}}=1$. Apart from some ringing artifacts due to the sharp cutoff in $\nu$ (hence, a sharp cutoff at $k_{\mathrm{max}}=\nu_{\mathrm{max}} k_0$), we recover the exact result for $\delta(x)$ one obtains when directly applying eq.~(\ref{deltaX}). We show that agreement in Fig.~\ref{fig:cos}. Indeed, we checked that increasing the maximum $\nu$ does improve the agreement between the curves. 

Additionally, as a separate check of convergence, we multiplied $\delta_{k,\Sigma}$ by a window function equal to $$W(k)=[\mathrm{ sinc}(\pi k/k_{\mathrm{max}})]^2,$$ which emulates the smoothing that occurs in Particle Mesh N-body codes that use cloud-in-cell assignment to recover the density. And indeed, after we apply that smoothing window, the resulting $1+\mathrm{IFT}[W\delta_{k,\Sigma}]$ is indistinguishable from the exact result (see  Fig.~\ref{fig:cos}).
So, once again, we confirm that eq.~(\ref{ksigmaIssue2}) is valid, and therefore \textit{once} $\delta_{k,\Sigma}(k)$ converges for a given $k$, that result matches the exact result for $\delta(k)$ (since their real space counterparts match). 

At this point, it is worth reminding the reader that summing in $\nu$ in eq.~(\ref{toy3kS}) up to a maximum $\nu$ is \textit{not} equivalent to summing in $n$ of $\delta_{k,n}$ (eq.~(\ref{toy3kn})) up to a maximum $n$. So, the real space $\delta_{x,\Sigma}=\sum_n \delta_{x,n}$ still fails to converge for any $A\geq C_{\mathrm{LL}}$ (including for $A=2$ shown in Fig.~\ref{fig:cos}) as we discussed above.

\subsection{Understanding the breakdown of Eulerian perturbation theory in real space}

In this section, we develop progressively more general analyses of the breakdown of $\delta_{x,\Sigma}$. Section~\ref{sec:saddle} uses Laplace's method, valid when $\log|s(q)|$ is well approximated by a symmetric quadratic near its maximum (or local maxima), and arrives at the breakdown condition $e\,|s(0)\,s''(0)|\gtrsim 1$ at that maximum (eq.~\ref{test}). Section~\ref{sec:saddle22} explores toy models where Laplace's approximation does not apply (i.e.\ $\log|s|$ at its maximum is skewed or has a leading power in $q$ higher than quadratic), and finds that the first breakdown of $\delta_{x,\Sigma}$ then occurs at locations $x$ where $s'(q)|_{q=x}\neq 0$, requiring a more general analysis. Section~\ref{sec:saddle1} provides such an analysis: we find that $\delta_{x,\Sigma}$ is controlled by a family of advection derivative-displacement combinations of which $|s\,s''|$ is just one.

\subsubsection{Breakdown condition using Laplace's method\label{sec:saddle}}

In the toy models explored above, we find the curious relationship (see equations~\ref{limitToy1}, \ref{peakRat1}, \ref{limitToy2}, \ref{peakRat2} and  \ref{peakRat3}):
\be\label{guess}
\lim\limits_{n\to\infty}\left|\frac{\delta_{k,n+1}\left(k_{\mathrm{peak},n+1}\right)}{\delta_{k,n}\left(k_{\mathrm{peak},n}\right)}\right|^2=\lim\limits_{n\to\infty}\left|\frac{\delta_{x,2n+2}(x=0)}{\delta_{x,2n}(x=0)}\right|\ ,
\ee
where $k_{\mathrm{peak},n}$ is the location of the peak of the magnitude of $\delta_{k,n}$, and $x=0$ corresponds to the location of an extremum of $s(q)|_{q=x}$.
From the equation above we see  that $\delta_{x,n}(x=0)$, which is an integral in $k$ over $\delta_{k,n}$, can be related to the peak value of $\delta_{k,n}$. That is a good indication that we should be able to  obtain the above equation using Laplace's method for general $s(q)$. We proceed to do that below.

Laplace's method relies on expanding a function around its maximum, in this case $s(q)$ around its maximum. Let us see why that maximum matters by focusing on eq.~(\ref{deltaK0}) for the overdensity in Fourier space. From that equation, we extract the result for  $\delta_{k,n}(k)$ for easy reference:
\be\label{deltaK0c}
\delta_{k,n}(k)= \int\limits_{-\infty}^\infty  dq e^{-ikq}\frac{(-ik)^n}{n!}s(q)^n\ .
\ee
From eq.~(\ref{deltaXn1}),  the $n$-th order overdensity is just the IFT of $\delta_{k,n}(k)$:
\be\label{deltaXn1c}
\delta_{x,n}(x)&=& \int\limits_{-\infty}^\infty \frac{dk}{2\pi}e^{i k x}\int\limits_{-\infty}^\infty  dq e^{-ikq}\frac{(-ik)^n}{n!}s(q)^n=\nonumber\\
&=&
\int\limits_{-\infty}^\infty \frac{dk}{2\pi}\int\limits_{-\infty}^\infty  dq e^{-ikq} \frac{(-ik)^n}{n!}s(q+x)^n\ ,
\ee
where in the second equality we shifted the origin of the $q$ integral.
To investigate the convergence of the sum over $\delta_{x,n}$, we need to find the behavior of  $|\delta_{x,n}|$ as $n\to\infty$. At large $n$, 
we can assume that values close to the maximum value of $|s|$ (call that maximum $|s|_\mathrm{max}$) matter as long as the oscillating component of the integrand does not lead to cancellations (see next Section~\ref{sec:saddle22} for counterexamples). Indeed, in the discussion leading up to eq.~(\ref{disc}), we already highlighted that as $n\to \infty$, the envelope of the integrand above given by  $s(q)^n$ is being ``sharpened'' to become a more and more spiky function of $q$. For $s$ away from the maximum, we have $(s/|s|_\mathrm{max})^n\to 0$ for large $n$ and $(s/|s|_\mathrm{max})^n$ pointwise converging to 1 at the location of the maximum. 

Then without loss of generality, we can move the origin of $x$ at the global extremum of $s(x)$, and to simplify the analysis, we can assume the extremum is a local positive maximum. In that case, near the origin, we write the Laplace approximation:
\be\label{us}
\log s(q)\equiv u(q)\approx u(0)-\frac{1}{2} q^2 u''(0)\ ,
\ee
 where we expanded $u(q)$ to second order. Plugging this approximation to $s(q)$ into eq.~(\ref{deltaK0c}), we find:
\be\label{r1}
\delta_{k,n}\approx\frac{(-ik)^n}{n!}e^{ u(0)n}\sqrt{\frac{2\pi}{u''(0)n}}e^{-\frac{k^2}{2u''(0)n}}\ .
\ee
In the above equation, and in what follows, we use $\approx$ to indicate that the result is exact only in the Laplace approximation of eq.~(\ref{us}).

Plugging eq.~(\ref{r1}) into eq.~(\ref{deltaXn1}), we can easily evaluate the inverse Fourier transform to find:
\be\label{r3}
\lim\limits_{n\to\infty}\left|\frac{\delta_{x,2n+2}(x=0)}{\delta_{x,2n}(x=0)}\right|\approx e\, |s(0)\, s''(0)|\ ,
\ee
where  we expressed $u$ with $s$ using eq.~(\ref{us}). Now, let us see how that ratio compares with the behavior of $\delta_{k,n}$ at its peak.

The magnitude of  $\delta_{k,n}$  peaks at $k_{\mathrm{peak},n}\approx n\sqrt{u''(0)}$ (which solves $\delta_{k,n}'(k)=0$). Plugging in this peak location into eq.~(\ref{r1}), and again expressing $u$ with $s$ using eq.~(\ref{us}), we find:
\be\label{r2}
\lim\limits_{n\to\infty}\left|\frac{\delta_{k,n+1}\left(k_{\mathrm{peak},n+1}\right)}{\delta_{k,n}\left(k_{\mathrm{peak},n}\right)}\right|^2\approx e\, |s(0)\, s''(0)|\ .
\ee
Comparing the above equation with eq.~(\ref{r3}), we see that at least in the approximation of eq.~(\ref{us}), we confirm eq.~(\ref{guess}). 

Thus, the peak magnitude of $\delta_{k,n}$ will grow with $n$ when
\be\label{test}
e\, |s(0)\, s''(0)|\gtrsim1\ .
\ee
This is exactly the regime when $\delta_{x,\Sigma}=\sum_n\delta_{x,n}$ fails to converge as  per the ratio test in eq.~(\ref{r3}).

The combination $|s\,s''|$ is a known parameter controlling the convergence of Eulerian PT (cf.~\cite{Tassev_Zaldarriaga_2012,Tassev_2014,Porto_Senatore_Zaldarriaga_2014}). What the calculation above shows is the explicit identification of this combination, evaluated at the maximum of $|s|$, as the parameter that controls the DC failure via the growth of the Fourier-space sliding bump for each realization of $s(q)$ (i.e.\ before ensemble averaging over the initial conditions), \textit{subject to the assumption} (\ref{us}) that $\log|s|$ is well approximated by a symmetric quadratic near its maximum. Section~\ref{sec:saddle22} demonstrates that this assumption is restrictive; the more general analysis in Section~\ref{sec:saddle1}  identifies the broader family of advection parameters that control the breakdown otherwise.

As one final check of the Laplace approximation, let us apply the approximate convergence test based on Laplace's method (eq.~\ref{test}) to our toy models in order to solve for the threshold in $A$ ($A_{x,\Sigma}^{(\mathrm{Laplace})}$) beyond which Eulerian perturbation theory fails in real space. We compare that result with the exact results for $A_{x,\Sigma}$ we found before. We find:
\be
A_{x,\Sigma}^{(\mathrm{Laplace})}=3\sqrt{\frac{3}{2 e}}\approx2.23 \textrm{  for the Lorentzian model; exact value $= 2$}\nonumber\\
A_{x,\Sigma}^{(\mathrm{Laplace})}= \exp(-1/2) \textrm{  for the Gaussian model, which recovers the exact value}\nonumber\\
A_{x,\Sigma}^{(\mathrm{Laplace})}= \exp(-1/2)\approx 0.607 \textrm{  for the Cosine model;  exact value = $C_{\mathrm{LL}}\approx 0.663$}\ .\nonumber\\
\ee
We can see that for the toy models we consider, the approximate values we find using Laplace's method, $A_{x,\Sigma}^{(\mathrm{Laplace})}$, are within $\sim10\%$ of the exact values\footnote{The fact that for the Gaussian model $A_{x,\Sigma}^{(\mathrm{Laplace})}$ recovers the exact value should not be surprising given that Laplace's method approximates the displacement as a Gaussian.}, which gives us further confidence in our convergence analysis so far.

\subsubsection{Breakdown condition beyond the Laplace approximation\label{sec:saddle22}}
The previous results: eq.~(\ref{guess}) and eq.~(\ref{test}) require that the Laplace approximation is valid, i.e. a \textit{symmetric quadratic} approximation near the maximum of $\log(s)$ captures the behavior of the integrals for $\delta_{k,n}$ and $\delta_{x,n}$. That symmetric quadratic behavior was common to all our toy models, so we got consistent results. However, one can devise toy models, where a Laplace approximation no longer gives us correct results, for example:
\be
s(q)&\propto& \exp\left(-(q k_0)^4\right)\nonumber\\
s(q)&\propto& q\exp\left(-(q k_0)^2\right)\nonumber\\
&\mathrm{\dots}&
\ee
We explored numerically $\delta_{x,n}$ resulting from the displacements above using eq.~(\ref{deltaXn1}) at high $n$, and we found out that the convergence of $\delta_{x,\Sigma}$ breaks down first at locations that \textit{do not} coincide with vanishing $s'(q)$. This is indeed allowed because
when we considered the pointwise convergence of $s^n$ in Section~\ref{sec:saddle}, we did not take into account any cancellations that can easily arise by the oscillating integrands in eq.~(\ref{deltaK0c},\ref{deltaXn1c}). 

Therefore, instead of the convergence condition in eq.~(\ref{test}), both $s'$ as well as higher derivative terms can matter in non-trivial combinations. To explore those further, below we are going to stop working with toy models and focus on the general expressions for $\delta_{x,\Sigma}$ and for $\delta_{\mathrm{rec}}$, the latter of which matches $\delta_{x,\Sigma}$ order by order (cf. eq.~(\ref{deltaXn2})). 

\subsubsection{Breakdown condition as a set of advection parameters\label{sec:saddle1}}

One way of finding the general regime of validity of $\delta_{x,\Sigma}$ is to focus on it being given as the expansion in $s$ of $\delta_{\mathrm{rec}}$. Using eq.~(\ref{deltaXn1}) and eq.~(\ref{deltaXn2}) we can see that $\delta_{x,\Sigma}$ fails to converge when
\be
|s'(x-s(x-s(x-s(\cdots))))|\geq1\ ,
\ee
once the nested argument is itself expanded in $s$,
or, more generally, when $\delta_{\mathrm{rec}}$ expanded in $s$ fails to converge.
For the toy models discussed above as well as in the Laplace approximation, we focused on extrema of $s(q)|_{q=x}$, where we observed the first breakdown of $\delta_{x,\Sigma}$. To lowest non-zero order in $s$, the above inequality translates into:
\be\label{test2}
|s(x)s''(x)|\gtrsim 1 \text{ where $s'(x)=0$}\ .
\ee
To an order of magnitude, this recovers the Laplace-method bound (\ref{test}). The difference of a factor of $e$ in eq.~(\ref{test}) that further restricts the bound comes from the symmetric-quadratic Gaussian approximation of $\log|s|$ accounting for higher orders in $s$ that the lowest-order  expansion in eq.~(\ref{test2}) does not.

Eq.~(\ref{test2}) is just the lowest-order specialization at an extremum. In general (cf.\ Section~\ref{sec:saddle22}) the first breakdown of $\delta_{x,\Sigma}$ as the displacement amplitude is scaled up can occur at other $x$ and can be controlled by other combinations of the displacement and its derivatives. In particular, Taylor-expanding the nested-$s$ argument of $\delta_{\mathrm{rec}}$ in powers of $s$ at fixed $x$ produces a series whose coefficients are derivative-displacement combinations of $s$ at $x$: $s'$, $s\,s''$, $s^2 s'''$, $s\,s'\,s''$, and so on at higher orders. We call these \textit{advection parameters}, as they arise from the iterative solution (eq.~\ref{its1}) of the equation $x=q+s(q)$, which advects particles from their initial to their final positions in Eulerian space\footnote{The same parameters arise from distributing the derivatives $\partial^ns^n$ in eq.~(\ref{deltaXn1}), although in that case it is harder to see their advection origin.}.

Appendix~\ref{app:advection-form} re-organizes $\delta_{x,\Sigma}$ as a sum directly over derivatives of $s$ (rather than as derivatives of $s^n$ as in eq.~(\ref{deltaXn1}), in which the advection structure is hidden), so that each advection parameter appears as the controlling parameter of a separate partial sum. That partial sum diverges once its advection parameter reaches $\mathcal{O}(1)$; in all our toy models $\delta_{x,\Sigma}$ itself then fails to converge at that $x$, although in principle signed cancellations between partial sums could let the full series survive somewhat longer (see Appendix~\ref{app:advection-form}). Which advection term reaches $\mathcal{O}(1)$ first, and at which $x$, depends on $s(q)$: for the symmetric-quadratic-peak case all our toy models share, it is $|s\,s''|$ at the peak (where $|\delta_L|=0$), giving eq.~(\ref{test2}); in the skewed/non-quadratic cases of Section~\ref{sec:saddle22}, other combinations dominate at other locations.

Because $\delta_{x,n}$ is obtained from $\delta_{k,n}$ via the DC-violating exchange of Section~\ref{sec:euleriandelta}, this entire family of convergence-controlling parameters is equivalently the family of parameters that govern the DC failure. We therefore have two equivalent ways of diagnosing the breakdown of EPT identified in this paper:
\begin{itemize}
\item In Fourier space, by the appearance of a sliding bump in the partial sums $\delta_{k,\Sigma}^{(N_{\mathrm{max}})}$ that escapes to higher $k$ as $N_{\mathrm{max}}\to\infty$ (sections~\ref{sec:saddle}, \ref{sec:saddle22}).
\item In real space, by any one of the advection parameters becoming $\mathcal{O}(1)$ at some $x$. For a detailed analysis of the radius of convergence of the sum $\delta_{x,\Sigma}$ in terms of those parameters, see Appendix~\ref{app:advection-form}.
\end{itemize}
Both pictures easily extend to 3D (next subsection). Either picture makes clear that the DC obstruction is unavoidable in any approach that constructs $\delta(\bm x)$ as an order-by-order expansion in $\bm s(\bm q)$. This is why BBGKY+ZA, which never expands $\delta$ in $\bm s$ in the first place, avoids the entire family of obstructions by construction. We turn to that construction in the next two sections, after first repeating our analysis in 3D.

\subsection{Expanding the overdensity in 3D}
The logic in 1D transparently translates to 3D. We start with the Eulerian density and then turn to the Lagrangian one. In 3D the Eulerian position is
\be
\bm x=\bm q+\bm s(\bm q)\ ,
\ee
and the exact overdensity is the Jacobian-weighted sum over all streams:
\be
1+\delta(\bm x)=\sum\limits_{\substack{{\bm{\tilde q}} \mathrm{\ such\ that}\\ 
		\bm x=\bm{\tilde{ q}}+\bm{s}(\bm{\tilde{ q}})}}\frac{1}{\lvert\det\left[dx_i(\bm{\tilde q})/d\tilde q_j\right]\rvert}
=\sum\limits_{\substack{{\bm{\tilde q}} \mathrm{\ such\ that}\\ 
		\bm x=\bm{\tilde{ q}}+\bm{s}(\bm{\tilde{ q}})}}\frac{1}{\lvert\det\left[\delta_{ij}+\partial_i s_j(\bm{\tilde{ q}})\right]\rvert}\ .
\ee
Here $\delta_{ij}$ is the Kronecker delta, and $\partial_i s_j(\bm q)\equiv \partial s_j(\bm q)/\partial q_i$.

The Fourier-space expansion remains straightforward. Exactly as in 1D, once the $q$-integral is written first and the exponential is expanded at fixed $\bm k$, one obtains
\be
\delta_{k,\Sigma}(\bm k)=\sum\limits_{n=1}^\infty \int  d^3q\, e^{-i\bm k\cdot \bm q}\frac{(-i)^n}{n!}\big(\bm k\cdot \bm s(\bm q)\big)^n\ .
\ee
The important point is the same as in 1D: given the boundedness of $\bm s(\bm q)$, for a fixed $\bm k$ the above sum is simply the Taylor series of an exponential with infinite radius of convergence, so the issue is not the expansion in $\bm s$ itself.

The failure of the expansion appears when one tries to obtain the Eulerian density order by order in real space. The exact same DC-violating exchange needs to be done in 3D as in 1D, with the direct 3D analogue of the DC-obstructed eq.~(\ref{deltaXn1}) being
\be\label{xSigma3d}
\delta_{x,\Sigma}(\bm x)=\sum\limits_{n=1}^\infty \frac{(-1)^n}{n!}\big[\underbrace{\partial_{a}\partial_{b}\partial_{c}\dots}_{\text{$n$ terms}}\big]\big(\underbrace{s_a(\bm x)s_b(\bm x)s_c(\bm x)\dots}_{\text{$n$ terms}}\big)\ ,
\ee
where repeated component indices are summed over and, as in 1D, $\partial_a s_b(\bm x)$ is shorthand for $\partial s_b(\bm q)/\partial q_a|_{\bm q=\bm x}$ rather than a derivative of $\bm s(\bm q(\bm x))$ with respect to Eulerian position.

There is also a direct 3D analogue of the recurrence construction:
\be\label{rec3d}
\delta_{\textrm{rec}}(\bm x)\equiv -1+ \frac{1}{\det\left[\delta_{ij}+\frac{\partial}{\partial q_i} s_j\Big(\bm q=\bm x-\bm s(\bm x-\bm s(\bm x-\bm s(\cdots)))\Big)\right]}\ .
\ee
This quantity makes sense only in the single-stream regime when a fixed point to the recursion in the denominator exists. Expanding eq.~(\ref{rec3d}) in powers of $\bm s$ reproduces $\delta_{x,\Sigma}$ (eq.~\ref{xSigma3d}) order by order, exactly as in 1D (see Footnote~\ref{ftnt:rec}).

The 1D estimate in eq.~(\ref{test2}) and the higher order advection parameters re-appear in 3D. Indeed, the corresponding structure in 3D is visible directly in eq.~(\ref{xSigma3d}): the order-$n$ term in the Eulerian real-space expansion is built from $n$ factors of $\bm s$ acted on by $n$ partial derivatives, and expanding eq.~(\ref{rec3d}) in powers of $\bm s$ produces the same hierarchy of derivative combinations, studied extensively in Appendix~\ref{app:advection-form}.

Thus, the DC-violating exchange leads to $\delta(\bm x)$ having multiple parameters controlling its convergence: not only the linear overdensity $|\delta_L|=|\bm\nabla\cdot\bm s|$ and the advection of that density by the displacement field $|\bm s\cdot\bm\nabla\delta_L|$ \cite{Tassev_Zaldarriaga_2012,Tassev_2014,Porto_Senatore_Zaldarriaga_2014}, but also the full family of displacement-derivative combinations as in 1D (but with appropriate index structure). Carrying over our results from 1D, it is therefore not surprising that Eulerian expansions can fail well before shell crossing, whenever any one of these higher-derivative combinations becomes $\mathcal{O}(1)$.

Turning to the Lagrangian-space density $\delta_q(\bm q)\equiv\delta(\bm x=\bm q+\bm s(\bm q))$, the 3D analogue of eq.~(\ref{deltaQ2}) is
\be\label{deltaQ23d}
1+\delta_q(\bm q)=\sum\limits_{\substack{\tilde{\bm q}\mathrm{\ such\ that}\\\tilde{\bm q}+\bm s(\tilde{\bm q})=\bm q+\bm s(\bm q)}}\frac{1}{\lvert\det[\delta_{ij}+\partial_i s_j(\tilde{\bm q})]\rvert}
=\int d^3q'\int\frac{d^3k}{(2\pi)^3}e^{i\bm k\cdot(\bm q+\bm s(\bm q)-\bm q'-\bm s(\bm q'))}\ .
\ee
Expanding $e^{i\bm k\cdot[\bm s(\bm q)-\bm s(\bm q')]}$ in $\bm s$ is the Taylor series of an exponential: absolutely converging with infinite radius of convergence, valid in the multi-stream regime, but non-uniformly converging in $\bm k$. Performing the same DC-violating exchange as in 1D (the $\bm k$-integral for each $n$ written via $i^n k_{a_1}\cdots k_{a_n}e^{i\bm k\cdot(\tilde{\bm q}-\bm q')}=\partial_{\tilde q_{a_1}}\cdots\partial_{\tilde q_{a_n}}e^{i\bm k\cdot(\tilde{\bm q}-\bm q')}$, followed by the $\bm q'$-integral against the resulting derivatives of $\delta_D(\tilde{\bm q}-\bm q')$) yields the 3D analogue of eq.~(\ref{deltaQ1}),
\be\label{deltaQ13d}
\delta_q(\bm q)\mathclap{\hspace{1.35em}\times}=\sum\limits_{n=1}^\infty\frac{1}{n!}\partial_{\tilde q_{a_1}}\cdots\partial_{\tilde q_{a_n}}\left\{\big[s_{a_1}(\bm q)-s_{a_1}(\tilde{\bm q})\big]\cdots\big[s_{a_n}(\bm q)-s_{a_n}(\tilde{\bm q})\big]\right\}\bigg|_{\tilde{\bm q}=\bm q}\ ,
\ee
with repeated component indices summed; we assumed $\bm s$ infinitely differentiable and the integrand absolutely converging in $\bm q'$, in parallel with the 1D step following eq.~(\ref{deltaQ1}). Each factor $[s_{a_j}(\bm q)-s_{a_j}(\tilde{\bm q})]$ vanishes at $\tilde{\bm q}=\bm q$, so a non-zero contribution requires that each of the $n$ factors receive at least one of the $n$ derivatives; thus each factor is acted upon by exactly one derivative. The action of $\partial_{\tilde q_{a_i}}$ on its target factor gives $-\partial_{a_i} s_{a_j}(\tilde{\bm q})|_{\tilde{\bm q}=\bm q}=-M_{a_i a_j}$, with $M_{ij}\equiv\partial_i s_j(\bm q)$. Each derivative assignment is therefore a permutation $\pi$ of $\{1,\ldots,n\}$ (i.e., an element of the symmetric group $S_n$), contributing $(-1)^n\prod_i M_{a_i\,a_{\pi(i)}}$, so eq.~(\ref{deltaQ13d}) reduces to
\be\label{deltaQ_perm_3d}
\delta_q(\bm q)\mathclap{\hspace{1.35em}\times}=\sum_{n=1}^\infty\frac{(-1)^n}{n!}\sum_{\pi\in S_n}\prod_{i=1}^n M_{a_i\,a_{\pi(i)}}\ ,
\ee
with all $a_i$ summed, so that the result is a scalar, and the way the indices pair up is dictated by $\pi$.  Summing the $a_i$'s contracts these pairs and threads the matrix factors into products of closed ``chains'' of indices (i.e. into products of $\mathrm{tr}(M^k)$).  After some non-illuminating combinatorics, and after adding back the $n=0$ term and summing all $n\geq 0$ we obtain the exponential generating function (with $I$ the $3\times 3$ identity matrix),
\be\label{detformula3d}
1+\delta_q(\bm q)\mathclap{\hspace{1.35em}\times}=\exp\!\Big[\sum_{k\geq 1}\frac{(-1)^k}{k}\mathrm{tr}(M^k)\Big]=\exp\!\big[-\mathrm{tr}\ln(I+M)\big]=\frac{1}{\det[I+M]}\ ,
\ee
where the second equality uses the matrix-logarithm series $\ln(I+M)=\sum_{k\geq 1}(-1)^{k+1}M^k/k$, which converges for $\|M\|<1$ --- the 3D analogue of the 1D condition $|s'|<1$. Once the absolute value is restored by hand (as in 1D), eq.~(\ref{detformula3d}) is the 3D analogue of eq.~(\ref{LPTstart}):
\be\label{LPTstart3d}
\delta_q(\bm q)\mathclap{\hspace{1.35em}\times}=\frac{1}{\lvert\det[\delta_{ij}+\partial_i s_j(\bm q)]\rvert}-1\equiv\delta_{q,\mathrm{ss}}(\bm q)\ .
\ee
As in 1D, the convergence radius is finite and is inherited \textit{not} from the $\bm s$-expansion (whose radius is infinite, see right after eq.~\ref{deltaQ23d}) but from the DC-violating exchange. Note also that the exchange led to losing all multi-stream information, resulting in an expression valid only in the single stream. Shell crossing here is the locus $\det(\delta_{ij}+\partial_i s_j)=0$, at which $\delta_{q,\mathrm{ss}}$ itself diverges; severely underdense configurations sit at the opposite end of the convergence radius, as in 1D (Footnote~\ref{ftnt:single}).

\section{The Zel'dovich approximation and the Vlasov-Poisson equation}\label{sec:prelims}

The rest of the paper aims to avoid the DC failure identified in the previous section by working with phase-space cumulants directly, never expanding $\delta$ in $\bm s$. The recipe uses the BBGKY hierarchy together with the Zel'dovich approximation (ZA). In this section we review the ZA and rewrite the Vlasov--Poisson system in the phase-space variables it naturally suggests, before turning to the BBGKY hierarchy itself in Section~\ref{sec:BBGKY}.

\subsection{The Zel'dovich approximation}
For the choice of phase-space variables used below, we follow HH very closely. We nevertheless restate the construction here in the notation of the present paper.

In the Zel'dovich approximation (ZA) \cite{zeldovich}, the Eulerian comoving position of a CDM particle is written as
\be
\bm{x}(\bm{q},\eta)=\bm{q}+D(\eta)\bm{s}(\bm{q})\ .
\ee
Here $\bm q$ labels the Lagrangian position, $\eta$ is conformal time, $\bm s$ is the stochastic displacement field fixed by the initial conditions, and $D$ is the linear growth factor\footnote{The growing solution of $d(a\partial_\eta D)/d\eta=4\pi G \bar\rho_M a^3 D$ for $D$ is the linear growth factor, which is normalized to $D(\eta_0)=1$ today (at $\eta_0$). Here $G$ is the gravitational constant, $\bar \rho_M(\eta)$ is the average matter density, and $a$ is the cosmological scale factor. }. 

The stochastic displacement is most conveniently expressed through the early-time overdensity field. At some initial time $\eta_I$ in the linear regime, we treat $\delta(\eta_I)$ as Gaussian and define
\be
\delta_L\equiv \frac{\delta(\eta_I)}{D(\eta_I)}\ .
\ee
Its power spectrum is
\be\label{pl}
\la\delta_{L}(\bm{k})\delta_L(\bm{k}')\ra=(2\pi)^3\d(\k+\k')P_{L}(k)\ ,
\ee
where $\d(\bm k)$ is the Dirac delta function and angular brackets denote ensemble averages over realizations of the initial conditions. Here $\bm{k}$ is the Fourier wavevector, and $P_L$ is the linear overdensity power spectrum today. 

For brevity, we use the same symbol for a quantity in real and Fourier space, distinguishing the two only through their arguments. Thus, in the ZA, the displacement is fixed by linear theory and therefore is related to $\delta_L$ by
\be\label{s}
\bm s(\bm x)=-\partial_{\bm x}\partial_{\bm x}^{-2}\delta_L(\bm x)\nonumber \\
\bm{s}(\bm{k})=i\frac{\bm{k}}{k^2}\delta_L(\bm{k}) \ ,
\ee
where we wrote $\bm s$ both in configuration space and in Fourier space\footnote{In this paper, we use the normalization $$\bm{s}(\bm{k})=\int d^3 q\, e^{-i \bm{q}\cdot\bm{k}} \bm{s}(\bm{q})\ ,$$which differs from that of HH by a factor of $(2\pi)^3$.\label{ft:fourier}}. 

We next turn to the one-particle distribution function. Let $\f(\bm{x},\mathrm{\bf v},\eta)$ denote the invariant phase-space density written in terms of the velocity variable proportional to the conjugate momentum, with the irrelevant particle-mass factors dropped. That velocity is
\be\label{vc}
\mathrm{\bf{v}}\equiv a \frac{d\bm{x}}{d\eta}\ ,
\ee
where $a$ is the cosmological scale factor. Up to the same omitted mass factor, the corresponding phase-space measure is $d^3x\,d^3\mathrm{v}$.
 
For what follows, that velocity is not the most convenient one. In the ZA,
$$\frac{d\bm{x}}{d\eta}= \partial_\eta D\bm{s}(\bm{q})\ ,$$
so a more useful choice is the rescaled variable
\be\label{dxdt}
\bm{v}\equiv (a\, \partial_\eta D)^{-1}\mathrm{\bf{v}}\ ,
\ee
for which the ZA gives simply $\bm{v}=\bm{s}(\bm{q})$. Thus $\bm v$ is time-independent in the ZA, and the free-streaming part of the later equations takes a particularly simple form.

To accompany the rescaling of the velocity, we also rescale the distribution function, tracking the determinant of the velocity Jacobian so that phase-space volumes remain invariant\footnote{The main difference from HH is that here we use $D$ as the time coordinate for $f$, rather than $\eta$. Further notational differences from HH that affect the explicit form of the displacement covariance are collected in footnote~\ref{HH_diff1}.\label{HH_diff}}
\be\label{f_def}
f(\bm{x},\bm{v},D)\equiv\left (a\partial_\eta D\right)^{3}\f\big(\bm{x},a\partial_\eta D\bm{ v},\eta(D)\big)\ .
\ee
The factor of $\left(a\partial_\eta D\right)^3$ guarantees that the differential particle number measured in either set of variables is the same,
\be\label{dNvx}
\f d^3x\,d^3\mathrm{v}=fd^3x\,d^3v\ .
\ee

In double Fourier space this becomes
\be\label{f_equiv_in_fourier}
f(\k,\o,D)&=&\int d^3x\,d^3v e^{-i(\k\cdot\bm{x}+\o\cdot\bm{v})}f(\bm{x},{\bm v},D)=\int d^3x\,d^3\mathrm{v} e^{-i(\k\cdot\bm{x}+\mathrm{\bf w}\cdot\mathrm{\bf v})}\f\big(\bm{x},\mathrm{\bf v},\eta(D)\big)\nonumber\\
&=&\f\left(\k,\mathrm{\bf w}=\frac{\o}{a\partial_\eta D},\eta(D)\right)\ ,
\ee
where $\o\equiv a\partial_\eta D\,\mathrm{\bf w}$ is the wavevector conjugate to $\bm v$, while $\mathrm{\bf w}$ is conjugate to $\mathrm{\bf v}$.

At fixed $\bm x$, the ZA may contain several streams, and summing over all of them gives the exact ZA phase-space distribution
\be\label{f_z}
f(\bm{x},\bm{v},D)=\int d^3q\delta_D(\bm{v}-\bm{s}(\bm{q}))\delta_D\big(\bm{x}-\bm{q}-D\bm{s}(\bm{q})\big)\ .
\ee
Fourier transforming eq.~(\ref{f_z}) with respect to both $\bm x$ and $\bm v$ gives
\be\label{f_kw}
\nonumber f(\bm{k},\bm{w},D)&=&\int d^3 x d^3 v\, e^{-i (\bm{k}\cdot\bm{x}+\bm{w}\cdot\bm{v})}  f(\bm{x},\bm{v},D)  \\
&=&\int d^3q e^{-i\bm{k}\cdot\bm{q}}e^{-i\bm{s}(\bm{q})\cdot(D\bm{k}+\bm{w})}\ .
\ee
In this representation, differentiating with respect to $D$ only acts on the combination $D\bm k+\bm w$, so the ZA distribution satisfies
\be\label{Z_eom}
\dot{f}-\k\cdot \partial_{\o}f=0\ ,
\ee
where a dot indicates a partial derivative with respect to the growth factor, $D$.

\subsection{The Vlasov-Poisson equation in double-Fourier space}\label{sec:zeldovich_IC}

We now return to the exact CDM dynamics governed by the Vlasov-Poisson system \cite{Bernardeau_2002,Valageas_2004}. The rewriting below again follows HH and \cite{Valageas_2004} very closely. Our purpose here is to rewrite the Vlasov-Poisson system in the form that will be used in the rest of this paper.

Because of the definition (\ref{dxdt}), the ZA corresponds to $d\bm v/d\eta=0$ for a test particle. Deviations from the ZA therefore appear directly as the failure of that simple free streaming. The exact equation of motion for the original velocity variable is
\be\label{dvvdt}
\frac{d \mathrm{\bf v}}{d\eta}=- a \partial_{\bm x} \phi\ ,
\ee
where $\phi$ is the Newtonian gravitational potential. For modes well inside the Hubble horizon, $\phi$ obeys
\be\label{poisson}
\partial_{\bm x}^2\phi=\frac{d(a\partial_\eta D)}{d\eta}\frac{\delta}{aD}\ .
\ee
Transforming to the variable $\bm v$, and using $d/dD=(\partial_\eta D)^{-1}d/d\eta$, we obtain
\be\label{dvdt}
\frac{d\bm{v}}{dD}&=&-\frac{d\ln\left(a\partial_\eta D\right)}{dD}\,\bm{v}-\frac{1}{(\partial_\eta D)^2}\partial_{\bm x} \phi\nonumber\\
&=&-\frac{d\ln\left(a\partial_\eta D\right)}{dD}\,\left[\bm{v} +\frac{\partial_{\bm x}\partial_{\bm x}^{-2}\delta(\bm x,\eta)}{D}\right]\ .
\ee
The term proportional to $\bm v$ is precisely the part that cancels the ZA acceleration at early times, so to first order in $s$ one indeed recovers $d\bm v/dD=0$, as intended when defining (\ref{dxdt}). This is analogous in spirit to COLA, which rewrites the dynamics relative to an LPT trajectory and solves for the residual acceleration in that comoving frame \cite{Tassev_Zaldarriaga_Eisenstein_2013}.

The invariant distribution $\f$ obeys the usual Vlasov equation
\be\label{vlasov_std0}
\partial_\eta \f + \partial_{\bm x}\cdot\left(\f\frac{d\bm x}{d\eta}\right)+ \partial_{\mathrm{\bf v}}\cdot\left(\f\frac{d\mathrm{\bf v}}{d\eta}\right)=0\ .
\ee
Integrated over a finite phase-space cell, this is simply particle-number conservation. Using eqs.~(\ref{vc}) and (\ref{dvvdt}), it reduces to
\be\label{vlasov_std}
\partial_\eta \f+\frac{1}{a}\mathrm{\bf v}\cdot\partial_{\bm{x}}\f-a\partial_{\bm{x}} \phi\cdot\partial_{\mathrm{\bf v}}\f=\frac{d\f}{d\eta}=0\ .
\ee

Rewriting the same equation in terms of the transformed distribution $f$, which satisfies eq.~(\ref{dNvx}), gives
\be\label{Vlasov_real}
&\dot f& + \partial_{\bm x}\cdot\left(f\frac{d\bm x}{d D}\right)+ \partial_{{\bm v}}\cdot\left(f\frac{d{\bm v}}{d D}\right)=\nonumber\\
&=&\dot f+ \bm v\cdot\partial_{\bm x}f-\frac{d\ln\left(a\partial_\eta D\right)}{d D}\,\left[\bm{v} +\frac{\partial_{\bm x}\partial_{\bm x}^{-2}\delta(\bm x,D)}{D}\right]\cdot\partial_{\bm v}f-3\frac{d\ln\left(a\partial_\eta D\right)}{d D}f=0\ ,\nonumber\\
&&
\ee
where we used eq.~(\ref{dvdt}) and, to remind the reader, a dot is a derivative with respect to $D$. Moreover, we freely switched the time dependence of $\delta$ for $D$. The same result follows by inserting the definition (\ref{f_def}) directly into eq.~(\ref{vlasov_std}).

The density field is obtained from $f$ by integrating over velocities:
\be\label{fdelta}
1+\delta (\bm{x},\eta)=\int d^3\mathrm{ v} \f(\bm{x},\mathrm{\bf v},\eta)=\int d^3v f(\bm{x},\bm{v},D(\eta))= f(\bm{x},\bm{w}=0,D(\eta))\ .
\ee
The last equality identifies the  velocity integral with $f$ evaluated at $\bm w=0$. Note that in the last equality, we did not Fourier transform $f$ in $\bm{x}$, but only in $\bm v$.

Transforming eq.~(\ref{Vlasov_real}) to double Fourier space and using eq.~(\ref{fdelta}) yields
\begin{eqnarray}\label{VP_master}
	\dot f&-&\k\cdot\partial_{\o}f-\\\nonumber
	&-&\frac{d\ln(a\partial_\eta D)}{d D}\left[D^{-1}\int\frac{ d^3k'}{(2\pi)^3}\frac{\o\cdot\k'}{k'^2}\Big[f(\k',\o=0,D)-(2\pi)^3\delta_D(\bm k')\Big]f(\k-\k',\o,D)-\o\cdot\partial_{\o}f\right]=0\ .
\end{eqnarray} 
The term proportional to $f$ in eq.~(\ref{Vlasov_real}) cancels in double Fourier space. Up to a change of variables, the resulting equation is equivalent to eq.~(7) in \cite{Valageas_2004}.

%*********************************************************************************************------------------------------------

The integrand in eq.~(\ref{VP_master}) contains the kernel $\k'/k'^2$ multiplied by the zero-mode of $f$, which is singular at $\k'=0$ and must be regulated. Following \cite{Valageas_2004}, we adopt the Jeans-swindle prescription
\be\label{jeans}
\frac{\k'}{k'^2}\d(\k')\equiv \frac{\k'}{k'^2+0^+}\d(\k')=0\ ,
\ee
which removes the spatially uniform component. A practical consequence is that a homogeneous universe, for which $f(\k,\o,D)=(2\pi)^{3} \d(\k)$, gives an identically vanishing force term: the Vlasov-Poisson equation then reduces to $\dot f=0$, which confirms $f(\k,\o,D)=(2\pi)^{3} \d(\k)$ as the homogeneous solution. The ZA distribution, by construction, satisfies the free-streaming equation (\ref{Z_eom}). Therefore, what drives the non-linear gravitational dynamics beyond ZA is entirely the bracketed term in eq.~(\ref{VP_master}).

\section{BBGKY hierarchy}\label{sec:BBGKY}

Having rewritten the Vlasov-Poisson equation in the ZA-adapted phase-space variables, we now derive the hierarchy obeyed by the moments and cumulants of the CDM phase-space density $f$. This is the starting point for the second part of the paper, and it makes explicit what initial data and closure input BBGKY+ZA requires. The standard BBGKY hierarchy \cite{peebles} is usually written for equal-time correlators. Here, following \cite{Valageas_2004} and  HH, we first formulate the unequal-time hierarchy directly in the variables $(\bm k,\bm w,D)$. The equal-time hierarchy is then recovered by setting the growth-factor arguments equal at the end.

\subsection{Operator notation}

Following HH, we use DeWitt notation, with index placement carrying no significance here. A label such as $a$ stands for the full set of continuous variables $(\bm k_a,\bm w_a,D_a)$, and repeated labels imply integration over all of them. Thus $f_a\equiv f(\bm k_a,\bm w_a,D_a)$, and a repeated label in an expression such as $L_{ab}f_b$ means
\be
L_{ab}f_b\equiv \int dD_b\, d^3k_b\, d^3w_b\, L_{ab}f_b\ .
\ee
Throughout this section the letters $a,b,c,\ldots$ are reserved for DeWitt labels in the sense above; $i,j,\ldots$ continue to denote Cartesian spatial-component indices, with the usual Einstein summation convention. Where both kinds of indices appear on the same object, as in $\beta^a_i$ or $\psi_{ij}(\bm q^a)$, the Latin DeWitt label is written as a superscript and the Cartesian one as a subscript to keep the two clearly separated.

In this notation, $L$ is the linear streaming operator in the ZA-adapted variables, while $K$ is the quadratic kernel generated by the gravitational interaction through the Poisson equation. With this notation, the quadratic Vlasov-Poisson equation, eq.~(\ref{VP_master}), can be written compactly as
\be\label{VP_ops}
\L{a}{b} f_b=K_{abc}f_b f_c\ ,
\ee
where, with no implied summation inside the definitions themselves,
\be
\label{L}
L_{ab}\equiv \mathcal{D}_a[\delta_{ab}]\  ,\ \mathrm{where}\\
\mathcal{D}_a\equiv \partial_{D_a}-\bm{k}_a\cdot \partial_{\bm{w}_a} +\frac{d\ln(a(\eta_a)\partial_\eta D(\eta_a))}{d D_a}\,\o_a\cdot\partial_{\o_a}\  ,\nonumber\\
\nonumber \delta_{ab}\equiv \delta_D(\bm{k}_a-\bm{k}_b)\d(\bm{w}_a-\bm{w}_b)\d(D_a-D_b)\ .
\ee

The vertex $K_{abc}$ is defined to be symmetric in its last two indices:
\be
K_{abc}&\equiv& \frac{1}{2} (2\pi)^{-3}\frac{d\ln\left(a(\eta_a)\partial_\eta D(\eta_a)\right)}{d D_a}D^{-1}(\eta_a)\times\\
&&\times\left[\frac{\k_b\cdot\o_a}{k_b^2}\d(\o_b)\d(\o_a-\o_c)\d(\k_a-\k_b-\k_c)\d(D_a-D_c)\d(D_a-D_b)  +\ (b \leftrightarrow c)\right]\ .\nonumber
\label{K}
\ee
The normalization of $K$ differs by a factor of $(2\pi)^6$ from HH because of the different Fourier transform convention (see Footnote~\ref{ft:fourier}).

By direct inspection, $L$ and $K$ obey the sign-reflection relations
\be\label{symm}
\L{a}{b}=\L{-a,}{-b}, \ \mathrm{and}\ K_{abc}=K_{-a,-b,-c}\ ,
\ee
with the convention that placing a minus sign on a subscript reverses all Fourier wavevectors carrying that label, while leaving the growth factor unchanged. Concretely, starting from $\delta_{ab}=\d(\k_a-\k_b)\d(\o_a-\o_b)\d(D_a-D_b)$ one obtains $\delta_{a,-b}=\d(\k_a+\k_b)\d(\o_a+\o_b)\d(D_a-D_b)$. Throughout, a comma inside a subscript is used purely as a punctuation mark between labels and carries no derivative meaning.

\subsection{Unequal-time moments and cumulants}

Multiplying eq.~(\ref{VP_ops}) by additional factors of $f$ and then ensemble averaging yields the hierarchy for unequal-time moments. Because the equation of motion is quadratic, the $n$-th moment couples to the $(n+1)$-st one. The first few equations are
\be
L_{ax}\langle f_x\rangle&=&K_{axy}\langle f_xf_y\rangle\\
L_{ax}\langle f_xf_b\rangle&=&K_{axy}\langle f_xf_yf_b\rangle\nonumber\\
L_{ax}\langle f_xf_bf_c\rangle&=&K_{axy}\langle f_xf_yf_bf_c\rangle\ .\nonumber
\ee

Re-expressing the moments in terms of cumulants gives
\be\label{cum_BBGKY}
L_{ax}\langle f_x\rangle_{\mathrm{c}}&=&K_{axy}\langle f_x f_y \rangle_{\mathrm{c}}\\
L_{ax}\langle f_x f_b\rangle_{\mathrm{c}}&=&K_{axy}\langle f_x f_y f_b\rangle_{\mathrm{c}}\ + \ 2K_{axy}\langle f_y\rangle_{\mathrm{c}} \langle f_x f_b\rangle_{\mathrm{c}}\nonumber\\
L_{ax}\langle f_x f_bf_c\rangle_{\mathrm{c}}&=&K_{axy}\langle f_x f_y f_bf_c\rangle_{\mathrm{c}}\ + \ 2K_{axy}\Big[\langle f_y \rangle_{\mathrm{c}} \langle f_x f_b f_c\rangle_{\mathrm{c}}+\langle f_x f_b \rangle_{\mathrm{c}} \langle f_y f_c\rangle_{\mathrm{c}}\Big]\nonumber\\
L_{ax}\langle f_x f_bf_cf_d\rangle_{\mathrm{c}}&=&K_{axy}\langle f_x f_y f_bf_cf_d\rangle_{\mathrm{c}}\ +\nonumber\\
&+& 2K_{axy}\Big[\langle f_y \rangle_{\mathrm{c}} \langle f_x f_b f_cf_d\rangle_{\mathrm{c}}+\langle f_x f_b \rangle_{\mathrm{c}} \langle f_y f_c f_d\rangle_{\mathrm{c}}+\langle f_x f_c \rangle_{\mathrm{c}} \langle f_y f_b f_d\rangle_{\mathrm{c}}+\langle f_x f_d \rangle_{\mathrm{c}} \langle f_y f_b f_c\rangle_{\mathrm{c}}\Big]\ .\nonumber
\ee

Here we used the symmetry $K_{axy}=K_{ayx}$, together with $K_{axy}\langle f_x \rangle_{\mathrm{c}} \langle f_y\rangle_{\mathrm{c}}=0$. The latter term is proportional to the gradient of the mean Newtonian potential and therefore vanishes for homogeneous $\langle f\rangle_{\mathrm{c}}$, cf. eq.~(\ref{jeans}).

The general pattern is easiest to write in terms of unordered index sets. For the $n$-th cumulant one finds
\be\label{gen_cum_BBGKY}
L_{ax}\langle f_x f^{|I|}_I\rangle_{\mathrm{c}}=K_{axy}\langle f_x f_y f^{|I|}_I\rangle_{\mathrm{c}}\ + \ \sum_{\substack{I_1 \subseteq I\vspace{3pt}\\ \text{distinct } I_1\vspace{3pt}\\ I_2 = I \setminus I_1}}K_{axy}\langle f_x f^{|I_1|}_{I_1}\rangle_{\mathrm{c}} \langle f_y f^{|I_2|}_{I_2}\rangle_{\mathrm{c}}\ ,
\ee
where $I=\{b,c,d,\cdots\}$ denotes the unordered set of the remaining $(n-1)$ labels, and $\langle f_x f^{|I|}_I\rangle_{\mathrm{c}}\equiv\langle f_x f_bf_cf_d\cdots\rangle_{\mathrm{c}}$. Here $|I|$ denotes the number of elements of the set; in this case, $|I|=n-1$.  The sum runs over all distinct unordered bipartitions $I=I_1\cup I_2$ with $I_1\cap I_2=\varnothing$. Note that the coefficient of $2$ in front of $K_{axy}$ in equations~(\ref{cum_BBGKY}) arises when $I_1$ in one term of the sum in  eq.~(\ref{gen_cum_BBGKY}) equals $I_2$ in another term, and the symmetry  $K_{axy}=K_{ayx}$ is taken into account.

Equation~(\ref{gen_cum_BBGKY}) makes the closure problem explicit: the $n$-th cumulant couples both to the $(n+1)$-point cumulant (first term on right-hand side of eq.~\ref{gen_cum_BBGKY}) and to products of lower-order cumulants. Thus, solving the hierarchy up to the $n$-th cumulant involves providing a closure relation for the $(n+1)$-point cumulant.

\subsection{Equal-time cumulants}

The equal-time hierarchy (e.g. \cite{peebles}) follows from the unequal-time hierarchy above by setting all growth-factor arguments equal. Writing the BBGKY hierarchy directly for the equal-time cumulants would then give a system that, after a physically motivated truncation, could be integrated forward in time.

A reasonable first step in investigating the hierarchy and its predictive power would be to close the hierarchy at the level of the equation for the second cumulant. Closure and initial conditions are two distinct inputs, and BBGKY+ZA supplies both from the Zel'dovich approximation, but at different orders. The closure replaces only the phase-space cumulant one order above the truncation by its exact Zel'dovich expression at all times: in the equation for the second cumulant, that would be the three-point function. The initial conditions, in contrast, require the full set of equal-time ZA cumulants up to that order at some initial $D$, since each cumulant of the evolved hierarchy needs its own initial conditions. Together these two ingredients define a concrete evolution problem for the equal-time cumulants. We therefore need the ZA one-, two-, three-, and (if one wants to go to higher orders) higher-point phase-space correlators -- the lower-order ones serving as initial data and the highest one supplying the closure -- and we turn to those now.

\section{Correlation Functions in the ZA}\label{sec:corrZA}
\subsection{One-point function}

Section~\ref{sec:BBGKY} showed that BBGKY+ZA needs the ZA phase-space correlators to supply the initial conditions and the closure input for the BBGKY hierarchy. We now turn to calculating those correlation functions.

The 1-point correlator is the mean phase-space density. In eq.~(\ref{f_kw}), the stochastic dependence enters only through a linear functional of the Gaussian displacement field, so the Gaussian cumulant theorem applies directly:
\be\label{gaussian_expansion}
\langle\exp (g)\rangle=\exp\left(\langle g\rangle_{\mathrm{c}}+\frac{1}{2}\langle g^2\rangle_{\mathrm{c}}\right)\ ,
\ee
where $\langle\rangle_{\mathrm{c}}$ denotes cumulants and $g$ is a Gaussian random variable. All connected cumulants of $g$ beyond the second vanish. Therefore, to make progress, we need the covariance of the displacement field:
\be\label{psi}
\psi_{ij}(\bm q)\equiv\langle s_i(\bm q') s_j(\bm q-\bm q')\rangle=\int \frac{d^3k}{(2\pi)^3}\frac{k_ik_j}{k^4}P_L(k)e^{i\bm k\cdot \bm q}\ ,
\ee
where we used eq.~(\ref{s}) and eq.~(\ref{pl}). Statistical homogeneity implies that the correlator depends only on the separation $\bm q$, not on the reference point $\bm q'$. It is also even under $\bm q\to-\bm q$.

Because $\psi_{ij}(\bm q)$ is an isotropic rank-two tensor built from $\bm q$, its tensorial decomposition is entirely fixed by $\delta_{ij}$ and $\hat q_i\hat q_j$, and the full information content lives in their two scalar prefactors. We introduce the combinations\footnote{\label{HH_diff1}Note, that $\gamma$ in this paper equals $3\gamma $ in HH; and $\zeta$ in this paper equals $\chi+\gamma$ in HH. The coefficients before the integrals in eq.~(\ref{useful}) are different from those in HH as here we used a different rescaling for $P_L$ in eq.~(\ref{pl}).}
\be\label{useful}
&\sigma_v^2\equiv \frac{1}{6\pi^2}\int\limits_0^\infty dk P_L(k)\nonumber\\
&\zeta(q)\equiv \frac{1}{6\pi^2}\int\limits_0^\infty dk P_L(k)[j_0(k q)+j_2(k q)]\nonumber\\
&\gamma(q)\equiv \frac{1}{2\pi^2}\int\limits_0^\infty dk P_L(k)j_2(kq)\ ,
\ee
with $j_l$ being the spherical Bessel function of the first kind. With these definitions, the covariance (\ref{psi}) takes the compact form
\be\label{psigamma}
\psi_{ij}(\bm q)=\delta_{ij}\zeta(q)-\hat q_i\hat q_j \gamma(q)\ ,
\ee
with the limit $\psi_{ij}(\bm 0)=\delta_{ij}\sigma_v^2$, identifying $\sigma_v$ as the  root-mean-square (rms) displacement along a single axis.

Substituting eq.~(\ref{psigamma}) into eq.~(\ref{f_kw}) and performing the Gaussian average gives the one-point function:
\be\label{fbarZ}
\langle f(\bm{k},\bm{w},D)\rangle=\langle f(\bm{k},\bm{w},D)\rangle_{\mathrm{c}}=(2\pi)^3\d(\bm{k})\exp\left[-\frac{\sigma_v^2 w^2}{2}\right]\ .
\ee

Only the homogeneous mode survives in $\bm k$, while the dependence on $\bm w$ remains Gaussian, with width fixed by the one-dimensional rescaled velocity (eq.~\ref{dxdt}) dispersion $\sigma_v$.

\subsection{Two-point function\label{sec:2ptf}}

For the two-point function the same Gaussian identity applies to the product of two copies of $f$. The self-contractions of each copy of the displacement reproduce the factors involving $\sigma_v^2$, while the separation-dependent cross-contraction is encoded by $\psi_{ij}(\bm q)$. One finds
\be\label{2pfz}
\langle f(\bm{k},\bm{w},D)\,f(\bm{k}',\bm{w}',D')\rangle=(2\pi)^3\d(\bm{k}+\bm{k}')e^{-\frac{1}{2}(\beta^2+\beta'^2)\sigma_v^2}\int d^3 q\,e^{-i\bm{q}\cdot\bm{k}}\exp\left[-\beta_i\beta'_j\psi_{ij}(\bm q)\right]\ ,\nonumber\\
\ee
where
\be\label{beta}
\bm \beta(\bm k,\bm w,D)\equiv \bm w+D\bm k\ ,
\ee
and $\bm\beta'\equiv \bm \beta(\bm k',\bm w',D')=\bm w'+D'\bm k'=\bm w'-D'\bm k$, where the last equality uses the Dirac delta in eq.~(\ref{2pfz}).

Subtracting the disconnected product gives the connected two-point cumulant:
\begin{eqnarray}\label{2pfz_simple}
&\langle f(\bm{k},\bm{w},D)\,f(\bm{k}',\bm{w}',D')\rangle_{\mathrm{c}}=\langle f(\bm{k},\bm{w},D)\,f(\bm{k}',\bm{w}',D')\rangle-\langle f(\bm{k},\bm{w},D)\rangle_{\mathrm{c}}\langle f(\bm{k}',\bm{w}',D')\rangle_{\mathrm{c}}=\nonumber\\&= (2\pi)^3\d(\bm{k}+\bm{k}')e^{-\frac{\sigma_v^2}{2}(\beta^2+\beta'^2)} \mathcal{P}(\bm k,\bm\beta,\bm \beta')\ ,
\end{eqnarray}
where
\be\label{Pcurly}
\mathcal{P}(\bm k,\bm l,\bm m)\equiv\int d^3 q\, e^{-i\bm{q}\cdot\bm k}\left[ e^{-l_im_j\psi_{ij}(\bm q)}-1\right]=\nonumber\\
=\int d^3 q\,e^{-i\bm{q}\cdot\bm k}\left[ \ e^{-{\bm l}\cdot{\bm m} \, \zeta(q)+({\bm l}\cdot {\bm {\hat q}})\,({\bm m}\cdot {\bm {\hat q}}) \gamma(q)}\ -1\right]\ .
\ee
The subtraction by $1$ removes the disconnected contribution (eq.~\ref{fbarZ}). We keep $\bm k$, $\bm l$, and $\bm m$ formally independent, since later manipulations act on different arguments separately. The same quantity $\mathcal{P}$ enters the higher cumulants below. Appendix~\ref{app:P} derives the representation for $\mathcal{P}$ used in the numerical evaluation, so from this point onward we write the cumulants in terms of $\mathcal{P}$.

Using eq.~(\ref{fdelta}), the connected density two-point function is obtained by setting $\bm w=\bm w'=0$:
\be\label{delta2}
\langle\delta(\bm{k},\eta)\delta(\bm{k}',\eta)\rangle_{\mathrm{c}}= \langle f(\bm{k},\bm{w}=0,D(\eta))\,f(\bm{k}',\bm{w}'=0,D(\eta))\rangle_{\mathrm{c}}\ .
\ee

It is useful to verify that eq.~(\ref{Pcurly}) reproduces the standard linear-theory two-point function in the linear regime. Expanding the exponent in $P_L$, to first order one finds \cite{Valageas_2004}:
\begin{eqnarray}\label{2pfz_simple_expanded}
		\langle f(\bm{k},\bm{w},D)\,f(\bm{k}',\bm{w}',D')\rangle_{\mathrm{c}}\approx\ (2\pi)^3\d(\bm{k}+\bm{k}')P_L(k)\left[-\frac{\left(\bm{\beta}\cdot\bm{k}\right)\, \left(\bm{\beta'}\cdot\bm{k}\right)}{k^4}\right]\ .
\end{eqnarray}
Evaluating the above expression at equal times and at vanishing $\bm w,\bm w'$ reduces $\bm\beta\cdot\bm k$ and $\bm\beta'\cdot\bm k$ to $\pm Dk^2$, leaving $\langle\delta(\bm{k},D)\delta(\bm{k}',D)\rangle\approx(2\pi)^3\delta^{(3)}(\bm{k}+\bm{k}')P_L(k)D^2$, which recovers the linear-theory matter power spectrum, eq.~(\ref{pl}).

Beyond the linear limit, eq.~(\ref{2pfz_simple}) at equal times and $\bm w=\bm w'=0$ reduces to a closed-form relation between $\mathcal{P}$ and the (fully nonlinear) Zel'dovich density power spectrum, $P_{Z}$. With $\bm\beta=D\bm k$ and $\bm\beta'=-D\bm k$ (so $\beta^2+\beta'^2=2D^2k^2$), eq.~(\ref{2pfz_simple}) reads $\langle\delta(\bm k)\delta(\bm k')\rangle_{\mathrm{c}}=(2\pi)^3\delta^{(3)}(\bm k+\bm k')\,P_{Z}(k)$ with
\be\label{PZ_check}
P_{ Z}(k,D)\;=\;e^{-\sigma_v^2D^2k^2}\,\mathcal{P}(\bm k,D\bm k,-D\bm k)\ ,
\ee
where we restored the time dependence of $P_Z$ through the growth factor as a second argument.
At this configuration $\hat{\bm l}=\hat{\bm k}$, $\hat{\bm m}=-\hat{\bm k}$, and $|\bm l|=|\bm m|=Dk$, so the magnitude product is $p\equiv lm=D^2k^2$. Equation~(\ref{PZ_check}) provides a stringent, fully nonlinear cross-check of any numerical implementation of $\mathcal{P}$: at this single geometry   $\mathcal{P}$ (scaled by the exponential prefactor above) must reduce to the independently-computed $P_{Z}(k)$. We carry out this check in Section~\ref{sec:Pnumresults}.

Note that in writing eq.~(\ref{PZ_check}), one can equivalently set first $D=0$ in $\bm \beta$ and $\bm \beta'$ \textit{and then} evaluate $\bm w=-\bm w'=D\bm k$ to get the same exact result. In other words, sampling the velocity wavevector $\bm w$ along $\bm k$ is equivalent to sampling the position wavevector $\bm k$ at different times (corresponding to $D$). Therefore, even when focusing on $\mathcal{P}$ of today, for this parallel configuration ($\bm w$'s along $\bm k$), finding the values of $\mathcal{P}$ at large velocity wavevectors ($w\gg k$) is equivalent to sampling  $P_Z$ into the future ($D\gg 1$). That is not surprising given that in the ZA, the rescaled velocity of particles is simply given by their displacement (see eq.~\ref{dxdt} and right after it). Indeed, from  eq.~(\ref{f_kw}) we can confirm that the above discussion applies realization by realization as well:
\be
f(\bm k,\bm w=0,D)=f(\bm k,\bm w=D\bm k,0)\ .
\ee

\subsection{Three-point function}

The two-point function involved a single kernel $\mathcal{P}$. In deriving the three-point function as products and convolutions of $\mathcal{P}$, we would need two close relatives of $\mathcal{P}$: $\mathcal{Q}$, the Fourier transform of the pair factor $e^{-l_im_j\psi_{ij}(\bm q)}$, which differs from $\mathcal{P}$ only by an explicit zero-mode; and $\tilde{\mathcal{Q}}$, the inverse Fourier transform of $\mathcal{Q}$, which simply returns the pair factor itself. 

Let $f_a\equiv f(\bm k^a,\bm w^a,D^a)$, with superscripts used only as labels and not as vector-component indices. Then we can write the 3-point function as:
\be
\langle f_af_bf_c\rangle=\int d^3q^{a}\,d^3q^{b}\,d^3q^{c}\, e^{-i\left(\bm k^a\cdot \bm q^a+\bm k^b\cdot \bm q^b+\bm k^c\cdot \bm q^c\right)}e^{-\frac{1}{2}\sigma_v^2 \left[(\beta^a)^2+(\beta^b)^2+(\beta^c)^2\right]}\times\nonumber\\
\times\exp\left[-\beta^a_i\beta^b_j\psi_{ij}(\bm q^a-\bm q^b)-\beta^a_i\beta^c_j\psi_{ij}(\bm q^a-\bm q^c)-\beta^b_i\beta^c_j\psi_{ij}(\bm q^b-\bm q^c)\right]\ .
\ee
Translation invariance allows one absolute Lagrangian coordinate to be factored out. Shifting $\bm q^L\to \bm q^L+\bm q^a$ for $L\neq a$ and then integrating over $\bm q^a$ gives:
\be\label{3pt_dD}
\langle f_af_bf_c\rangle=(2\pi)^3\delta_D(\bm k^a+\bm k^b+\bm k^c)e^{-\frac{1}{2}\sigma_v^2 \left[(\beta^a)^2+(\beta^b)^2+(\beta^c)^2\right]}\int d^3q^{b}\,d^3q^{c}\, e^{-i\left(\bm k^b\cdot \bm q^b+\bm k^c\cdot \bm q^c\right)}\times\nonumber\\
\times\exp\left[-\beta^a_i\beta^b_j\psi_{ij}(\bm q^b)-\beta^a_i\beta^c_j\psi_{ij}(\bm q^c)-\beta^b_i\beta^c_j\psi_{ij}(\bm q^b-\bm q^c)\right]\ ,\nonumber\\
\ee
where we used that $\psi_{ij}(-\bm q)=\psi_{ij}(\bm q)$.

We now define the Fourier transform of a single pair factor before subtracting its disconnected part:
\be\label{Qcurly}
\mathcal{Q}(\bm k,\bm l,\bm m)\equiv \int d^3 q\, e^{-i\bm{q}\cdot\bm k} e^{-l_im_j\psi_{ij}(\bm q)}=\nonumber\\
=\mathcal{P}(\bm k,\bm l,\bm m)+(2\pi)^3\delta_D(\bm k)\ .
\ee
The last equality follows immediately from eq.~(\ref{Pcurly}), so $\mathcal{Q}$ differs from $\mathcal{P}$ only by the explicit zero-mode term.

The reason for introducing $\mathcal{Q}$ is that its inverse Fourier transform reproduces the pairwise Gaussian factor itself:
\be\label{Qtilde}
\tilde {\mathcal{Q}}(\bm y,\bm l,\bm m)\equiv \int \frac{d^3k}{(2\pi)^3}\, e^{i\bm y\cdot \bm k}\mathcal{Q}(\bm k,\bm l,\bm m)=e^{-l_im_j\psi_{ij}(\bm y)}\ .
\ee
Again, the three vector arguments are treated as independent.

Replacing the exponentials in eq.~(\ref{3pt_dD}) by $\tilde{\mathcal{Q}}$ factors and Fourier transforming back once gives
\be\label{3pt_Qt}
\langle f_af_bf_c\rangle=(2\pi)^3\delta_D(\bm k^a+\bm k^b+\bm k^c)e^{-\frac{1}{2}\sigma_v^2 \left[(\beta^a)^2+(\beta^b)^2+(\beta^c)^2\right]}\int d^3q^{b}d^3q^{c} e^{-i\left(\bm k^b\cdot \bm q^b+\bm k^c\cdot \bm q^c\right)}\times\nonumber\\
\times\tilde{\mathcal{Q}}(\bm q^b,\bm\beta^a,\bm\beta^b)
\tilde{\mathcal{Q}}(\bm q^c,\bm\beta^a,\bm\beta^c)
\tilde{\mathcal{Q}}(\bm q^c-\bm q^b,\bm\beta^b,\bm\beta^c)
\nonumber\\
\ee
or, after one more Fourier transform,
\be\label{3pt_Q}
\langle f_af_bf_c\rangle=(2\pi)^3\delta_D(\bm k^a+\bm k^b+\bm k^c)e^{-\frac{1}{2}\sigma_v^2 \left[(\beta^a)^2+(\beta^b)^2+(\beta^c)^2\right]}\times\nonumber\\
\times\int \frac{d^3k'}{(2\pi)^3} 
{\mathcal{Q}}(\bm k^b+\bm k',\bm\beta^a,\bm\beta^b)
{\mathcal{Q}}(\bm k^c-\bm k',\bm\beta^a,\bm\beta^c)
{\mathcal{Q}}(\bm k',\bm\beta^b,\bm\beta^c)
\nonumber\\
\ee
The connected piece is obtained by expanding each $\mathcal{Q}$ as $\mathcal{P}+(2\pi)^3\delta_D$ and subtracting the disconnected pieces. This gives
\be\label{3pt_Qc}
\langle f_af_bf_c\rangle_{\mathrm{c}}=(2\pi)^3\delta_D(\bm k^a+\bm k^b+\bm k^c)e^{-\frac{1}{2}\sigma_v^2 \left[(\beta^a)^2+(\beta^b)^2+(\beta^c)^2\right]}\times\nonumber\\
\times\Big\{\mathcal{P}(\bm k^b,\bm\beta^a,\bm\beta^b)\mathcal{P}(\bm k^c,\bm\beta^a,\bm\beta^c)+\mathcal{P}(\bm k^a,\bm\beta^a,\bm\beta^b)\mathcal{P}(\bm k^c,\bm\beta^b,\bm\beta^c)+\nonumber\\
+\mathcal{P}(\bm k^a,\bm\beta^a,\bm\beta^c)\mathcal{P}(\bm k^b,\bm\beta^b,\bm\beta^c)+\int \frac{d^3k'}{(2\pi)^3} 
{\mathcal{P}}(\bm k^b+\bm k',\bm\beta^a,\bm\beta^b)
{\mathcal{P}}(\bm k^c-\bm k',\bm\beta^a,\bm\beta^c)
{\mathcal{P}}(\bm k',\bm\beta^b,\bm\beta^c)
\Big\}
\nonumber\\
\ee
The last integral is a convolution in the intermediate wavevector $\bm K$ of
\footnote{Here we are following the convolution convention: $A(\bm K)*B(\bm K)\equiv\int d^3k'A(\bm k')B(\bm K-\bm k')$}
$\mathcal{P}(\bm k^b+\bm K,\bm\beta^a,\bm\beta^b)\mathcal{P}(\bm K,\bm\beta^b,\bm\beta^c)$ with $\mathcal{P}(\bm K,\bm\beta^a,\bm\beta^c)$, evaluated at $\bm K=\bm k^c$.

\subsection{N-point function}

Nothing conceptually new happens when considering higher $n$-point functions. For an arbitrary number of factors,
\be
\langle f_af_bf_c\cdots\rangle=\int \prod\limits_{I}\left[d^3q^{I} e^{-i\bm k^I\cdot \bm q^I}e^{-\frac{1}{2}\sigma_v^2 (\beta^I)^2}\prod\limits_{
	J> I}e^{-\beta^I_i\beta^J_j\psi_{ij}(\bm q^J-\bm q^I)}\right]\ ,
\ee
where $I$ and $J$ run over the labels $a,b,c,\cdots$. By the notation $J>I$, we imply that no unordered pair of indices $(I,J)$ should be repeated (and clearly, $I\neq J$).

Choosing one label, say $a$, as an anchor and integrating over $\bm q^a$ after the translation $\bm q^L\to \bm q^L+\bm q^a$ for $L\neq a$ again extracts the overall Dirac delta:
\be
\langle f_af_bf_c\cdots\rangle=(2\pi)^3\delta_D\left(\sum\limits_{I}\bm k^I\right)e^{-\frac{1}{2}\sigma_v^2 \left[\sum\limits_{I}(\beta^I)^2\right]}\times\\
\nonumber
\int \prod\limits_{I\neq a}\left[d^3q^{I} e^{-i\bm k^I\cdot \bm q^I}e^{-\beta^a_i\beta^I_j\psi_{ij}(\bm q^I)}\left(\prod\limits_{\substack{J\neq a\\
		J> I}}e^{-\beta^I_i\beta^J_j\psi_{ij}(\bm q^J-\bm q^I)}\right)\right]\ .
\ee

Introducing $\tilde{\mathcal{Q}}$ for each remaining pair factor gives
\be
\langle f_af_bf_c\cdots\rangle=(2\pi)^3\delta_D\left(\sum\limits_{I}\bm k^I\right)e^{-\frac{1}{2}\sigma_v^2 \left[\sum\limits_{I}(\beta^I)^2\right]}\times\\
\nonumber
\int \prod\limits_{I\neq a}\left[d^3q^{I} e^{-i\bm k^I\cdot \bm q^I}\tilde{ \mathcal{Q}}(\bm q^I,\bm \beta^a,\bm \beta^I)\left(\prod\limits_{\substack{J\neq a\\
		J> I}}\tilde{ \mathcal{Q}}(\bm q^J-\bm q^I,\bm \beta^I,\bm \beta^J)\right)\right]\ .
\ee
The connected $n$-point cumulants can then be obtained from the products of $\tilde{ \mathcal{Q}}$ above by rewriting those first as a product of $\mathcal{Q}$'s (using Fourier transforms) and then those as $\mathcal{P}$ and Dirac-delta pieces, and then retaining only the connected terms. The bookkeeping becomes combinatorial, but no new ingredient beyond $\mathcal{P}$ is required.

\subsection{Integrating the phase-space 2-point function $\mathcal{P}$}

The analytic calculation of the phase-space 2-point function kernel $\mathcal{P}$ is presented in full in Appendix~\ref{app:P}. Here we summarize those results, starting from the symmetry reduction of $\mathcal{P}$. It depends on three vectors $\bm k, \bm l, \bm m$ which carry nine components. Rotational invariance removes three degrees of freedom (DoF), leaving six DoF. Here, however, $\mathcal{P}$ depends on $\bm l$ and $\bm m$ only through the tensor product $l_i m_j$, so only the product of their magnitudes appears. Writing $k\equiv |\bm k|$, $l\equiv |\bm l|$, $m\equiv |\bm m|$, and $p\equiv lm$, the dependence reduces to five independent DoF. We choose them to be $k$, $p$, the cosine $\mu_{lm}$, and the two combinations $s\equiv\mu_{kl}^2+\mu_{km}^2$ and $t\equiv\mu_{kl}\mu_{km}$, with $\mu_{kl}\equiv \bm{\hat k}\cdot \bm{\hat l}$, and similarly for $\mu_{km}$ and $\mu_{lm}$. 
The scalar $s$ used here is unrelated to the displacement of Section~\ref{sec:approx}. For fixed $\mu_{lm}$, we first leave the factor $e^{-\mu_{lm}p\zeta(q)}$ unexpanded:
\be\label{FINALP_exp}
\mathcal{P}(\bm k,\bm l,\bm m)=\left(\frac{2\pi}{k}\right)^{3/2}\sum\limits_{L=0}^{\infty}\sum\limits_{M=L}^{\infty} \mathcal{C}_{LM}(s,t,\mu_{lm})\mathcal{G}_{LM}(k,p,\mu_{lm})\nonumber\\ 
\textrm{with $s\equiv\mu_{kl}^2+\mu_{km}^2$, $t\equiv\mu_{kl}\mu_{km}$.}\nonumber\\
\ee
with $\mathcal{C}_{LM}$ defined further down, and:
\be\label{Glmn}
\mathcal{G}_{LM}(k,p,\mu_{lm})\equiv\int\limits_0^\infty dq\, k \mathcal{H}_{M}(q,p,\mu_{lm})J_{2L+1/2}(k q)\ ,
\ee
where $J_\nu$ denotes the Bessel function of the first kind. Above we defined 
\be\label{Hlmn}
\mathcal{H}_{M}(q,p,\mu_{lm})\equiv  q^{3/2} e ^{-\mu_{lm}p\,\zeta(q)}i_M\Big(\gamma(q) p\Big)- q^{3/2}\delta_{M0}\ ,
\ee
with $i_n(x)$ being the modified spherical Bessel function of the first kind, defined as $i_n(x)\equiv \sqrt{\pi/(2 x)}I_{n+1/2}(x)$ with $I_n(x)$ being the modified Bessel function of the first kind. The subtraction term proportional to $\delta_{M0}$ contributes only when $L=0$ in eq.~(\ref{FINALP_exp}) (that is enforced by $\mathcal{C}_{LM}$ being zero for $M=0$ and $L>0$, see below) and ensures that the disconnected piece of the two-point function is removed.

The expression in  eq.~(\ref{FINALP_exp}) is adapted to calculations organized in slices of constant $\mu_{lm}$. An \textit{alternative} form follows by expanding the $\zeta$-dependent exponential in eq.~(\ref{Hlmn}) in Legendre modes of $\mu_{lm}$, which fully separates the angular and radial dependence:
\be\label{FINALP}
\mathcal{P}(\bm k,\bm l,\bm m)=\left(\frac{2\pi}{k}\right)^{3/2}\sum\limits_{L=0}^{\infty}\sum\limits_{M=L}^{\infty}\sum\limits_{N=0}^{\infty}
(2N+1)\mathrm{P}_N(-\mu_{lm})\mathcal{C}_{LM}(s,t,\mu_{lm})\mathcal{E}_{LMN}(k,p)\nonumber\\ 
\textrm{with $s\equiv\mu_{kl}^2+\mu_{km}^2$, $t\equiv\mu_{kl}\mu_{km}$.}\nonumber\\
\ee
Here $\mathrm{P}_N(y)$ denotes the Legendre polynomial of order $N$.
The function $\mathcal{E}$ depends only on $k$ and the product of the magnitudes of $\bm l$ and $\bm m$ ($p= l\,m$), and is defined as:
\be\label{Elmn}
\mathcal{E}_{LMN}(k,p)\equiv\int\limits_0^\infty dq\, k \mathcal{F}_{MN}(q,p)J_{2L+1/2}(k q)\ ,
\ee
where $\mathcal{F}_{MN}$ is given by:
\be
\mathcal{F}_{MN}(q,p)\equiv q^{3/2} i_M\Big(\gamma(q) p\Big)i_N\Big(\zeta(q)p\Big)-q^{3/2}\delta_{M0}\delta_{N0}\ .
\ee
Similar to above, the subtraction term proportional to $\delta_{M0}\delta_{N0}$ contributes only when $L=0$ and ensures that the disconnected piece of the two-point function is removed.

Inspecting the expression for $\mathcal{P}$ in eq.~(\ref{FINALP}), we can see that the entire non-trivial angular dependence is packed into the polynomial $\mathcal{C}_{LM}$ (which is cosmology independent, see below), while $\mathcal{E}_{LMN}$ contains cosmology-dependent radial Hankel transforms. Similarly,  eq.~(\ref{FINALP_exp}) uses $\mathcal{C}_{LM}$ but does couple angular and radial dependence through the cosmology-dependent Hankel transformed quantity, $\mathcal{G}_{LM}$.

Now, let's turn to the polynomial $\mathcal{C}_{LM}$. It is defined using the angle-averaged quantity $\tilde S_{LM}$ from eq.~(\ref{SLM1}) through:
\be
&\mathcal{C}_{LM}\big(s=\mu_{kl}^2+\mu_{km}^2,\ t=\mu_{kl}\mu_{km},\ \mu_{lm}\big)\equiv\tilde{\mathcal{C}}_{LM}(\mu_{kl},\mu_{km},\mu_{lm})\nonumber\\
&= (-1)^{L}(2M+1)(4L+1)\tilde S_{LM}(\bm k,\bm l,\bm m)\ ,
\ee
where $\tilde{\mathcal{C}}_{LM}$ is explicitly given in eq.~(\ref{CLM_Tilde}).

Collecting the algebra of Appendix~\ref{app:P}, we find:
\be\label{CLM_FINAL}
&\mathcal{C}_{LM}(s=\mu_{kl}^2+\mu_{km}^2,\ t=\mu_{kl}\mu_{km},\ \mu_{lm})=\nonumber\\
&=\sum\limits_{U=0}^L\sum\limits_{V=0}^{\lfloor\frac{M}{2}\rfloor}\sum\limits_{x,y,z}^{''}\sum\limits_{d=0}^{\lfloor \frac{y-x}{4}\rfloor} \left\{\frac{(-1)^{L + U + V}(2M+1)(4L+1)(M - 2V)!(4L - 2U - 1)!!( 2M - 2V-1)!!
}{\big( 2(L + M - U) - 4V+1\big)!! (2U)!! (2V)!!(2u)!!(2v)!!(2w)!!x!y!z!}T_{C,d}\right\}\times{t}^{x+2d}{s}^{C-2d}\left({\mu_{lm}}\right)^z
\nonumber\\
&\mathrm{with\ }T_{C> 0,d}\equiv (-1)^d \frac{C}{C - d} \binom{C - d}{d}\ \ \mathrm{and}\ \ T_{00}\equiv 	1\ .
\nonumber\\
\ee
The double primes indicate that, for fixed $L$, $M$, $U$, and $V$, only a finite set of terms survives. The allowed tuples are those for which $u,v,w,x,y,z$ are all non-negative integers with the first three defined as shown in the first three equations below:
\be
u = L-U - \frac{x + y}{2}\nonumber\\
v = \frac{M-2V - x - z}{2}\nonumber\\
w = \frac{M-2V - y - z}{2}\nonumber\\
y\geq x\geq 0\nonumber\\
z\geq0\nonumber\\
C=\frac{y-x}{2}\ .
\ee
Note that $C$ is guaranteed to be an integer because $x$ and $y$ must have the same parity in order for $u$ to be an integer.

Some reference expressions for $\mathcal{C}_{LM}$ are shown in Table~\ref{tab:lmc_values}. Figures~\ref{fig:L=1}, \ref{fig:L=3} and \ref{fig:L=10} show the angular polynomial $\mathcal{C}_{LM}$ defined in eq.~(\ref{CLM_FINAL}) for fixed $L=1$, $L=3$, and $L=10$, for a sample of values of $M$. They are included here for reference. Since these angular functions are independent of the cosmology-dependent radial sector, they can be precomputed and stored for fast retrieval when the formalism is applied to a particular cosmology.
\begin{table}[h!]
	\centering
	\begin{tabular}{|c|c|c|}
		\hline
		$L$ & $M$ & $\mathcal{C}_{LM}(s,t,z)$\\
		\hline\hline
		0 & 0 & 1 \\
		\hline
		0 & 1 & $z$ \\
		1 & 1 & $z-3t$ \\
		\hline
		0 & 2 & $-2 + z^2$ \\
		1 & 2 & $\frac{5}{14}\big(2 - 3s - 12tz + 4z^2\big)$ \\
		2 & 2 & $\frac{3}{14}\big(1 - 5s + 35t^2 - 20tz + 2z^2\big)$ \\
		\hline
		0 & 3 & $-2z + z^3$ \\
		1 & 3 & $8t - z - \frac{5sz}{2} - 5tz^2 + \frac{5z^3}{3}$ \\
		2 & 3 & $\frac{3}{22}\Big(5\big(-6 + 7s\big)t + 3\big(3 - 10s + 35t^2\big)z - 60tz^2 + 6z^3\Big)$ \\
		3 & 3 & $\frac{5}{66}\Big(-21\big(1 - 3s + 11t^2\big)t + 3\big(1 - 7s + 63t^2\big)z - 42tz^2 + 2z^3\Big)$ \\
		\hline
	\end{tabular}
	\caption{Table of $\mathcal{C}_{LM}$ for $L\leq M\leq 3$. Note that for $L>M$, $\mathcal{C}_{LM}=0$. Here $s=\mu_{kl}^2+\mu_{km}^2$, $t=\mu_{kl}\mu_{km}$, $z= \mu_{lm}$.}
	\label{tab:lmc_values}
\end{table}

\subsection{Choice of variables for $\mathcal{P}$ for numerics and plots}\label{sec:Psubstitution}

The variables $s=\mu_{kl}^2+\mu_{km}^2$ and $t=\mu_{kl}\mu_{km}$ entering $\mathcal{C}_{LM}$ in eq.~(\ref{CLM_FINAL}) are convenient for the angular algebra, but they hide the constraint imposed on $(\mu_{kl},\mu_{km},\mu_{lm})$ by the triangle inequality (eq.~\ref{triangle_ineq}). For the numerical implementation of $\mathcal{P}$ and for plotting purposes it is useful to switch to a pair of dimensionless variables $(X,Z)$ whose allowed domain is independent of $\mu_{lm}$. We define those two variables via:
\be\label{Psub_def}
s&=&(1-\mu_{lm})(1-X)+2\,\mu_{lm}\,Z\ \nonumber\\ t&=&\tfrac{1}{2}(\mu_{lm}-1)(1-X)+Z\ .
\ee
For any fixed $\mu_{lm}\in[-1,1]$, the triangle inequality eq.~(\ref{triangle_ineq}) is equivalent to $X\geq0$, $Z\geq0$, and $X+Z\leq1$.

Substituting eq.~(\ref{Psub_def}) into $\mathcal{C}_{LM}(s,t,\mu_{lm})$ and then into eq.~(\ref{FINALP}) (equivalently eq.~\ref{FINALP_exp}) turns $\mathcal{P}$ into the following expansion
\be\label{Pexp_XZ}
\mathcal{P}(\bm k,\bm l,\bm m)=\sum_{a,b\geq 0} c_{ab}(\mu_{lm},p,k)\,(1-\mu_{lm})^{a}\,X^{a}\,Z^{b}\ ,
\ee
with $p= l m$ as usual, and the coefficients $c_{ab}$ depending on $\mu_{lm}$, $p$, $k$, and on the cosmology through $P_L$.
Note that every power $X^a$ is multiplied by the prefactor $(1-\mu_{lm})^{a}$: the $a$-th order in $X$ vanishes at $\mu_{lm}=1$ and/or when $X=0$ for $a\geq 1$.

In the linear (and even mildly nonlinear) regime, $\mathcal{P}$ is dominated by a small number of terms in eq.~(\ref{Pexp_XZ}). In the fully linear limit of  low $k$ \textit{and} low $\sqrt{p}$, both small relative to $1/\sigma_v$ (see eq.~\ref{2pfz_simple_expanded}), the expansion collapses to
\be\label{Plin_limit}
\mathcal{P}(\bm k,\bm l,\bm m)\;\approx\;-\frac{p}{k^{2}}\left[\tfrac{1}{2}(\mu_{lm}-1)(1-X)+Z\right] P_{L}(k)=c_{00}^{\mathrm{lin}}+c_{10}^{\rm lin}(1-\mu_{lm})X+c_{01}^{\rm lin}\,Z\ ,
\ee
with $c_{00}^{\rm lin}=-\tfrac{p}{2k^{2}}(\mu_{lm}-1)P_{L}(k)$, $c_{10}^{\rm lin}=-\tfrac{p}{2k^{2}}P_{L}(k)$, and $c_{01}^{\rm lin}=-\tfrac{p}{k^{2}}P_{L}(k)$. All $c_{ab}$ with $a+b\geq2$ vanish at linear order. The square bracket equals $\mu_{kl}\mu_{km}$ (eq.~\ref{Psub_def}), so eq.~(\ref{Plin_limit}) reduces to $\mathcal{P}\approx-(p/k^2)\,\mu_{kl}\mu_{km}\,P_L(k)$. In the next sections we show our numerical pipeline and results for $\mathcal{P}$, and we test the linear approximation above.

\subsection{Numerical implementation of $\mathcal{P}$}\label{sec:Pnumerics}

We implemented both eq.~(\ref{FINALP_exp}) and eq.~(\ref{FINALP}) for the numerical evaluation of $\mathcal{P}$. The latter separates the angular dependence from the radial integrals, but on our hardware we find the evaluation of eq.~(\ref{FINALP_exp}) to a fixed precision to be faster (in terms of wall clock time). The pipeline used to produce the figures in this paper follows the substituted-variable form of Section~\ref{sec:Psubstitution}. The cosmology-dependent inputs $\zeta(q)$ and $\gamma(q)$ are reconstructed from the linear total-matter power spectrum $P_L(k)$ evaluated with the Planck 2018 cosmological parameters \cite{Planck2020}.

The pipeline splits into three stages: (i) An \textit{angular} stage that assembles the cosmology-independent angular polynomials at high arbitrary precision and stores the coefficient table
\be\label{CLM_ab_def}
\mathcal{C}_{LM}(X,Z;\,\mu_{lm})=\sum_{a,b\geq 0}\mathcal{C}_{LM,\,ab}(\mu_{lm})\,X^{a}\,Z^{b}
\ee
to disk per $\mu_{lm}$-grid value indexed by $(L,M,a,b)$. In the above equation $\mathcal{C}_{LM}$ is that given by eq.~(\ref{CLM_FINAL}) with the substitution of eq.~(\ref{Psub_def}). The coefficients $\mathcal{C}_{LM,\,ab}$ are simply defined as the coefficients in front of $X^aZ^b$ after that substitution is performed.
 (ii) The next stage is a \textit{radial} stage that evaluates the Hankel transforms in eqs.~(\ref{Glmn}, \ref{Elmn}) using FFTLog~\cite{Hamilton_2000,Simonovic_2018} run at two bias values, $\pm 0.9$, and stitched on the $k$-grid. (iii) Then we follow with a \textit{contraction} stage that amounts to performing the sums in either eq.~(\ref{FINALP_exp}) or eq.~(\ref{FINALP}), which contracts the angular tables against the radial kernel tables with compensated long-double summation. Because the angular tables depend only on $(L,M,\mu_{lm})$, they are cosmology-independent and computed once across runs; only the radial stage needs re-running when the cosmology changes. Convergence is checked per-$(k,\mu_{lm},p)$ at $\mathcal{O}(1\%)$ relative tolerance from the contributions of the sum of the last few $L$ and $M$ terms. Implementation details (precision settings, the angular-cutoff tiers, the Planck-taper window, and the bias-splicing matching condition) are collected in Appendix~\ref{app:numerics}, with the cross-check of the bias-splicing against direct integration shown in Appendix~\ref{app:interp}. With this pipeline and those implementation details, we are able to make the plots of the next section to within $\mathcal{O}(1\%)$ accuracy for the ranges of variables shown.

\subsection{Numerical results}\label{sec:Pnumresults}

We test the pipeline of Section~\ref{sec:Pnumerics} against two analytic identities for $\mathcal{P}$. We then find that the figures indicate a closed-form approximation to $\mathcal{P}$ of broader validity.

\begin{figure}[p]
\centering
\includegraphics[width=0.9\textwidth]{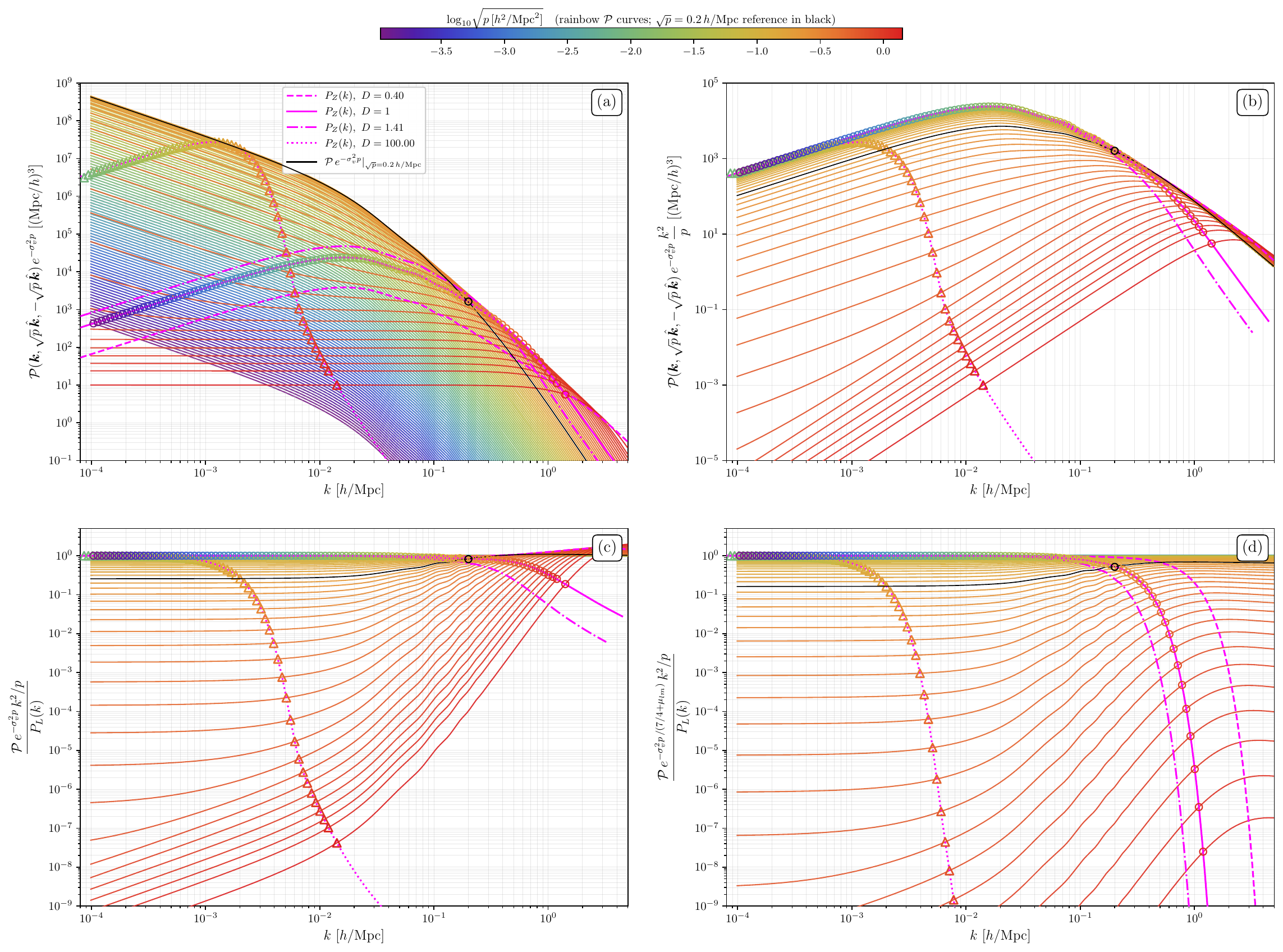}
\caption{\small The four panels rescale the Zel'dovich identity, eq.~(\ref{PZ_check}), into the normalization of the top-left panel of Fig.~\ref{fig:Plin_limit_check_X0}, one step at a time. We use the figure as a check of eq.~(\ref{PZ_check}) (see discussion in main text). Each of the four panels shows rescalings of $\mathcal{P}(\bm k,\sqrt p\,\hat{\bm k},-\sqrt p\,\hat{\bm k})$ for $112$ values of $\sqrt p$ log-uniformly spaced in $[1.04\times 10^{-4},\,1.41]\,h/\mathrm{Mpc}$, color-coded by $\log_{10}\sqrt p$ via the shared colorbar at the top. The $\sqrt p=0.2\,h/\mathrm{Mpc}\approx \sigma_v^{-1}$ reference curve is drawn in black, on top of the rainbow set of curves.  The shared legend (in panel~(a)) lists the magenta overlays $P_Z(k,D)$ at the values of $D$ shown. In each panel, the magenta curves are rescaled by the panel-specific factor (as shown on the  respective ordinate label) evaluated at $p=D^2 k^2$ for that $D$.  From eq.~(\ref{PZ_check}) we can see that the $P_Z(k,D)$ curves can be thought of as cross-sections of the shown $\mathcal{P}e^{-\sigma^2_vp}$ in the $(p,k)$ plane at constant  $D=\sqrt{p}/k$. The values of $D$ are chosen as follows: $D=0.4$ corresponds to redshift of $\approx 2$ when $P_Z$ is most enhanced relative to $P_L$ at small scales (explaining the curves going above 1 at high $k$ in panel \emph{(c)}). At later times $P_Z$ gets suppressed. Present day corresponds to $D=1$, and for that $P_Z$, we circled the locations where $\sqrt{p}=k$ on the $\mathcal{P}$ curves and see that those locations are locations through which the $P_Z$ curve passes, thus confirming eq.~(\ref{PZ_check}). We did a similar check for all  $P_Z$'s shown, but the only other equivalent markers we show are for $D=100$ (open triangles marking locations where $\sqrt{p}=Dk$ on the $\mathcal{P}$ curves) to avoid visual overload. The large  $D$ values are into the future: one relatively close to today ($D=1.41$), showing the further suppression of $P_Z$ relative to $P_L$, and one with large $D=100$, showing a pronounced suppression. The four panels show: \emph{(a)} the raw $\mathcal{P}\,e^{-\sigma_v^2 p}$; \emph{(b)} multiplied by $k^2/p$, removing the $p/k^2$ coefficient of the linear $c_{00}^{\rm lin}$ (see eq.~\ref{Plin_limit}); \emph{(c)} further divided by the linear power spectrum $P_L(k)$ of today; \emph{(d)} with $e^{-\sigma_v^2 p}$ replaced by $e^{-\sigma_v^2 p/(7/4+\mu_{lm})}$ (here, $\mu_{lm}=-1$  so the factor equals $e^{-\sigma_v^2 p\,\cdot\,4/3}$). The \emph{(d)} panel's plotted quantity (here on a log scale) is scaled to match the top-left panel of Fig.~\ref{fig:Plin_limit_check_X0} (there shown on a linear scale). The scales affect $P_Z$ in the following way: panel \emph{(a)} shows $P_Z$; \emph{(b)} divides by $D^2$ which collapses the magenta curves in the linear regime to $P_L$; \emph{(c)} divides by $P_L$ which makes the magenta curves asymptote to 1 in the linear regime at low $k$; \emph{(d)} the rescaling factor stretches the spacing between the curves at large $p$ (or equivalently large $Dk$). The horizontal asymptotes at high $p$ and low $k$ in panel \emph{(a)} are the asymptotically-caustic-generated tail of $P_Z$, as shown in Appendix~\ref{app:asympt}. Indeed, at fixed large $\sqrt p=Dk$, the low-$k$ end of those  curves is the $D\gg 1$ limit. There, for $\sqrt p\,\sigma_v\gg1$, $P_Z$ becomes a power law in $\sqrt p=Dk$ alone (see the open triangles at highest $k$'s), with no separate $k$ or $D$ dependence -- which is why the asymptotes of $\mathcal{P}$ at constant $p$ are flat in $k$ and as we see in detail in Appendix~\ref{app:asympt}, caustic in nature (but see Section~\ref{sec:asympt:summary}).}
\label{fig:zeldovich_check}
\end{figure}

\textbf{Numerically testing the Zel'dovich identity, eq.~(\ref{PZ_check}).}\quad Figure~\ref{fig:zeldovich_check} confirms the Zel'dovich identity, eq.~(\ref{PZ_check}), to better than $\mathcal{O}(1\%)$ across the entire displayed $\sqrt p$ range. Since eq.~(\ref{PZ_check}) sets $\hat{\bm l}=\hat{\bm k}=-\hat{\bm m}$, it has $\mu_{lm}=-1$, $\mu_{kl}=1$, and $\mu_{km}=-1$, which by eq.~(\ref{Psub_def}) means $X=Z=0$. With $X=0$ every $a\geq 1$ term in eq.~(\ref{Pexp_XZ}) drops out, and with $Z=0$ every $b\geq 1$ term drops out, so the check directly probes a single coefficient in eq.~(\ref{Pexp_XZ}), namely $c_{00}(-1,p,k)$.

The Figure shows both sides of the equation with various rescalings in each panel (see the Figure caption for details). The multiple rainbow curves correspond to $\mathcal{P}(\bm k,\sqrt{p}\hat{\bm k},-\sqrt{p}\hat{\bm k})e^{-\sigma_v^2 p}$, with the locations $\sqrt{p}=k$ circled. We expect $P_Z(k)$ to pass through those same locations when evaluated at $D=1$, and indeed it does (solid magenta curve) in all panels.

\begin{figure}[t!]
\centering
\includegraphics[width=0.95\textwidth]{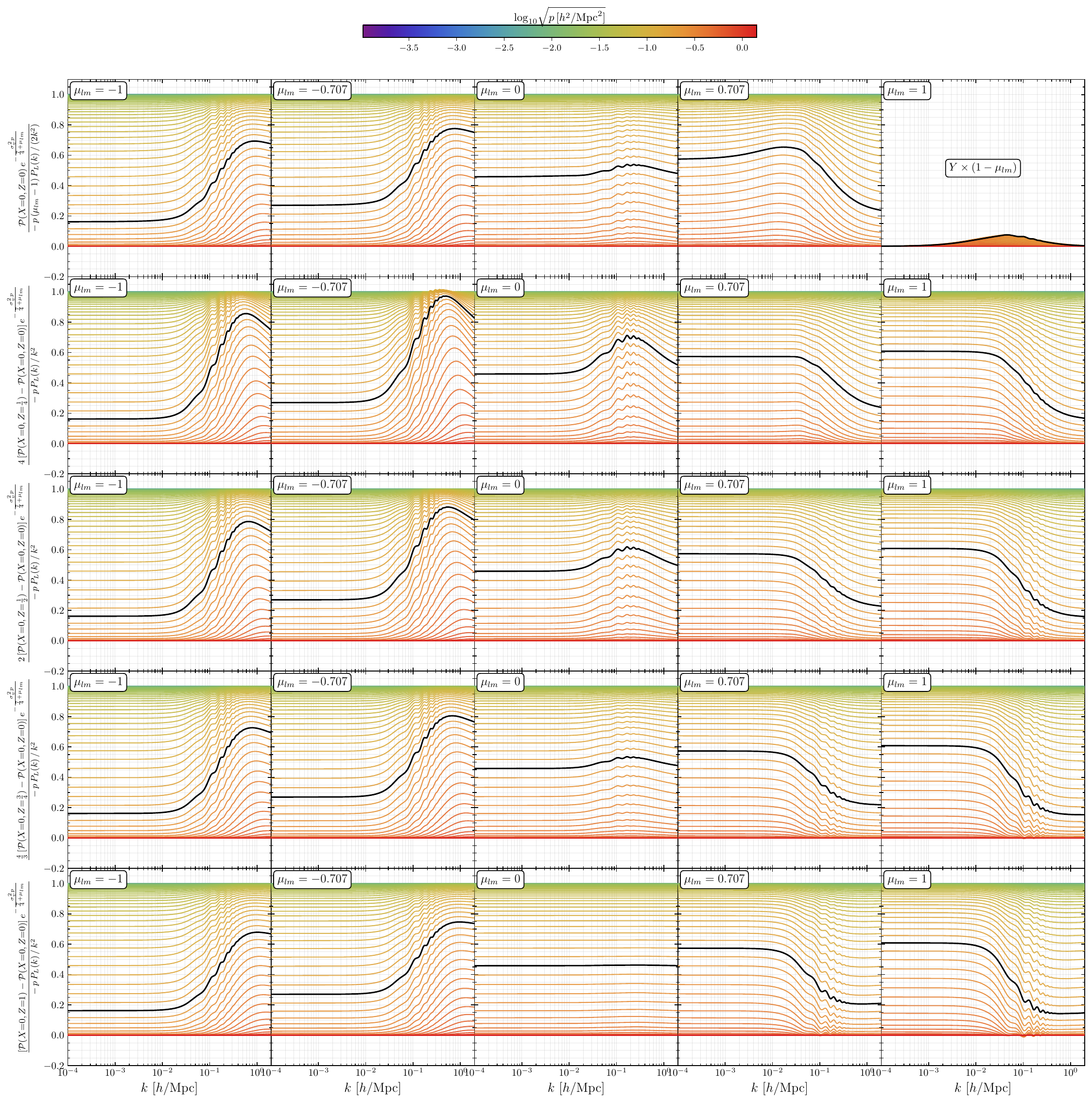}
\caption{Numerical check of the linear-$P_L$ limit, eq.~(\ref{Plin_limit}). $\mathcal{P}$ is fully non-linear (evaluated using eq.~\ref{Pexp_XZ}). Here we evaluate it at $X=0$ where only the $a=0$ terms of the sum over $X^aZ^b$ in eq.~(\ref{Pexp_XZ}) contribute. \emph{Top row:} $\mathcal{P}(X{=}0,Z{=}0)/c_{00}^{\rm lin}$ ratio scaled by $e^{-\sigma_v^2 p/(7/4+\mu_{lm})}$ which compresses the high-$\sqrt p$ curves into the panel range. \emph{Rows 2--5:} $[\mathcal{P}(X{=}0,Z)-\mathcal{P}(X{=}0,Z{=}0)]/(Zc_{01}^{\rm lin})$ ratio  scaled the same way and evaluated at $Z=1/4$, $1/2$, $3/4$, $1$. As per  eq.~(\ref{Plin_limit}), all curves should asymptote to 1 in the linear regime, which the figure confirms (for the top-right panel, see below). Columns scan $\mu_{lm}\in\{-1,-0.7,0,+0.7,+1\}$. The top-right cell ($\mu_{lm}=+1$) uses a \emph{different} normalization than the rest of the top row: dropping the $(1-\mu_{lm})$ factor from $c_{00}^{\rm lin}$ (i.e.\ denominator $pP_L(k)/(2k^2)$ instead of $p(1-\mu_{lm})P_L(k)/(2k^2)$) avoids the $0/0$ that the original normalization would produce there. That change in the denominator is flagged by a ``$Y\times(1-\mu_{lm})$'' box in the panel ($Y$ being the expression in the label of the ordinate). The values in that panel sit near zero at low $p$ as predicted by linear theory. Similar to Fig.~\ref{fig:zeldovich_check}, each of the $112$ curves is at fixed $\sqrt p\in[1.04\times 10^{-4},\,1.41]\,h/\mathrm{Mpc}$, color-coded by $\log_{10}\sqrt p$; the $\sqrt p=0.2\,h/\mathrm{Mpc}\approx 1/\sigma_v$ reference is in black.}
\label{fig:Plin_limit_check_X0}
\end{figure}

\begin{figure}[h!]
\centering
\includegraphics[width=0.95\textwidth]{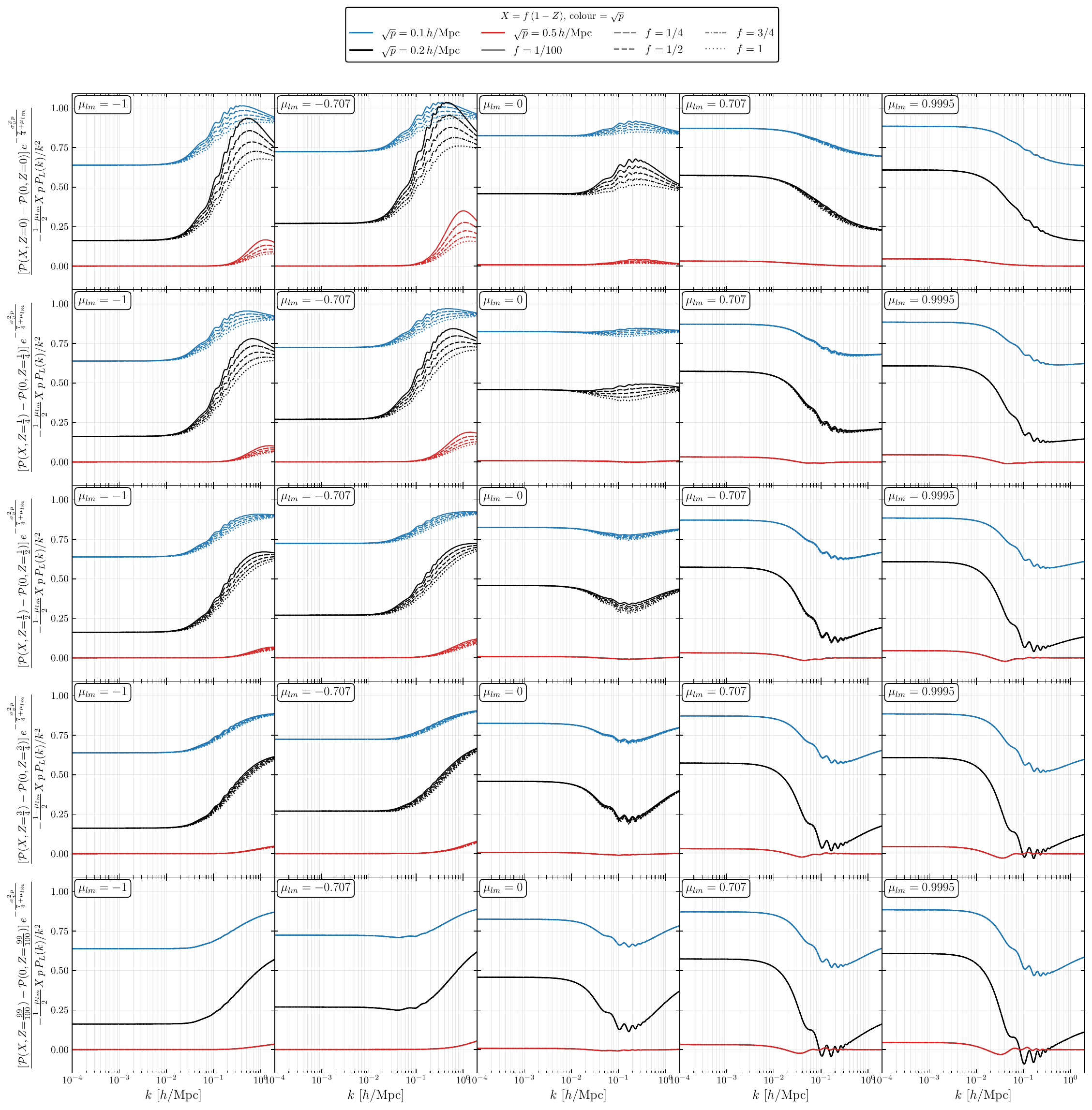}
\caption{Numerical check of the $X$-dependence of the linear-$P_L$ limit, eq.~(\ref{Plin_limit}); this is the linearization of Fig.~\ref{fig:Plin_normed_Xpieces}, obtained by replacing $\mathcal{P}_*$ with its linear value $pP_L(k)/k^2$ (eq.~\ref{Plin_limit}). Each panel plots the $X$-slope ratio $[\mathcal{P}(X,Z)-\mathcal{P}(X{=}0,Z)]/[-p\,(1-\mu_{lm})\,X\,P_L(k)/(2k^2)]$, scaled by $e^{-\sigma_v^2 p/(7/4+\mu_{lm})}$ -- the same ad hoc range compression factor used in Fig.~\ref{fig:Plin_limit_check_X0}. Since eq.~(\ref{Pstar_lowk}) gives $\mathcal{P}(X,Z)-\mathcal{P}(X{=}0,Z)=-\tfrac{1-\mu_{lm}}{2}X\,\mathcal{P}_*$ wherever the identity holds, this (unscaled) ratio $\to1$ at low $\sqrt p\,\sigma_v$ (previous figure) and deviates from $1$ as nonlinearity grows. \emph{Rows} fix $Z\in\{0,1/4,1/2,3/4,99/100\}$ ($Z=1$ is omitted, since there $X\leq X_{\max}=1-Z=0$ leaves no $X$ freedom). Within each row, $X$ is swept as a fraction $f$ of its triangle-allowed maximum $X_{\max}=1-Z$, i.e.\ $X=f(1-Z)$ with $f\in\{1/100,1/4,1/2,3/4,1\}$ (encoded by line style, with $f=1/100$ drawn solid; $f=0$, i.e.\ $X=0$ is not shown as it results in 0 for the fraction shown). Line color encodes $\sqrt p\in\{0.1,0.2,0.5\}\,h/\mathrm{Mpc}$. Columns scan $\mu_{lm}\in\{-1,-0.7,0,+0.7,+0.9995\}$; because the $(1-\mu_{lm})^a$ prefactor of eq.~(\ref{Pexp_XZ}) makes the $X$-dependence vanish identically at $\mu_{lm}=+1$, the rightmost column  uses the nearest output value $\mu_{lm}<1$ (here $\mu_{lm}\simeq0.9995$).}
\label{fig:Plin_limit_check_Xpieces}
\end{figure}

\textbf{Testing the low-$p$, linear-$P_L$ limit, eq.~(\ref{Plin_limit}).}\quad Figures~\ref{fig:Plin_limit_check_X0} and~\ref{fig:Plin_limit_check_Xpieces} check $\mathcal{P}$ as a ratio to its linear counterpart given by eq.~(\ref{Plin_limit}) for a variety of $\mu_{lm}$, $X$ and $Z$ values. The figures confirm that at low $\sqrt p\sigma_v$ and low $k\sigma_v$ the linear approximation is valid for all values of $\mu_{lm}$, $X$ and $Z$. That can be seen by the collapse of all low-$p$ curves to the asymptotic value of $1$. The details can be found in the Figure captions.

\begin{figure}[t!]
\centering
\includegraphics[width=0.95\textwidth]{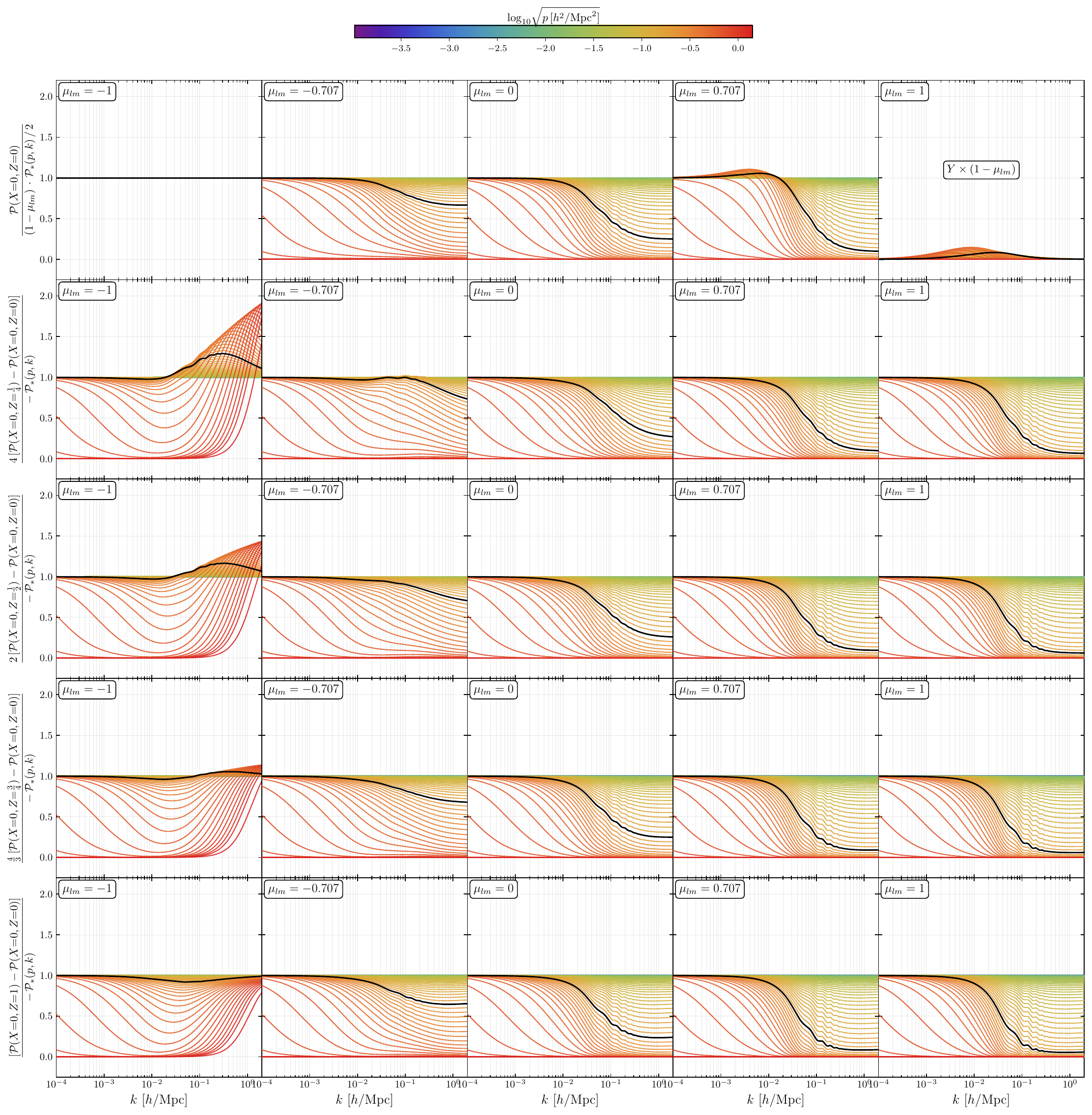}
\caption{Numerical check of the $\mathcal{P}_*$-normalized  identity, eq.~(\ref{Pstar_lowk}). $\mathcal{P}$ is fully non-linear (evaluated using eq.~\ref{Pexp_XZ}). Here we evaluate it at $X=0$ where only the $a=0$ terms of the sum over $X^aZ^b$ in eq.~(\ref{Pexp_XZ}) contribute. \emph{Top row:} $\mathcal{P}(X{=}0,Z{=}0)/[\tfrac{1-\mu_{lm}}{2}\mathcal{P}_*]$ ratio. \emph{Rows 2--5:} $[\mathcal{P}(X{=}0,Z)-\mathcal{P}(X{=}0,Z{=}0)]/(-Z\mathcal{P}_*)$ ratio evaluated at $Z=1/4$, $1/2$, $3/4$, $1$. No exponential prefactor is needed (in contrast with Fig.~\ref{fig:Plin_limit_check_X0}). Similar to Fig.~\ref{fig:Plin_limit_check_X0}, at low $p$, the curves asymptote to 1 for all $k$. What we found interesting as a numerical result is that here all curves asymptote to 1 at low $k$ for \emph{every} $\sqrt p$ as well, which is in contrast with Fig.~\ref{fig:Plin_limit_check_X0}, where that is no longer true at high $p$ (for the top-right panel, see below). Columns scan $\mu_{lm}\in\{-1,-0.7,0,+0.7,+1\}$. The values plotted in the top-left cell ($\mu_{lm}=-1$) are identically $1$ by construction, since $\mu_{lm}=-1$ is the value used in the definition of $\mathcal{P}_*$. The top-right cell ($\mu_{lm}=+1$, $Z=0$) is singular because the $\tfrac{1-\mu_{lm}}{2}$ factor in its denominator vanishes there; in that panel we instead plot the ratio multiplied by $(1-\mu_{lm})$ (equivalently, denominator $\tfrac12\mathcal{P}_*$ rather than $\tfrac{1-\mu_{lm}}{2}\mathcal{P}_*$), and we flag that change by a ``$Y\times(1-\mu_{lm})$'' box in the panel. The displayed quantity sits near zero at low $p$ as expected from eq.~(\ref{Pstar_lowk}). Similar to Fig.~\ref{fig:Plin_limit_check_X0}, each of the $112$ curves is one $\sqrt p\in[1.04\times 10^{-4},\,1.41]\,h/\mathrm{Mpc}$, color-coded by $\log_{10}\sqrt p$; the $\sqrt p=0.2\,h/\mathrm{Mpc}$ reference is in black.}
\label{fig:Plin_normed_X0}
\end{figure}

\begin{figure}[h!]
\centering
\includegraphics[width=0.95\textwidth]{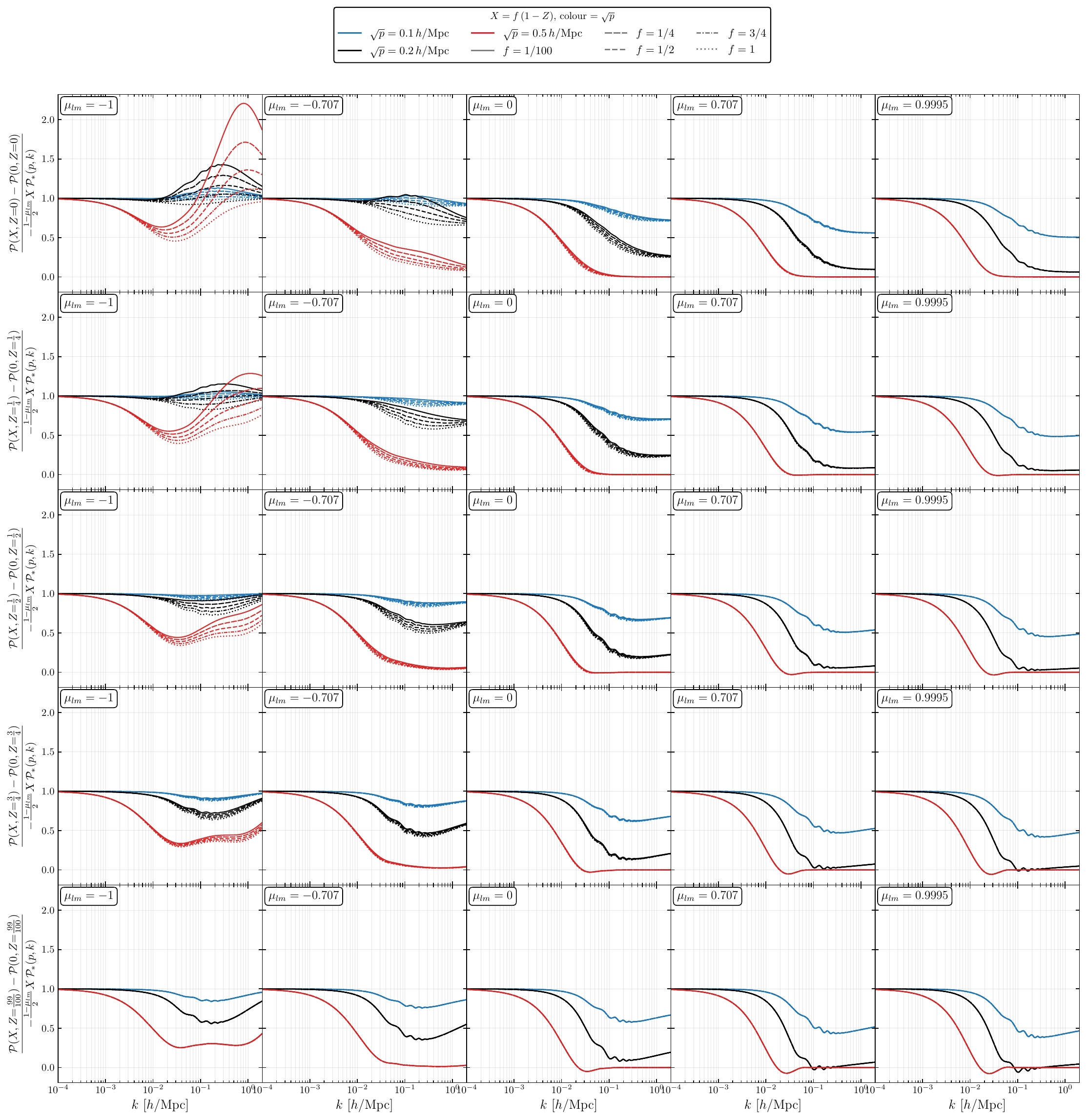}
\caption{Numerical check of the $X$-dependence of the $\mathcal{P}_*$-normalized identity, eq.~(\ref{Pstar_lowk}). Each panel plots the $X$-slope ratio $[\mathcal{P}(X,Z)-\mathcal{P}(X{=}0,Z)]/[-\tfrac{1-\mu_{lm}}{2}X\,\mathcal{P}_*]$. Eq.~(\ref{Pstar_lowk}) gives $\mathcal{P}(X,Z)-\mathcal{P}(X{=}0,Z)\approx-\tfrac{1-\mu_{lm}}{2}X\,\mathcal{P}_*$ for all $k$ at low $p$ and for low enough $k$ for all $p$. Therefore, the ratio shown should asymptote to $1$ in those regimes for every $f$, departing from $1$ only in the deep non-linear regime (high $k$, high $p$). \emph{Rows} fix $Z\in\{0,1/4,1/2,3/4,99/100\}$ ($Z=1$ is omitted, since there $X_{\max}=1-Z=0$). Within each row $X$ is swept as $X=f(1-Z)$, a fraction $f$ of the triangle-allowed maximum $X_{\max}=1-Z$, with $f\in\{1/100,1/4,1/2,3/4,1\}$ (encoded by line style, with $f=1/100$ drawn solid). Line color encodes $\sqrt p\in\{0.1,0.2,0.5\}\,h/\mathrm{Mpc}$. Columns scan $\mu_{lm}\in\{-1,-0.7,0,+0.7,+0.9995\}$; since the $(1-\mu_{lm})^a$ prefactor of eq.~(\ref{Pexp_XZ}) makes the $X$-dependence vanish identically at $\mu_{lm}=+1$, the rightmost column instead uses the nearest output value $\mu_{lm}<1$ (here $\mu_{lm}\simeq0.9995$). }
\label{fig:Plin_normed_Xpieces}
\end{figure}

\textbf{Extended Zel'dovich identity, eq.~(\ref{Pstar_lowk}).}\quad Figures~\ref{fig:Plin_normed_X0} and~\ref{fig:Plin_normed_Xpieces} are similar to Figures~\ref{fig:Plin_limit_check_X0} and~\ref{fig:Plin_limit_check_Xpieces}, but instead of showing the ratio of $\mathcal{P}$ to its linear counterpart, we take the ratio of $\mathcal{P}$ to a reference $\mathcal{P}_*$ defined by
\be\label{Pstar_def}
\mathcal{P}_*(p,k)\;\equiv\;\mathcal{P}(\mu_{lm}{=}-1,\,X{=}0,\,Z{=}0,\,p,\,k)
\;=\;P_Z\!\left(k,\,D=\tfrac{\sqrt p}{k}\right)\,e^{\sigma_v^2\,p}\,,
\ee
where the second equality follows from the Zel'dovich identity, eq.~(\ref{PZ_check}), with $|\bm l|=|\bm m|=Dk$ (cf.\ eq.~\ref{Pcurly}). Note, that here $\mathcal{P}$ is not taking the standard $\bm k,\bm l,\bm m$ as arguments but their reparametrized counterparts: $\mu_{lm},X,Z,p,k$. Inspecting Figures~\ref{fig:Plin_normed_X0} and~\ref{fig:Plin_normed_Xpieces}, we see that we can generalize eq.~(\ref{Plin_limit}) to
\be\label{Pstar_lowk}
\mathcal{P}(\mu_{lm},X,Z,p,k)\;\approx\;\mathcal{P}_*(p,k)\,\Bigl[\tfrac{1-\mu_{lm}}{2}(1-X)-Z\Bigr]\,,
\ee
valid for \emph{all} $\sqrt p$ at low enough $k\sigma_v$ as well as for all $k$ at low enough $\sqrt{p}\sigma_v$. Using $\tfrac{1-\mu_{lm}}{2}(1-X)-Z=-\mu_{kl}\mu_{km}$ (eq.~\ref{Psub_def}), eq.~(\ref{Pstar_lowk}) is equivalent to
\be\label{Pstar_lowk1}
\mathcal{P}(\bm k,\bm l,\bm m)\;\approx\;-P_Z\!\left(k,\,D=\tfrac{\sqrt{lm}}{k}\right)\,e^{\sigma_v^2\,lm}\,\bigl(\hat{\bm l}\cdot\hat{\bm k}\bigr)\bigl(\hat{\bm m}\cdot\hat{\bm k}\bigr)\,,
\ee
which, plugged into eq.~(\ref{2pfz_simple}), gives a compact closed-form approximation to the connected phase-space two-point cumulant:
\be\label{2pfz_simple1}
\langle f(\bm k,\bm w,D)\,f(\bm k',\bm w',D')\rangle_{\mathrm c}\;\approx\;-(2\pi)^3\,\delta_D(\bm k+\bm k')\,e^{-\frac{\sigma_v^2}{2}(|\beta|-|\beta'|)^2}\,P_Z\!\left(k,\,D''=\tfrac{\sqrt{\beta\beta'}}{k}\right)\,\bigl(\hat{\bm\beta}\cdot\hat{\bm k}\bigr)\bigl(\hat{\bm\beta'}\cdot\hat{\bm k}\bigr)\,,\nonumber\\
\ee
with $\bm \beta$ defined in eq.~(\ref{beta}) and $\bm\beta'$ defined in the text right after that.
Equation~(\ref{2pfz_simple1}) is one of the central results of this section. In the linear limit it reduces to eq.~(\ref{2pfz_simple_expanded}). It reproduces eq.~(\ref{PZ_check}) identically for $\bm\beta=-\bm\beta'=D\bm k$. The figures indicate that eq.~(\ref{2pfz_simple1}) is valid for all $p$, $\mu_{lm}$, $X$ and $Z$ as long as $k\sigma_v$ is small enough as well as for all $k$ for small $p$. Equation~(\ref{2pfz_simple1}) collapses the raw 11 DoF's of its left-hand side into the closed-form approximation on the right-hand side\footnote{See the discussion at the end of Section~\ref{sec:2ptf} for an explanation why for large $\beta$ and $\beta'$, we need future values  (large $D''$) of $P_Z$.}. This approximation, together with the checks of this section, gives us confidence in our numerical scheme for calculating the CDM phase-space two-point function.

\section{Summary}\label{sec:summary}

The paper has two parts and a shared motivation. We isolate one manipulation in the perturbative treatment of cosmological large-scale structure (LSS) that causes standard perturbation theories (PTs) to eventually fail to converge with the growth of structure. Then we introduce the first steps toward a numerical implementation of an alternative that avoids that manipulation.

\paragraph{Failures in Dominated Convergence.} The standard perturbative approaches to the study of LSS involve expanding the CDM overdensity $\delta$ in powers of the CDM particle displacement $\bm s$. Depending on the flavor of the PT, that expansion may be done in the Poisson equation (Lagrangian PT) or in all equations (Eulerian PT). One can think of that expansion as approximating mass conservation since $\delta[\bm s]$ can be understood simply as displacing the particles and then doing a Particle Mesh assignment of each particle's mass on a grid to recover the overdensity. In the case of LPT, the expansion corresponds to perturbatively applying conservation of the active gravitational mass; in EPT -- perturbatively applying conservation of both the inertial and the active gravitational mass. Either way, we show that with the growth of structure, those approximations eventually fail not because of the expansion of $\delta$ in $\bm s$ itself, but because of further manipulations that those PTs require, which violate Lebesgue's dominated-convergence (DC) theorem.

To make that statement more precise, let us first focus on EPT. The Fourier-space PT series for $\delta(\bm k)$ in powers of $\bm s$ converges absolutely for all $\bm k$. A breakdown appears when one insists on obtaining $\delta(\bm x)$ order by order in real space (a necessary step in calculating composite operators such as $\delta(\bm x)\bm v(\bm x)$ at a fixed order): this requires exchanging an infinite sum (or a limit) with a Fourier integral, and depending on the shape of the displacement, the resulting partial sums can carry an exponentially large sliding bump at high $k$ whose integral does not vanish. The above exchange is then forbidden by Lebesgue's dominated-convergence (DC) theorem, and thus leads to failure to converge.

We show that this DC failure is why the order-by-order convergence of the Eulerian $\delta(\bm{x})$, expanded in $\bm s$, is controlled by a family of advection (derivative-displacement) combinations of $\bm s$. In 1D this family is $|s\,s''|$, $|s^2 s'''|$, $|s\,s'\,s''|$, $\ldots$ (familiar from e.g. \cite{Tassev_Zaldarriaga_2012,Tassev_2014,Porto_Senatore_Zaldarriaga_2014}), alongside the standard $|\delta_L|=|s'|$; in 3D the same family appears with appropriately contracted indices. As the displacement amplitude for a particular realization of $s(q)$ is scaled up, the first divergence of $\delta_{x,\Sigma}$ occurs at the $x$ where either $|s'|$ or any one of these advection parameters first reaches $\mathcal{O}(1)$; which parameter and where depends on the shape of $s$.

The same DC obstruction occurs in the LPT expansion of the Lagrangian overdensity $\delta_q(\bm q)$ entering Poisson's equation. The expansion of $\delta_q(\bm q)$ in powers of $\bm s$ can again be done with an infinite radius of convergence, similar to the expansion of $\delta(\bm x)$ in EPT in Fourier space. However,  to separate the individual orders of  $\delta_q(\bm q)$ one needs to perform the same dominated-convergence-violating exchange as done in EPT. The resulting series can then be ``resummed'' to recover the familiar single-stream formula. Its validity region $|s'|<1$ is the resummation region of the DC-violating exchanged series, bounded by shell crossing on one side and severely underdense regions on the other. We find that the additional advection family of parameters, which control the DC violation of EPT, do not control the obstruction in LPT. That fact gives us a new way of thinking about the advantages of working in Lagrangian space \cite{Tassev_Zaldarriaga_2012,Tassev_2014,Porto_Senatore_Zaldarriaga_2014,Carlson_2012}, and helps us understand why when LPT keeps the density power spectrum unexpanded in the advection terms, that  is not just an ad hoc ``resummation'' trick, but rather a rigorous way to stave off the DC obstruction until a single parameter (as opposed to a whole family of parameters), $|s'|$ becomes large.

We should stress that we do not claim our analysis exhausts every way the full nonlinear problem can fail. However, the DC obstructions we identified per realization of $\bm s$ are inevitable when one expands $\delta[\bm s]$, and we do not expect them to disappear under ensemble averaging over initial conditions. Effective Field Theories of LSS in Lagrangian and Eulerian space can clearly avoid the problem by smoothing at short scales, and thus by eliminating the sliding bump in $\bm k$. But that means that their smoothing scale is controlled by the DC obstruction, independent of whether non-linearities beyond mass conservation are important or not.

\paragraph{BBGKY+ZA as a concrete numerical program.}  Motivated by the observed failure described above, we turned to a route that never expands $\delta$ in $\bm s$: instead we would like to evolve the connected cumulants of the phase-space distribution $f(\bm x,\bm v,D)$ under the BBGKY hierarchy, with initial conditions and closure supplied by the Zel'dovich approximation. In this paper we refer to this combination as BBGKY+ZA\footnote{To our knowledge, the prospect of using the full (unexpanded in the linear displacement) ZA phase-space $n$-point functions together with the BBGKY hierarchy to study the statistics of CDM was first mentioned in \cite{Tassev_2011}.}. The DC obstruction identified in the first part of this paper is absent from BBGKY+ZA by construction. Yet, other modes of failure from the full nonlinear dynamics may still arise.

To make progress with the BBGKY+ZA program, we derived the ZA one-, two-, three-, and $n$-point phase-space correlators, with the higher-$n$-point correlators ($n>2$) expressed as products and convolutions of a kernel $\mathcal{P}$, which is proportional to the ZA phase-space two-point function. The first obstacle, then, is that evaluating $\mathcal{P}$ requires integrating a wildly oscillating integrand.  The central technical result of this paper is a stable numerical representation of $\mathcal{P}$:
we replace the oscillating integrand with discrete sums, and show that arbitrary-precision arithmetic combined with FFTLog handles them successfully. In the end, the same lesson recurs across both halves of the paper: an absolutely convergent series can still hide enormous cancellations between exponentially large terms (for example, we kept more than $120$ significant figures for $\delta_{k,\Sigma}$ in the toy model in Fig.~\ref{fig:esc}) and it is arbitrary-precision arithmetic that renders such a series usable. We present first numerical results for $\mathcal{P}$.

\paragraph{Outlook.}  The natural next steps are to: (i) evaluate the exact ZA phase-space 3-point function using $\mathcal{P}$ and use that as closure in the BBGKY equation for the equal-time two-point cumulant; (ii) integrate the BBGKY hierarchy forward in time with the ZA phase-space cumulants supplying the initial conditions; and (iii) compare with simulations. This paper does not carry out that integration; it provides the building blocks. An open question is whether the ZA would be adequate as closure of the hierarchy, and whether carrying out the BBGKY+ZA program together with short-scale EFT-like regulation can reach the non-linear regime and beyond shell crossing.

$N$-body simulations reach that regime per realization, so statistics are accessible only by averaging over many runs. BBGKY acts directly on cumulants (ensemble-averaged, smooth functions of their arguments), so if the hierarchy can be controlled through shell crossing at the cumulant level, it would expose those statistics without the per-realization averaging cost. We make no claim to demonstrate such access here; this paper is one step toward it -- a stable numerical footing for BBGKY+ZA at the two-point level, together with the analytical ingredients needed to push the program forward. The DC obstruction is absent by construction, so -- unlike in EPT, LPT, and their EFTs -- no short-scale regulation is forced just to cure it; any regulation BBGKY+ZA eventually needs would address nonlinear physics beyond mass conservation instead.

The C++ code and analysis scripts that produced the figures of this paper are released under the GNU GPL v3 at \url{https://github.com/stassev/CDM-phase-space-kernel-ZA}.

\appendix

\section{A derivative expansion of $\delta_{x,\Sigma}$ in 1D and radius of convergence\label{app:advection-form}}

In Section~\ref{sec:saddle1} we noted that the convergence of $\delta_{x,\Sigma}$ at a given $x$ is controlled by a family of advection derivative-displacement combinations of $s$ at $x$: $|s'|$, $|s\,s''|$, $|s^2 s'''|$, $|s\,s'\,s''|$, and so on at higher orders. Writing $\delta_{x,\Sigma}$ as $\sum_n \partial_x^n(-s)^n/n!$ (eq.~\ref{deltaXn1}) does not display these advection combinations cleanly: the operator $\partial_x^n$ acting on the product $s^n$ generates them only after the Leibniz (product) rule is applied $n$ times. Here we re-organize $\delta_{x,\Sigma}$ as a sum directly over derivatives of $s$, making the role of each advection parameter visible. Later in the appendix, we discuss the radius of convergence of the resulting sum in 1D.

Applying the Leibniz rule to the derivative $\partial_x^n(-s)^n$ in eq.~(\ref{deltaXn1}) and collecting terms by their derivative content gives
\be\label{advection-form-1}
\delta_{x,\Sigma}(x) &=& \sum_{N=1}^{\infty}\; \sum_{a_1 > a_2 > \dots > a_N \ge 1} \;
\sum_{i_1, \dots, i_N \ge 1}\left\{
\binom{i_1+i_2+\dots+i_N+n_0}{i_1, i_2, \dots, i_N, n_0}
\prod_{m=1}^{N} \left(-\frac{X_{a_m}}{a_m!}\right)^{i_m}\right\} ,\nonumber\\
\ee
with
\be\label{advection-form-n0}
X_{a_m}&\equiv&\bigl(-s(x)\bigr)^{a_m-1}s^{(a_m)}(x), \ \ \mathrm{and }\ \ 
n_0 \equiv \sum_{m=1}^{N} (a_m-1)\,i_m\ ,
\ee
and the multinomial coefficient $\binom{\sum i_m + n_0}{i_1,\dots,i_N,n_0} \equiv (i_1+\dots+i_N+n_0)!/(i_1!\cdots i_N!\,n_0!)$. The superscript in parenthesis $(a_m)$ denotes the $a_m$-th derivative with respect to $q$ evaluated at $q=x$.

The structure of eq.~(\ref{advection-form-1}) is the desired one: every term is built from a finite product of advective combinations $s^{a_m-1}s^{(a_m)}$. The leading terms ($N=1$, single $a_1\equiv m$, $i_1\equiv i$) involve only one derivative type and produce $s^{m-1}\,s^{(m)}$ to powers $i$. For $m=1,2,3,\ldots$ this reproduces the family of advection parameters of Section~\ref{sec:saddle1}: $|s'|, |s\,s''|, |s^2 s'''|,\ldots$. Higher-$N$ contributions in eq.~(\ref{advection-form-1}) are ``cross'' terms involving products of derivatives of different orders; these produce the mixed combinations $|s\,s'\,s''|$, $|s^2\,s'\,s'''|$, and so on.

\subsection{Partial sum for $N=1$\label{app:N1}}
In what follows we manipulate sums only formally; sticking with the rest of the paper's convention would require writing $\mathclap{\hspace{1.35em}\times}=$ throughout, but that quickly becomes ugly, so we simply write standard equalities. With that in mind, let us denote the sum in eq.~(\ref{advection-form-1}) restricted to a fixed $N$ as $T_N$. Focusing on $T_{1}$ and replacing $a_1\to a$, $i_1\to i$, the multinomial coefficient simplifies to $\binom{a i}{i,(a-1)i} = (ai)!/(i!\,((a-1)i)!)$. Then further restricting the sum to fixed $a$ (call it $T_{N=1,a}$), the resulting sum over $i$ at fixed $a$ can be performed in closed form. For $a=1$, $2$, $3$, the resulting partial sums read
\be\label{advection-form-m1}
T_{1,1}(x) \equiv -1 + \frac{1}{1+s'(x)}\ ,
\ee
\be\label{advection-form-m2}
T_{1,2}(x) \equiv -1 + \frac{1}{\sqrt{1 - 2\, s(x)\, s''(x)}}\ ,
\ee
\be\label{advection-form-m3}
T_{1,3}(x) \equiv -1 + \frac{
\cosh\!\Big[\frac{1}{3}\,\sinh^{-1}\!\Big(\frac{3\, s(x)\, \sqrt{s'''(x)}}{2\sqrt{2}}\Big)\Big]
}{
\sqrt{1 + \frac{9}{8}\, s(x)^2\, s'''(x)}
}\ .
\ee
Eq.~(\ref{advection-form-m3}) remains real even when $s'''(x)<0$: the hyperbolic functions turn into the corresponding trigonometric ones, with the radical in the denominator leading to a radius of convergence $\frac{9}{8}\,s^2\,|s'''|<1$. For general $a\ge 3$, the closed form is a generalized hypergeometric function:
\be\label{advection-form-qfp}
T_{1,a}(x)+1 = {}_{a-1}F_{a-2}\!\Bigg(
\tfrac{1}{a},\, \tfrac{2}{a},\, \dots,\, \tfrac{a-1}{a}\;;\;
\tfrac{1}{a-1},\, \tfrac{2}{a-1},\, \dots,\, \tfrac{a-2}{a-1}\;;\;
\frac{(-a)^a}{a!\,(a-1)^{a-1}}\, s(x)^{a-1}\, s^{(a)}(x)
\Bigg).\nonumber\\
\ee
Eq. (\ref{advection-form-m3}) is the $a=3$ special case of eq.~(\ref{advection-form-qfp}). With the definitions above for $T_{1,a}$, we can finally write the formal equation
\be\label{advection-form-N1-sum}
\delta_{x,\Sigma}(x) = T_{1,1}(x) + T_{1,2}(x) + \sum_{a=3}^{\infty} T_{1,a}(x)+\sum\limits_{N=2}^{\infty}{T_N},
\ee
where $T_{N\ge 2}$ can be read off from eq.~(\ref{advection-form-1}).

Each $T_{1,a}$ in eq.~(\ref{advection-form-qfp}) has its own finite radius of convergence in the corresponding advection parameter $|X_a|=|s^{a-1}\,s^{(a)}|$, set by where the underlying hypergeometric series ceases to converge: $|s'|<1$ for $T_{1,1}$; $|2\,s\,s''|<1$ for $T_{1,2}$; analogous bounds at higher $a$. Each $T_{1,a}$ thus signals that its corresponding advection parameter must remain below $\mathcal{O}(1)$ for its partial sum alone to converge. These $T_{1,a}$ are, however, only the $N=1$ contributions to $\delta_{x,\Sigma}$. The full $\delta_{x,\Sigma}$ also contains all $N\ge 2$ contributions from eq.~(\ref{advection-form-1}) --- mixed-derivative terms involving products of multiple advection parameters such as $|s\,s'\,s''|$, $|s^2\,s'\,s'''|$, and so on entering $T_{N\ge 2}$. The actual convergence condition of $\delta_{x,\Sigma}$ at a given $x$ is therefore not directly given by the per-$a$ $T_{1,a}$ radii of convergence. The structure of $T_{1,a}$ does, however, make explicit \textit{why each advection parameter individually matters}: each one shows up at all orders in $i$ at $N=1$, and could block convergence on its own if it became $\mathcal{O}(1)$.

\subsection{General expression for the radii of convergence}

Now let us relax the $N=1$ restriction  in Appendix~\ref{app:N1} to the sum of eq.~(\ref{advection-form-1}). Let us focus on a partial sum for some $N$ and on a particular combination of derivatives $a_m$. Thus, we will focus on the convergence of the nested sums over $i_1,\dots,i_N$. Therefore, we are going to analyze the summand in curly brackets in eq.~(\ref{advection-form-1}); let us denote it by $U$. Let us also package the indices $i_k$ ($k=1,\cdots,N$) as a vector $\bm{i}$. As $I\equiv |\bm i|$ goes to infinity, the summand becomes largest in some unit direction $\bm{\hat i}$ for which we can write $\bm{i}=I \bm{\hat i}$ (at large $I$, the distinction between integer and continuous components of $\bm i$ becomes immaterial). Thus at large $I$ the summand $U$ is bound from above by:
\be\label{U}
|U|\le
\binom{I \tilde \sigma}{I \hat i_1, I \hat i_2, \dots, I \hat i_N, I \tilde n_0}
\prod_{m=1}^{N} \left(\frac{|X_{a_m}|}{a_m!}\right)^{I \hat i_m}\ ,
\ee
with
\be
\tilde n_0\equiv \frac{n_0}{I}= \sum_{m=1}^{N} (a_m-1)\,\hat i_m\nonumber\\
\tilde \sigma =\hat i_1+ \hat i_2+\dots+\hat i_N+\tilde n_0\ .
\ee
 
The sums over $i_1,\dots,i_N$ converge absolutely if as $I\to\infty$
\be
\left(|U| I^{N-1}\right)^{1/I}<1\ ,
\ee
where the factor of $I^{N-1}$ takes into account the fact that we have $N$ nested sums.
Using Stirling's approximation for the factorials in the multinomial coefficient in eq.~(\ref{U}), the exponentials in the approximation cancel and we obtain the following sufficient condition for convergence of the sums over the indices of $\bm i$:
\be\label{FINALconv1D}
\left(|U| I^{N-1}\right)^{1/I}\le \frac{{\tilde \sigma}^{\tilde \sigma}}{{\tilde n_0}^{\tilde n_0} \prod\limits_{m=1}^N {\hat i_m}^{\hat i_m}}\prod_{m=1}^{N} \left(\frac{|X_{a_m}|}{a_m!}\right)^{\hat i_m}< 1\ ,
\ee
which needs to be evaluated at $\bm{\hat i}$ which maximizes the left-hand side of the second inequality.

Two caveats are in order. First, eq.~(\ref{FINALconv1D}) was derived from $|U|$, with signs of $X_{a_m}$ stripped; it is therefore a sufficient condition for \textit{absolute} convergence of the \textit{partial sum} at fixed $(N,\{a_m\})$. The signed series $\delta_{x,\Sigma}$ may converge in a strictly larger region by cancellations between positive and negative terms. Second, eq.~(\ref{FINALconv1D}) controls only the inner sums over $i_1,\dots,i_N$ at fixed $(N,\{a_m\})$. Convergence of the outer sums (over $N$ and over the sorted tuples $a_1>\dots>a_N\ge 1$) is a separate question that we do not address here; it can in principle further restrict the convergence region or relax it (once signs are taken into account).

For $N=1$ the condition in eq.~(\ref{FINALconv1D}) becomes:
\be
\frac{|X_a|a^a}{a!(a-1)^{a-1}}\le 1\ .
\ee
That reproduces the convergence radii that can be read off from eqs.~(\ref{advection-form-m1}, \ref{advection-form-m2}, \ref{advection-form-m3}, \ref{advection-form-qfp}). Indeed, using the definition of $X_a$ from eq.~(\ref{advection-form-n0}), as $a\to 1$, we get $|X_1|=|s'|<1$; for $a=2$, we find $|2 s s''|<1$, etc. For higher $N$, eq.~(\ref{FINALconv1D}) gives constraints on mixed products of derivatives and displacements as expected at the end of Section~\ref{app:N1}. And that completes our analysis of the convergence radii in 1D.

\subsection{An aside: applying the Lagrange--B\"urmann formula to the overdensity\label{app:lagrange-burmann}}

The fixed-point equation $q(x) = x - s(q(x))$ that defines $q(x)$ in the single-stream regime (cf.~eq.~\ref{its1}) is exactly of the form to which the Lagrange--B\"urmann (LB) inversion formula applies. In this short subsection we use it to recover eq.~(\ref{deltaXn1}) and to derive a related but distinct representation for $\delta_{x,\Sigma}(x)$.

For our relation $q = x + (-1)\,s(q)$, the LB formula reads
\be\label{LB-master}
H(q(x))=H(x)+\sum_{n\ge1}\frac{(-1)^n}{n!}\,\frac{d^{n-1}}{dx^{n-1}}\!\left[s(x)^n\,H'(x)\right]\ ,
\ee
for any function $H$ analytic at $x$. Different choices of $H$ produce different equivalent series for $\delta_{x,\Sigma}(x)$.

\paragraph{Choice 1: $H(q)=q$ recovers $\delta_{x,\Sigma}$.}
With $H'=1$, eq.~(\ref{LB-master}) gives a series for $q(x)$ itself,
\be\label{LB-qstar}
q(x)=x+\sum_{n\ge1}\frac{(-1)^n}{n!}\,\frac{d^{n-1}}{dx^{n-1}}\bigl[s(x)^n\bigr]\ .
\ee
Differentiating both sides in $x$ and using $1+\delta(x)=q'(x)$ (which follows from differentiating $q+s(q)=x$),
\be\label{LB-deltaxsigma}
\delta(x)=\sum_{n\ge1}\frac{(-1)^n}{n!}\,\frac{d^{n}}{dx^{n}}\bigl[s(x)^n\bigr]=\sum_{n\ge1}\frac{1}{n!}\,\partial_x^n\bigl[(-s(x))^n\bigr]\ ,
\ee
which is precisely the original $\delta_{x,\Sigma}$ of eq.~(\ref{deltaXn1}). The same identity is recovered from the generating-function representation of $\delta_{x,\Sigma}$ given in the footnote below\footnote{$\delta_{x,\Sigma}$ admits a formal compact generating-function representation in terms of derivatives of $s$ (as opposed to derivatives of $s^n$):
	\be\label{advection-form-gf}
	\delta_{x,\Sigma}(x) = \sum_{n=1}^{\infty} \frac{1}{n!} \left.\frac{d^n}{dt^n}\Big( F(t,x)^n \Big)\right|_{t=0}\ ,\nonumber\\
	F(t,x) = \frac{s(x-ts(x))}{s(x)}= 1 + \sum_{a=1}^{\infty} t^a \left[-\frac{\bigl(-s(x)\bigr)^{a-1}\, s^{(a)}(x)}{a!}\right]\ .
	\ee
	Expanding $F(t,x)^n$ in $t$ and taking the $t^n$ coefficient, allows us to see that the above equation reproduce the sum in eq.~(\ref{advection-form-1}). This formal equality was confirmed by direct substitution for the first several powers in $s$.}; choice 1 is therefore not new.

\paragraph{Choice 2: $H(q)=-1+1/(1+s'(q))$ gives a Lagrangian-piece-plus-correction form.}
Now take
\be
H(q)\equiv -1 + \frac{1}{1+s'(q)}\ ,\qquad H'(q)=-\frac{s''(q)}{(1+s'(q))^2}\ .
\ee
Then $H(q(x))=\delta(x)$ (this is the exact single-stream relation $1+\delta=1/(1+s'(q(x)))$), while $H(x)=-s'(x)/(1+s'(x))\equiv\delta_q(x)$ is the Lagrangian-space single-stream overdensity (\ref{LPTstart}) evaluated at $q=x$ (i.e.\ as a local function of $s'(x)$, with $q$ relabeled as $x$). Substituting in eq.~(\ref{LB-master}) gives
\be\label{LB-deltax-split}
\delta(x)&=&\delta_q(x)+\sum_{n\ge1}\frac{(-1)^{n-1}}{n!}\,\frac{d^{n-1}}{dx^{n-1}}\!\left[\frac{s(x)^n\,s''(x)}{(1+s'(x))^2}\right]\nonumber\\
&=&\delta_q(x)+\sum_{n\ge1}\frac{(-1)^n}{n!}\,\frac{d^{n-1}}{dx^{n-1}}\bigl[s(x)^n\,\delta_q'(x)\bigr]\ ,
\ee
where in the second line we recognized $-s''(x)/(1+s'(x))^2 = \partial_x[1/(1+s'(x))] = \delta_q'(x)$. Note that $\delta_q$ and $s$ and their derivative are all evaluated at $q=x$, e.g.~$\delta_q'(x)=\partial_x\delta_q(x)=\partial_q\delta_q(q)|_{q=x}$.

Eq.~(\ref{LB-deltax-split}) is a different organization of $\delta_{x,\Sigma}(x)$ from eq.~(\ref{deltaXn1}). It separates the part of the overdensity that depends only on $s'(x)$ (the non-perturbative Lagrangian piece $\delta_q(x)=-s'/(1+s')$) from a series correction that quantifies the Eulerian-vs-Lagrangian mismatch. Notice that this correction starts at $n=1$ with leading order $s\,s''/(1+s')^2\sim s\,s''$ in $s$. Hence, if $s\,s''$ is small while $s'$ is large but bounded ($|s'|<1$), the correction is small and $\delta(x)\approx\delta_q(x)$. Conversely, $\delta(x)$ is clearly controlled by parameters such as $|s\,s''|$ beyond $|s'|$, which can cause it to fail well before $\delta_q(q)$ does. This is a manifestation of the general fact (cf.~Section~\ref{sec:saddle1}) that $|s\,s''|$ is an expansion parameter independent of $|s'|$.

\section{Integrating $\mathcal{P}$\label{app:P}}

In this appendix we start from the angular average $S(\bm k,\bm l,\bm m)$ in eq.~(\ref{angular_int1}) to find the result for the angular polynomials $\tilde{\mathcal{C}}_{LM}$ (eq.~\ref{CLM_Tilde}) and $\mathcal{C}_{LM}$ (eq.~\ref{CLM_FINAL}). Those carry nearly all of the $\hat{\bm k},\hat{\bm l},\hat{\bm m}$ dependence of $\mathcal{P}$ used in Section~\ref{sec:Pnumerics}.

To evaluate $\mathcal{P}$ as defined in eq.~(\ref{Pcurly}), and therefore the connected two-point function in eq.~(\ref{2pfz_simple}), we perform the integral over $d^3q$. Writing $k\equiv |\bm k|$, $l\equiv |\bm l|$, and $m\equiv |\bm m|$, the measure splits as $d^3q=q^2dq\, d\Omega_q$ into a radial part, $q^2dq$, and an angular part, $d\Omega_q$, over the direction of $\hat{\bm q}$. The radial integral will be done numerically, while the angular integral can be performed analytically. Keeping only the direction-dependent part of the integrand, and dividing by $4\pi$ for convenience, we define
\be\label{angular_int1}
S(\bm k,\bm l,\bm m)\equiv \int \frac{d\Omega_q}{4\pi} \ \exp\Big[-i{\bm q}\cdot {\bm k}+ ({\bm l}\cdot {\bm {\hat q}})\,({\bm m}\cdot {\bm {\hat q}}) \gamma(q)\Big]\ .
\ee

To evaluate the integral above, we first expand the integrand in a series. However, a direct power series need not converge rapidly numerically, and its radius of convergence may be finite since $lm\gamma(q)$ need not be a small parameter and $qk$ can be arbitrarily large.
Instead, we use the identities below:
\be\label{usefulExp}
e^{i x y}=\sum\limits_{l=0}^\infty i^l(2l+1)j_l(x)\mathrm{P}_l(y)\nonumber\\
e^{x y}=\sum\limits_{l=0}^\infty (2l+1)i_l(x)\mathrm{P}_l(y)\ ,
\ee
with $j_l(x)$ being the spherical Bessel function of the first kind, $i_l(x)$ being the modified spherical Bessel function of the first kind, and $\mathrm{P}_l(y)$ being the Legendre polynomials.

Using eq.~(\ref{usefulExp}), the angular average can be written as
\be\label{S1}
S= \sum\limits_{L= 0}^\infty\sum\limits_{M= L}^\infty (-1)^L (4L+1)(2M+1)j_{2L}(qk)i_{M}\big(\gamma(q) l m\big)\times\tilde  S_{LM}\ .
\ee

The constraint $M\ge L$ of the summation index $M$ comes from the fact that $\tilde S_{LM}$ defined below vanishes for $M<L$ --- a result we obtain further below (also, see footnote~\ref{ftntVanish}). In eq.~(\ref{S1}) we kept only even-order $j_{2L}$'s, since for odd orders the integral $\tilde S_{LM}$ vanishes. Here
\be\label{SLM1}
\tilde S_{LM}(\bm k,\bm l,\bm m)\equiv\int  \frac{d\Omega_q}{4\pi} \mathrm{P}_{2L}(\hat{\bm q}\cdot\hat{\bm k})\mathrm{P}_M\Big(\big(\hat{\bm q}\cdot\hat{\bm l}\big)\big(\hat{\bm q}\cdot\hat{\bm m}\big)\Big)\ ,
\ee
which is the integral appearing in eq.~(\ref{CLM_Tilde}).

Expanding both Legendre polynomials in powers of their arguments gives
\be\label{SLM2}
\tilde S_{LM}=2^{-2L-M}\sum\limits_{U=0}^{L}\sum\limits_{V=0}^{\lfloor \frac{M}{2}\rfloor}(-1)^{U+V}\binom{2L}{U}\binom{4L-2U}{2L}\binom{M}{V}\binom{2M-2V}{M} S'_{L-U,M-2V}\ ,\nonumber\\
\ee
where
\be\label{intOmega1}
S'_{AB}\equiv\int  \frac{d\Omega_q}{4\pi} \left(\hat{\bm q}\cdot\hat{\bm k}\right)^{2A}\left[\big(\hat{\bm q}\cdot\hat{\bm l}\big)\big(\hat{\bm q}\cdot\hat{\bm m}\big)\right]^{B}\ .
\ee

The remaining task is to average products of unit-vector components. For odd powers the result vanishes by parity, while for $2N$ unit vectors one finds\footnote{Equation (\ref{intd2q}) was obtained by using symmetry arguments, together with fully contracting the integrand with products of Kronecker deltas for the first few values of $N$. After that, the general pattern was obtained.}:
\be\label{intd2q}
\int \frac{d\Omega_q}{4\pi}\,  \left(\bm{\hat q}^{2N}\right)_{ijkl\dots}=\frac{1}{(2N+1)!!}\left[\underbrace{\delta_{ij}\delta_{kl}\dots}_{\textrm{$N$ terms}}+\textrm{chord\ diagram\ permutations}\right]\ ,
\ee
where $i,j,k,l,...$ are the indices of the $2N$ unit vectors, such that $\left(\bm{\hat q}^{2N}\right)_{ijkl\dots}\equiv \hat q_i\hat q_j\hat q_k\hat q_l\dots$ (for a total of $2N$ terms). The chord-diagram permutations are the inequivalent pairings of the indices, and their total number is $(2N-1)!!$.

Contracting eq.~(\ref{intd2q}) with the powers of $\hat{\bm k}$, $\hat{\bm l}$, and $\hat{\bm m}$ generates a finite but nontrivial set of combinatorial factors. Writing the result first for non-unit vectors makes that bookkeeping easier to trace. Using $\mu_{kl}\equiv \hat{\bm k}\cdot\hat{\bm l}$, and similarly for $\mu_{km}$ and $\mu_{lm}$, one obtains
\be\label{combinatorics}
\left(\delta_{ij}\delta_{kl}\dots+\textrm{chord\ diagram\ permutations}\right)\big(\bm k^{2A} \bm l^B\bm m^B\big)_{ijkl\dots}=\nonumber\\ 
=\sum \limits_{u,v,w,x,y,z\geq0}' (\mu_{kl}k l)^x(\mu_{km}k m)^y(\mu_{lm}l m)^z k^{2u}l^{2v}m^{2w}\times\nonumber\\ x!y!z!(2u-1)!!(2v-1)!!(2w-1)!!\times\nonumber\\ \binom{2A}{2u}\binom{B}{2v}\binom{B}{2w}\binom{2A-2u}{x}\binom{B-2v}{z}\binom{B-2w}{y}\nonumber\\
=\sum \limits_{u,v,w,x,y,z\geq0}' \mu_{kl}^x\mu_{km}^y\mu_{lm}^z  k^{2A}(l\, m)^B
\times\frac{(2A)!\,\left(B!\right)^2}{x!\,y!\,z!\,u!\,v!\,w!\,2^{u+v+w}}\ .
\ee
The prime indicates that the sum is restricted by
\be\label{restricted}
x+y+2u=2A\nonumber\\
x+z+2v=B\nonumber\\
y+z+2w=B\ ,
\ee
with all summation indices non-negative integers. Using these constraints, eq.~(\ref{intOmega1}) becomes
\be\label{SABprime}
S'_{AB}=\sum \limits_{u,v,w,x,y,z\geq0}' \frac{(2A)!\,\left(B!\right)^2}{\Big(2(A+B)+1\Big)!!\,x!\,y!\,z!\,u!\,v!\,w!\,2^{u+v+w}}\mu_{kl}^x\mu_{km}^y\mu_{lm}^z\ ,
\ee
with the same restrictions as in eq.~(\ref{restricted}).

Equations~(\ref{S1}), (\ref{SLM2}), and (\ref{SABprime}), together with the restrictions in eq.~(\ref{restricted}), provide the final answer for $S$.\footnote{\label{ftntVanish}Now that we have an explicit expression for $\tilde S_{LM}$, one can check explicitly that it vanishes for $M<L$, a fact used in restricting the sum over $M$ in eq.~(\ref{S1}). We checked that explicitly for all $M<L\leq 30$, and conjecture that it is valid for all $M<L$. Note also that the first two sums in eq.~(\ref{S1}), $\sum\limits_{L= 0}^\infty\sum\limits_{M= L}^\infty$, can be rearranged as $\sum\limits_{M= 0}^\infty\sum\limits_{L=0}^M$, so only one of the summation indices in evaluating $\mathcal{P}$ has an infinite range. In the implementation used here, however, we perform the summation over a finite range of $M$ first and then the Hankel transforms for each $L$.}
Substituting back into the two-point function gives
\begin{eqnarray}\label{2pfz_simple2}
\begin{aligned}
&\langle f(\bm{k},\bm{w},D)\,f(\bm{k}',\bm{w}',D')\rangle_{\mathrm{c}}=\\
&=\ (2\pi)^3\d(\bm{k}+\bm{k}')e^{-\frac{\sigma_v^2}{2}(\beta^2+\beta'^2)} \, \int\limits_{0}^\infty dq\, 4\pi q^2\left\{ \ e^{-{\bm \beta}\cdot{\bm \beta'} \, \zeta(q)}S(\bm k,\bm \beta,\bm \beta')\ -j_0(kq)\right\},
\end{aligned}
\end{eqnarray}
and therefore
\be\label{PcurlyLast}
\mathcal{P}(\bm k,\bm l,\bm m)=4\pi \int\limits_{0}^\infty dq \,q^2\left\{ \ e^{-{\bm l}\cdot{\bm m} \, \zeta(q)}S(\bm k,\bm l,\bm m)\ -j_0(kq)\right\}\ .
\ee
At this stage the only remaining angular dependence sits in the direction cosines. These are constrained by the triangle inequality:
\be\label{triangle_ineq}
\mu_{kl}^2+\mu_{km}^2+\mu_{lm}^2-2\mu_{kl}\mu_{km}\mu_{lm}\leq1\ .
\ee
We use the inequality above to perform one last substitution in our final result for $\mathcal{P}$ in Section~\ref{sec:Psubstitution}.

Substituting eq.~(\ref{S1}) directly into eq.~(\ref{PcurlyLast}) already yields the representation eq.~(\ref{FINALP_exp}) of the main text, in which the factor $e^{-\bm l\cdot \bm m\,\zeta(q)}$ is left unexpanded so that calculations can be organized at fixed $\mu_{lm}$. To reach the fully separated form eq.~(\ref{FINALP}), we next expand that remaining exponential using eq.~(\ref{usefulExp}), thereby splitting the magnitude product $lm$ from the directional cosine $\mu_{lm}$. This isolates the radial Hankel-transform piece from the purely angular polynomial piece.

To wrap up this appendix, we show how we arrived at eq.~(\ref{CLM_FINAL}) next. Let us define the angular polynomial
\be
&\tilde{\mathcal{C}}_{LM}(\mu_{kl},\mu_{km},\mu_{lm})\equiv (-1)^{L}(2M+1)(4L+1)\tilde S_{LM}(\bm k,\bm l,\bm m)\ .
\ee
Collecting the results of this appendix, one finds\footnote{Equation~(\ref{CLM_Tilde}) may also be reorganized as
\be\label{CLMfootnote}
&\tilde{\mathcal{C}}_{LM}(\mu_{kl},\mu_{km},\mu_{lm})=\nonumber\\
&=\sum\limits_{A=0}^{\lfloor\frac{M}{2}\rfloor}\sum\limits_{B=A}^{\lfloor\frac{M}{2}\rfloor}
\sum\limits_{N=2B-M}^{M-2B}
G_{L,M,A,B,|N|}
\Big[Y_{M-2A,N}(\hat{\bm l})Y^*_{M-2B,N}(\hat{\bm m}) + (\bm l \longleftrightarrow \bm m)\Big]\ ,
\nonumber\\
\ee
where the $z$ axis of the spherical harmonics is picked to coincide with $\bm{\hat k}$.
The symmetry under $\bm l\longleftrightarrow \bm m$ together with the fact that the terms $\pm N$ enter symmetrically (as the coefficients $G$ depend only on $|N|$) makes the expression manifestly real. No compact closed form for $G_{L,M,A,B,|N|}$ was found by the author, but this representation shows explicitly which spherical harmonics contribute. For numerical work, however, the polynomial form in the main text is the representation we use.}:
\be\label{CLM_Tilde}
&\tilde{\mathcal{C}}_{LM}(\mu_{kl},\mu_{km},\mu_{lm})=\nonumber\\
&=\sum\limits_{U=0}^L\sum\limits_{V=0}^{\lfloor\frac{M}{2}\rfloor}\sum\limits_{x,y,z}' \left\{\frac{(-1)^{L + U + V}(2M+1)(4L+1)(M - 2V)!(4L - 2U - 1)!!( 2M - 2V-1)!!
}{\big( 2(L + M - U) - 4V+1\big)!! (2U)!! (2V)!!(2u)!!(2v)!!(2w)!!x!y!z!}\right\}\times{\mu_{kl}}^x{\mu_{km}}^y{\mu_{lm}}^z\ ,
\nonumber\\
\ee
where
\be
u = L-U - \frac{x + y}{2}\nonumber\\
v = \frac{M-2V - x - z}{2}\nonumber\\
w = \frac{M-2V - y - z}{2}\ ,
\ee
and the primed sum in eq.~(\ref{CLM_Tilde}) is restricted to the finite set of non-negative integers $x,y,z$ for which $u$, $v$, and $w$ are themselves non-negative integers. The polynomial is symmetric under interchange of $\mu_{kl}$ and $\mu_{km}$, and we make that symmetry explicit. Using
\be
\sum\limits_{d=0}^{\lfloor \frac{C}{2}\rfloor}(-1)^d \frac{C}{C - d} \binom{C - d}{d} (\mu_{kl} \mu_{km})^{x + 2d} (\mu_{kl}^2 + \mu_{km}^2)^{C - 2d}=\mu_{kl}^{2C + x} \mu_{km}^x + \mu_{kl}^x \mu_{km}^{2C + x}\ ,\nonumber\\
\ee
for integer $C>0$ and integer $x\geq 0$, and pairing the terms related by $x\leftrightarrow y$, one arrives at eq.~(\ref{CLM_FINAL}).

\begin{figure}[ht!]
\centering
\includegraphics[width=\textwidth]{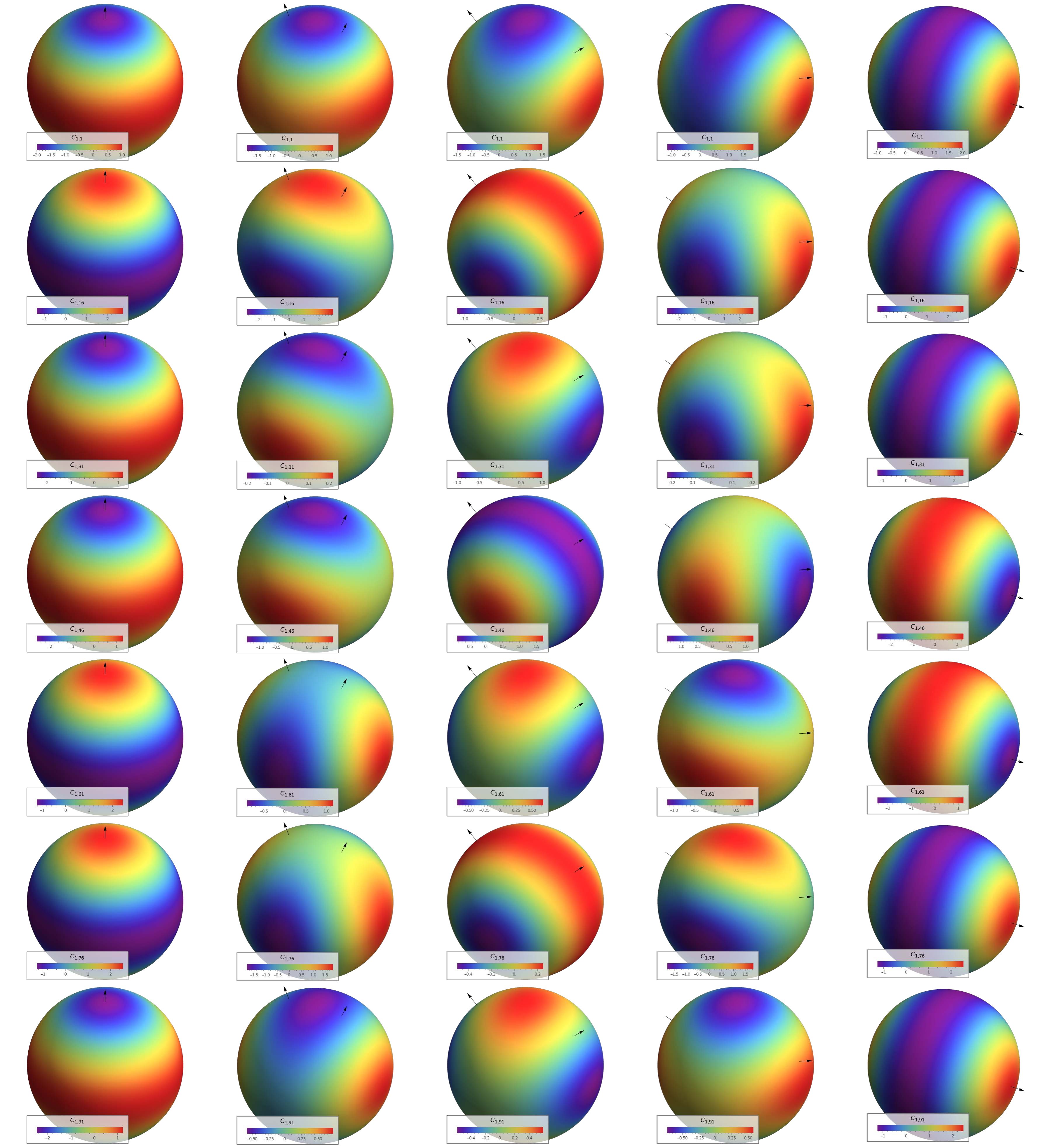}
\caption{As a reference, we show plots of the angular polynomial $\mathcal{C}_{LM}(s=\mu_{kl}^2+\mu_{km}^2,\ t=\mu_{kl}\mu_{km},\ \mu_{lm})$ from eq.~(\ref{CLM_FINAL}), or equivalently of $\tilde{\mathcal{C}}_{LM}(\mu_{kl},\mu_{km},\mu_{lm})$ (eq.~\ref{CLM_Tilde}). The panels are for fixed $L=1$ and each row corresponds to a fixed $M$ ($L$ and $M$ values can be read off from the $\mathcal{C}_{LM}$ label right above the colorbars: the first row corresponds to $L=M$ and each consecutive row increases $M$ by 15). 
Each column corresponds to a particular value of $\mu_{lm}$, with corresponding fixed unit vectors $\hat l$ and $\hat m$  shown as arrows (lying on the ``prime meridian''). The five columns, read left to right, are for $\mu_{lm}=1$, $\sqrt{2}/2$, $0$, $-\sqrt{2}/2$, $-1$, respectively. The ``north pole'' of the plots is chosen to be the direction midway between $\hat{\bm{l}}$ and $\hat{\bm{m}}$ on the spheres, which in the first column (for which $\hat{\bm{l}}=\hat{\bm{m}}$) corresponds to the intersection between the arrow and the surface of the sphere.  Each location (``latitude'' and ``longitude'') on the spheres corresponds to a particular direction of $\hat{\bm{k}}$ as seen from the center of the spheres. Each sphere is viewed from the same location with the same line of sight. Note that the color bars span different ranges in each figure and are not centered at zero; readers may need to zoom in to view them clearly in the electronic version.}
\label{fig:L=1}
\end{figure}

\begin{figure}[ht!]
\centering
\includegraphics[width=\textwidth]{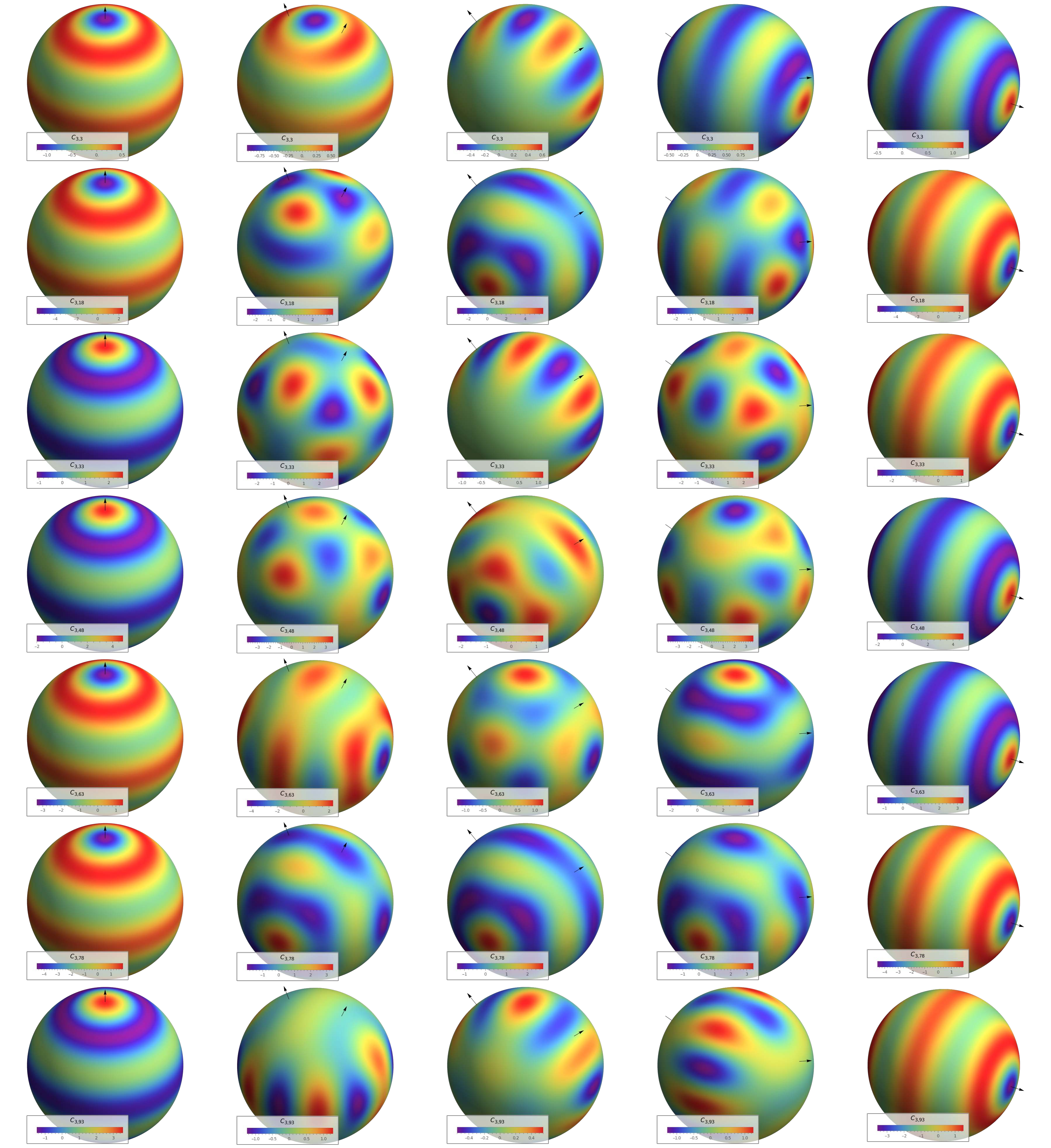}
\caption{Same quantity as in Fig.~\ref{fig:L=1}, but for fixed $L=3$.}
\label{fig:L=3}
\end{figure}

\begin{figure}[ht!]
\centering
\includegraphics[width=\textwidth]{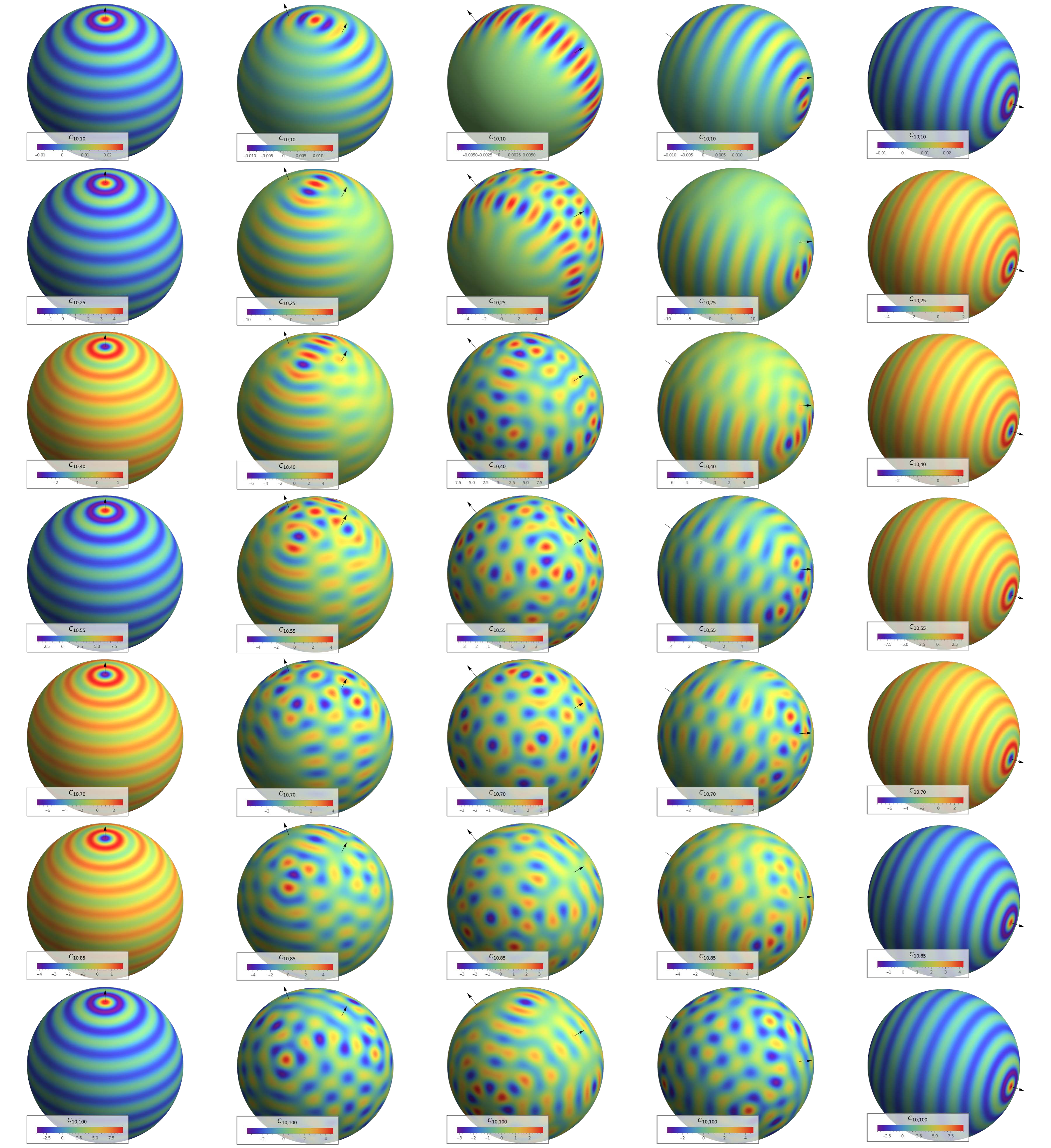}
\caption{Same quantity as in Fig.~\ref{fig:L=1}, but for fixed $L=10$.}
\label{fig:L=10}
\end{figure}

\section{Calculating $P_Z$}\label{app:PZ}

The Zel'dovich density power spectrum follows by setting $\bm w=\bm w'=0$ in the two-point function and subtracting the disconnected contribution. Using eq.~(\ref{delta2}) and then setting $\bm w=\bm w'=0$ in eq.~(\ref{2pfz}), one finds
\be
P_Z(\bm k,D)=\int d^3 q \ e^{-i{\bm q}\cdot {\bm k}} \exp\Big[-D^2k^2 \, \big(\sigma_v^2-\zeta(q)\big)-D^2 ({\bm k}\cdot {\bm {\hat q}})\,({\bm k}\cdot {\bm {\hat q}}) \gamma(q)\Big],
\ee
minus a disconnected piece proportional to $\delta_D(\bm k)$. Using $\mu\equiv \hat{\bm k}\cdot\hat{\bm q}$, this becomes
\be
P_Z=4\pi\int\limits_0^\infty dq \ q^2 \exp\Big[-D^2k^2 \, \big(\sigma_v^2-\zeta(q)\big)\Big] \int\limits_{-1}^{1}\frac{d\mu}{2}e^{-i\mu qk}e^{-(Dk)^2\mu^2\gamma(q)}\ ,
\ee
again minus the disconnected contribution. The inner $\mu$-integral is a Gaussian in $\mu$, and, if evaluated in closed form, returns an error function of complex argument, which is an oscillatory function of $q$ that makes the subsequent $q$-integration awkward for a wide range of $k$. It is therefore preferable to reorganize the $\mu$-integral into a Bessel-function expansion for which the $q$-integral can be evaluated with the Fast Hankel Transform (FHT).

We present two such reorganizations. The first is the standard Zel'dovich-power-spectrum expansion, which we include as the baseline expression against which numerical results are compared. The second is a new expansion that we construct as a cross-check of  the numerical value of $P_Z$.

\subsection*{Standard expansion}

Using the identity
\be
j_n(z)=\frac{z^n}{2^{n+1}n!}\int\limits_{-1}^1 d\mu \ e^{-i z \mu}(1-\mu^2)^n\ ,
\ee
the Gaussian-in-$\mu$ factor $\exp[-(Dk)^2\mu^2\gamma(q)]$ expands directly into spherical Bessel functions of order $n$, yielding
\be\label{PZ_std}
P_Z=4\pi\int\limits_0^\infty dq \,q^2\times\nonumber\\
\times\left\{ e^{-D^2k^2 \, \big(\sigma_v^2-\zeta(q)+\gamma(q)\big)}\sum\limits_{n=0}^\infty\left[\frac{2D^2\gamma(q)k}{q}\right]^nj_n(kq)\right.\nonumber\\
\left.-\,e^{-(Dk)^2\sigma_v^2}j_0(kq)\right\}\ .
\ee
The disconnected subtraction regularizes the $q$-integral. Each term in the sum is an FHT of a cosmology-dependent radial kernel against $j_n(kq)$, which we evaluate with FFTLog\footnote{See for example  \href{https://nbodykit.readthedocs.io/en/latest/_modules/nbodykit/cosmology/power/zeldovich.html}{here} for a representative implementation.}.

\subsection*{An alternative expansion}

Here we offer an alternative reorganization of the $\mu$-integral which serves as a cross-check of the expansion above, and, as it turns out, numerically lets us probe higher $k$ more easily. We expand the second exponent $\exp[-(Dk)^2\mu^2\gamma(q)]$ as a power series in $\mu^2$ and then use
\be
\int\limits_{-1}^1 \frac{d\mu}{2} e^{-i \mu a}\,\mu^{2n}=(-1)^n\frac{d^{2n}}{da^{2n}}\int\limits_{-1}^1 \frac{d\mu}{2} e^{-i \mu a}=(-1)^n\frac{d^{2n}}{da^{2n}} j_0(a)\equiv(-1)^nj^{(2n)}_0(a)\ ,
\ee
with $a\equiv qk$, and the superscript in parentheses denoting a derivative with respect to the argument. We then rewrite the derivative $j^{(2n)}_0(a)$ as a linear combination of spherical Bessel functions of the first kind via the recurrence
\be\label{recur_D}
j^{(n)}_m(a)=\frac{m}{2m+1} j^{(n-1)}_{m-1}(a)-\frac{m+1}{2m+1}j^{(n-1)}_{m+1}(a)\ ,
\ee
so that
\be
j^{(2n)}_0(a)=\sum\limits_{l=0}^n c_l(n)j_{2l}(a)\ ,
\ee
with numerical coefficients $c_l(n)$. Expanding both sides in power series in $a$ and matching coefficients gives
\be\label{superpos_DER}
j^{(2n)}_0(a)=\sum\limits_{l=0}^{n}\frac{(-1)^{l+n}\,2^l\,(4l+1)\,n!\,(2n-1)!!}{(n-l)!\,\left[2(n+l)+1\right]!!}
j_{2l}(a)\ .
\ee
Substituting back into the Gaussian $\mu$-integral gives
\be\label{final_Pz}
P_Z=4\pi\int\limits_0^\infty dq \, q^2 \times\nonumber\\
\times\left\{e^{-D^2k^2 \, \big(\sigma_v^2-\zeta(q)\big)}\sum\limits_{n=0}^\infty\frac{\left[(Dk)^2\gamma(q)\right]^n}{n!}j^{(2n)}_0(kq)\right.\nonumber\\
\left.-\,e^{-(Dk)^2\sigma_v^2}j_0(kq)\right\}\ ,
\ee
with $j^{(2n)}_0(kq)$ written as a linear combination of $j_{2l}(kq)$ via eq.~(\ref{superpos_DER}) and the disconnected piece restored explicitly.

The sum over $n$ in eq.~(\ref{final_Pz}) can be performed analytically once combined with eq.~(\ref{superpos_DER}):
\be
\sum\limits_{n=0}^\infty\frac{d^n}{n!}j^{(2n)}_0(a)=
\sum\limits_{n=0}^\infty\frac{d^n}{n!}
\sum\limits_{l=0}^{n}\frac{(-1)^{l+n}\,2^l\,(4l+1)\,n!\,(2n-1)!!}{(n-l)!\,\left[2(n+l)+1\right]!!}
j_{2l}(a)\nonumber\\
=
\sum\limits_{l=0}^{\infty}\sum\limits_{n=l}^\infty\frac{d^n}{n!}
\frac{(-1)^{l+n}\,2^l\,(4l+1)\,n!\,(2n-1)!!}{(n-l)!\,\left[2(n+l)+1\right]!!}
j_{2l}(a)
\nonumber\\
=
\sum\limits_{l=0}^{\infty}\frac{1}{2}(4l+1)\Gamma\left(l+\frac{1}{2}\right) \,_1\bm{F}_1\left(l+\frac{1}{2};2l+\frac{3}{2};-d\right)\,d^l\,j_{2l}(a)\ ,\nonumber\\
\ee
with $d\equiv (Dk)^2\gamma(q)$. Here $\,_1\bm{F}_1$ denotes the regularized confluent hypergeometric function, $\,_1\bm{F}_1(a;b;x)=\,_1F_1(a;b;x)/\Gamma(b)$, with $\,_1F_1$ the Kummer confluent hypergeometric function and $\Gamma$ the Gamma function. Using the above identity we obtain a new expansion for $P_Z$:
\be\label{final_Pz1}
P_Z=2\pi e^{-(Dk)^2\sigma_v^2}\sum\limits_{l=0}^{\infty}(4l+1)\Gamma\left(l+\frac{1}{2}\right)\int\limits_0^\infty dq \, q^2 \times\nonumber\\
\times\left\{e^{D^2k^2 \zeta(q)} \,_1\bm{F}_1\left(l+\frac{1}{2};2l+\frac{3}{2};-d\right)\,d^l\,j_{2l}(qk)-j_0(kq)\right\}\ .
\ee

Eqs.~(\ref{PZ_std}) and (\ref{final_Pz1}) are the two representations of $P_Z$ we use. In eq.~(\ref{PZ_std}) the spherical Bessel functions appear at all non-negative orders, whereas eq.~(\ref{final_Pz1}) contains only even orders explicitly. We evaluated both numerically\footnote{We evaluate the regularized confluent hypergeometric function in eq.~(\ref{final_Pz1}) using the FLINT library.} and confirmed that they agree. The standard expansion, eq.~(\ref{PZ_std}), is cheaper to evaluate --- no hypergeometric evaluation is needed. The second representation, eq.~(\ref{final_Pz1}), then serves two purposes: a numerical cross-check of $P_Z$ (which we performed), and more importantly, it allowed  us to push the numerical evaluation of $P_Z$ to higher $k$.

\section{Asymptotic regimes of the Zel'dovich power spectrum\label{app:asympt}}

Panel~(a) of Fig.~\ref{fig:zeldovich_check} displays $\mathcal{P}\exp(-\sigma_v^2p)$ as curves of fixed $\sqrt p=\tilde k$, where $\tilde k\equiv Dk$ is the rescaled wavevector. The curves develop horizontal plateaus at high $\sqrt p=\tilde k$  and low $k=\tilde k/D$ (both relative to $1/\sigma_v$), implying that the plateaus develop when $D\gg1$. The magenta curves  correspond to the power spectrum $P_Z$ evaluated at different times (or $D$'s). From eq.~(\ref{PZ_check}), we can see that we can understand the appearance of the plateaus if we understand the high $k$ tail of $P_Z$, and in particular, why that tail becomes dependent only on $\tilde k$, but not on $D$ or $k$ individually. Indeed, if $P_Z(k=\tilde k/D,D)$ in the $D\gg 1$ limit becomes just a function of $\tilde k=\sqrt{p}$ at high $\tilde k$, then $\mathcal{P}\exp(-\sigma_v^2p)$ would just be a function of $p$ at high $p$, but not of $k$ or $D$ separately and that would explain the observed plateaus. 

\begin{figure}[h!]
	\centering
	\includegraphics[width=\textwidth]{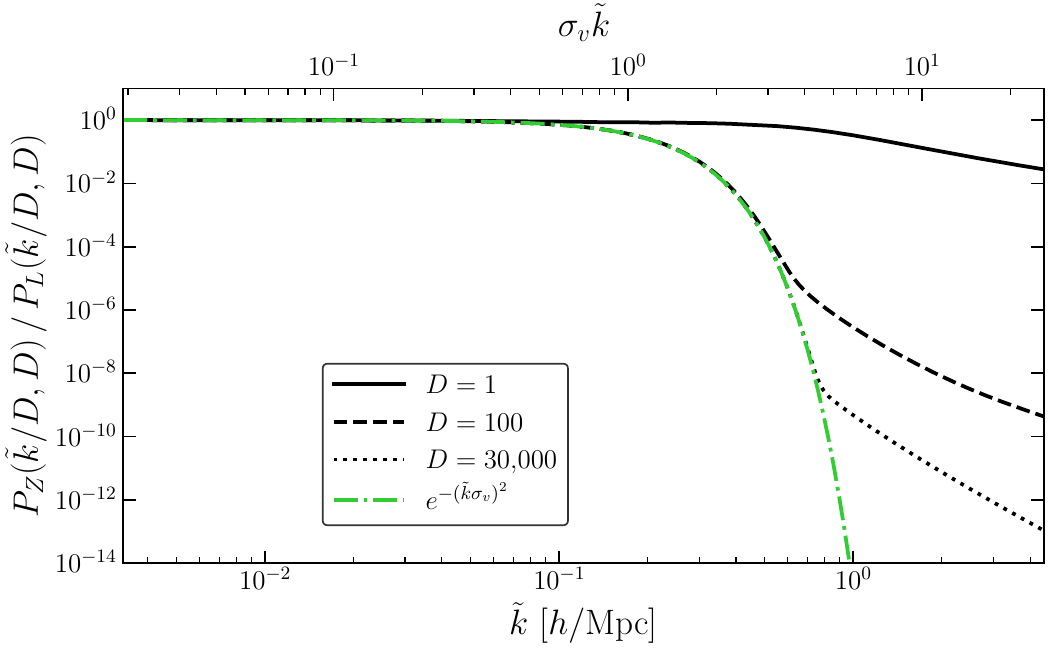}
	\caption{Ratio of the Zel'dovich matter power spectrum to the linear power spectrum, $P_Z(\tilde k/D, D)/P_L(\tilde k/D, D)$, at the rescaled wavevector $\tilde k\equiv Dk$ for growth factors $D\in\{1, 100, 30{,}000\}$. Both spectra are evaluated at $D$, with $P_L(k,D)=D^2P_L(k)$, and $P_L(k)$ being the linear power today ($D=1$). The free-streaming Gaussian smoothing kernel $\exp(-\sigma_v^2 \tilde k^2)$ (green dot-dashed curve) is shown for reference. At large $D$, in the regime $\tilde k\sigma_v\lesssim \mathcal{O}(1)$ the deviation of $P_Z$ from linear theory is captured by the free-streaming smoothing alone, $P_Z(\tilde k/D, D)\approx e^{-\sigma_v^2 \tilde k^2}\,D^2 P_L(\tilde k/D)$. At $D=1$ the deviation of $P_Z/P_L$ from the smoothing kernel shown becomes significant because the  approximation $\sigma_v^2(>\tilde k/D)\approx\sigma_v^2$ is no longer valid, and the correct smoothing scale  $\sigma_v(>\tilde k)$ is at much larger $\tilde k$ for $D=1$ (see ca. eq.~(\ref{pz-approx-smalltidlek})). The complementary regime $\tilde k\sigma_v\gtrsim 1$ is examined in Fig.~\ref{fig:pz_asympt_late_D_high_k}. See Appendix~\ref{app:asympt} for further discussion.}
	\label{fig:pz_asympt_late_D_low_k}
\end{figure}

\begin{figure}[h!]
	\centering
	\includegraphics[width=\textwidth]{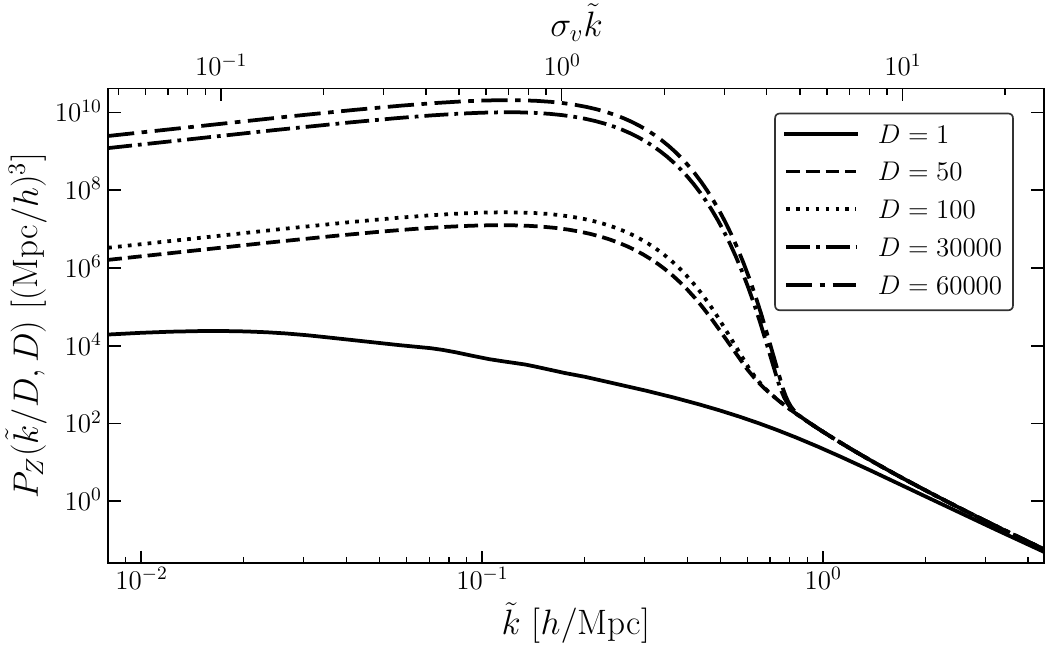}
	\caption{The Zel'dovich power spectrum $P_Z(\tilde k/D, D)$ versus $\tilde k$ for a set of representative growth factors $D\in\{1, 50, 100, 30{,}000, 60{,}000\}$ (with the $D$'s in each of the last two pairs picked to be a factor of 2 apart). For $\tilde k\gtrsim 1\,h/\mathrm{Mpc}\approx 5/\sigma_v$ all curves collapse onto a common, $D$-independent power-law tail with a slowly running exponent. At very high $\tilde k\sigma_v$, $P_Z$ eventually loses memory of the input $P_L$ at $\tilde k/D$ and becomes purely caustic-generated.  Note that the displayed window does not reach  into the purely caustic regime, for which $P_Z$ decays as $k^{-3}$ \cite{Schneider_Bartelmann_1995}. See Appendix~\ref{app:asympt} for further discussion.}
	\label{fig:pz_asympt_late_D_high_k}
\end{figure}

Thus, in this section we are going to study $P_Z(\tilde k/D,D)$ in the $D\gg1$ limit. In that limit, sweeping over $\tilde k$, three regimes appear separated by $\tilde k\sigma_v$:
\begin{itemize}
	\item $\tilde k\sigma_v\ll 1$ --- linear theory $P_Z\approx D^2\,P_L(\tilde k/D)$ (Fig.~\ref{fig:pz_asympt_late_D_low_k}).
	\item $\tilde k\sigma_v\sim 1$ ---  linear theory damped by a free-streaming Gaussian, $P_Z\approx e^{-\tilde k^2\sigma_v^2}\,D^2\,P_L(\tilde k/D)$ (Fig.~\ref{fig:pz_asympt_late_D_low_k}).
	\item $\tilde k\sigma_v\gg 1$ --- an asymptotically caustic-driven $D$-independent tail $\propto\tilde k^{-6/\beta}$ (Fig.~\ref{fig:pz_asympt_late_D_high_k}), with $\beta$ the effective small-$q$ exponent of $\Delta\psi_{ij}(\bm q)\equiv\psi_{ij}(\bm 0)-\psi_{ij}(\bm q)$. The strict $\beta=2$ limit recovers the universal $\tilde k^{-3}$ pancake caustic asymptote of \cite{Schneider_Bartelmann_1995,Konrad_2021,Konrad_2022}; the $\tilde k$ values in our window are dominated by $q$'s for which $\beta\in(1,2)$ (Section~\ref{sec:asympt:alphabeta}), giving a slope steeper than $-3$, which is not surprising given that we do not go to high enough $k$ \cite{Konrad_2022}.
\end{itemize}
Note that asymptotic analysis at high $k$ has been done for the ZA \cite{Schneider_Bartelmann_1995,Konrad_2021,Konrad_2022} but not in the $D\gg 1$ regime, which we study below. Also, in the discussion below we try to strike a pedagogical balance between detailed calculations at multiple orders \cite{Konrad_2021,Konrad_2022} and the large-$k$ asymptotics at lowest order \cite{Schneider_Bartelmann_1995}.

\subsection{Expressing \(P_Z\) using the rescaled wavevector \(\tilde k\)\label{sec:asympt:scaling}}

Starting from eqs.~(\ref{Pcurly}) and (\ref{PZ_check}), the substitution $\tilde{\bm k}\equiv D\bm k$ gives
\be\label{eq:asympt:PZ-factored}
P_Z\!\left(\frac{\tilde{\bm k}}{D}, D\right)=e^{-\tilde k^2\sigma_v^2}\!\int\! d^3q\, e^{-i\bm q\cdot\tilde{\bm k}/D}\left[e^{\tilde k_i\tilde k_j\,\psi_{ij}(\bm q)} - 1\right]\,.
\ee
The free-streaming Gaussian $e^{-\tilde k^2\sigma_v^2}$ is pulled out; the term in the square bracket asymptotes to $e^{\tilde k^2\sigma_v^2}-1$ with $q\to 0$ (where $\psi_{ij}\to\sigma_v^2\delta_{ij}$) and vanishes as $q\to\infty$ (where $\zeta,\gamma\to 0$, eq.~\ref{useful}). Equivalently, with $\Delta\psi_{ij}(\bm q)\equiv\psi_{ij}(\bm 0)-\psi_{ij}(\bm q)$ and $\Delta\psi_{ij}(\bm 0)=0$,
\be\label{eq:asympt:PZ-Deltapsi}
P_Z\!\left(\frac{\tilde{\bm k}}{D}, D\right)=\int\! d^3q\, e^{-i\bm q\cdot\tilde{\bm k}/D}\left[e^{-\tilde k_i\tilde k_j\,\Delta\psi_{ij}(\bm q)} - e^{-\tilde k^2\sigma_v^2}\right]\,.
\ee
The two non-trivial regimes follow from inspecting the exponents in the term in square bracket. Using eqs.~(\ref{useful}) and (\ref{psigamma}), we can see that $|\tilde k_i\tilde k_j\,\psi_{ij}(\bm q)|\le\tilde k^2\sigma_v^2$ (since the integrals over $P_L$ entering in $\psi$ through $\zeta$ and $\gamma$ are weighted by $j_l\le1$). We can use $k^2\sigma_v^2$ as a simple  heuristic to separate the regimes:
\begin{itemize}
	\item $\tilde k\sigma_v\ll 1$: $|\tilde k_i\tilde k_j\,\psi_{ij}(\bm q)|\le\tilde k^2\sigma_v^2\ll 1$ \textit{uniformly} in $\bm q$, so the term in the bracket in eq.~(\ref{eq:asympt:PZ-factored}) is Taylor-expandable for all $q$ (Section~\ref{sec:asympt:smallk}).
	\item $\tilde k\sigma_v\gg 1$: the term in the bracket in eq.~(\ref{eq:asympt:PZ-Deltapsi}) is sharply localized near $\bm q=\bm 0$ by the small-$q$ form of $\Delta\psi_{ij}$ (Section~\ref{sec:asympt:largek}).
\end{itemize}
In our analysis of $\tilde k\sigma_v\ll1$ we end up also capturing the transition at $\tilde k\sigma_v\sim 1$ quite well. Before working out the regimes, we record the asymptotic exponents of $\zeta$ and $\gamma$, which will be of use.

\subsection{Asymptotic exponents of $\zeta$ and $\gamma$\label{sec:asympt:alphabeta}}

The small- and large-$q$ behavior of $\Delta\psi_{ij}$ is set by the integrals of $\zeta$ and $\gamma$ over $P_L$ (eqs.~\ref{useful}, \ref{psigamma}). Let us write their asymptotics as power laws:
\be\label{eq:asympt:scaling-inputs}
\gamma(q)&\sim&\gamma_\infty\,q^{-\alpha}\,,\quad \zeta(q)\sim\zeta_\infty\,q^{-\alpha}\,, \qquad q\to\infty\,,\\
\gamma(q)&\sim&\gamma_0\,q^{\beta_\gamma}\,,\quad \zeta(q)-\zeta(0)\sim-\zeta_0\,q^{\beta_\zeta}\,, \qquad\!\! q\to 0\,,\nonumber
\ee
with $\gamma_{0,\infty}$ and $\zeta_{0,\infty}$ being dimensioned coefficients.
First, let us focus on the limits for $\gamma$ as given by eq.~(\ref{useful}). Both limits come from the \textit{kernel} $j_2(kq)$, which rises as $(kq)^2/15$ for $kq\ll1$ (its \textit{toe}), turns over near $kq\sim1$ (its \textit{peak}), and decays as $-\sin(kq)/(kq)$ for $kq\gg1$ (its oscillating \textit{tail}). The kernel thus weights $\gamma$ around $k\sim1/q$, and the only question is which part of $P_L$ sits at that $k$. When $P_L\to A_\omega\,k^{\omega}$ is a power law across that region (for some constant $A_\omega$ and $\omega$), substituting $u=kq$ gives
\be\label{eq:asympt:substitution}
\gamma(q)\sim\frac{A_\omega}{2\pi^2}\,q^{-(1+\omega)}\!\int_0^\infty\! du\,u^{\omega}\,j_2(u)\,,
\ee
a single power $q^{-(1+\omega)}$ fixed by the local slope $\omega$ at $k\sim1/q$. The $u$-integral is finite only for $-3<\omega<1$: the toe of $j_2$ gives an integrand $u^{\omega+2}$ (needs $\omega>-3$), whereas the tail gives $u^{\omega-1}\sin u$ (needs $\omega<1$). But $P_L$ is not a single power law --- it \textit{breaks} at the matter--radiation--equality turnover $k_{\rm eq}\sim10^{-2}\,h/\mathrm{Mpc}$, from $P_L\propto k^{n_s}$ (IR) to $P_L\propto k^{n_s-4}$ (UV, with log corrections \cite{BBKS_1986}). Whether eq.~(\ref{eq:asympt:substitution}) is the whole story therefore depends on where the $j_2$ peak $k\sim1/q$ falls relative to that break, i.e.\ it depends on $q\,k_{\rm eq}$. 

Turning to the limits of $\zeta(q)$, the two limits have to be treated separately because of the subtraction of the $q=0$ term ($e^{-\tilde k^2\sigma_v^2}$) in the square brackets in eq.~(\ref{eq:asympt:PZ-Deltapsi}). At $q\to 0$, we care about the limit of $\zeta(q)-\zeta(0)$ (at zeroth order in $q$, the $\zeta(q\to 0)$ contribution is subtracted away by the $e^{-\tilde k^2\sigma_v^2}$ term), which has kernel with a similarly quadratic toe: $j_0+j_2-1\simeq -u^2/10$. At $q\to \infty$, the term $e^{-\tilde k^2\sigma_v^2}$ no longer provides cancellations, so what matters is the tail of the kernel $j_0+j_2$ in the integrand of the full $\zeta(q)$. That tail decays more rapidly than $j_2$ alone by an extra power of $u$. Either way, for both limits, we have the relevant $\zeta$ scaling as $q^{-(1+\omega)}$ but this time the range of convergence is  $-3<\omega<2$, with the upper limit being lifted by the more rapidly decaying tail. The peak of the $\zeta$ contributions to $P_Z$ is again at $qk\sim1$ once the two limits are stitched together.

\paragraph{Large $q$ ($q\,k_{\rm eq}\gg1$): kernels peak in the IR, one term.}
The kernel peak $k\sim1/q$ sits well below the turnover $k_{\rm eq}$. Thus,  $P_L\to A\,k^{n_s}$, so $\omega=n_s$ in eq.~(\ref{eq:asympt:substitution}). The break in the $P_L$ power law maps to $u=q\,k_{\rm eq}\gg1$, out on the oscillating tail of the kernels of $\gamma$ and $\zeta$, so the UV side of $P_L$ is averaged away and never competes. Only the IR slope matters then, and since $-3<n_s<1$ ($n_s<1$ securing the upper end; and clearly the inequality for $\zeta$, $n_s<2$, is also satisfied) eq.~(\ref{eq:asympt:substitution}) provides us with the exact answer:
\be\label{eq:asympt:alpha-rigorous}
\alpha\;=\;1+n_s\;\approx\;1.965\,,
\ee
the latter being the value from our cosmology. We confirmed that scaling numerically for both $\gamma$ and $\zeta$.

\paragraph{Small $q$ ($q\,k_{\rm eq}\ll1$): kernels peak in the UV, two terms.}
The kernel peak $k\sim1/q$ now sits far above the turnover, where $P_L\propto k^{n_s-4}$, so the slope there is $\omega=n_s-4<-3$ and eq.~(\ref{eq:asympt:substitution}) for $\gamma$ diverges because of the lower limit of the integral. The same holds for $\zeta$ as well. Therefore, its result is no longer the full answer. For $u_{\rm eq}\equiv q\,k_{\rm eq}\ll1$ the \textit{bulk} of $P_L$ (near $k_{\rm eq}$, where its weight concentrates) now lies under the quadratic toes of the kernels of both $\gamma(q)$ and $\zeta(q)-\zeta(0)$. The \textit{peak} of the kernels and \textit{bulk} of $P_L$ give two contributions, which can be analyzed separately by splitting the $u$ integral in $\gamma$ and $\zeta$ at $u_{\rm split}$ such that $u_{\rm eq}\ll u_{\rm split}\ll1$.
\begin{itemize}
\item the \emph{peak} contribution (integral over $u>u_{\rm split}$) gives a contribution as per eq.~(\ref{eq:asympt:substitution}) with $\omega=n_s-4$, giving the non-analytic $\psi\sim q^{3-n_s}$. That contribution for both $\gamma$ and $\zeta$ is rendered finite by lifting the lower limit of the integral to $u_{\rm split}$.
\item the \emph{bulk} contribution (integral over $u<u_{\rm split}$), where $j_2(u)\simeq(u)^2/15$, gives the analytic piece for $\gamma$: $ \frac{q^2}{30\pi^2}\int_0^\infty dk\,k^2P_L(k)\propto q^2$, with the integral over the \emph{whole} $P_L$ (including IR, bulk, and UV up to $k=u_{\rm split}/(q)\gg k_{\rm eq}$) being convergent. The same analytic $q^2$ contribution holds for $\zeta(q)-\zeta(0)$ as well, as its toe is similarly quadratic. This piece loses any information about the local power of $P_L$ at $k\sim1/q$ and as we will see, will eventually give rise to the $k^{-3}$ caustic-generated power \cite{Schneider_Bartelmann_1995} of the Zel'dovich power spectrum at high $k$.
\end{itemize}
Together, the two contributions at small $q$ give\footnote{Note that for large $q$ there is no such bulk term: the bulk then sits on the oscillating tail of the kernel, which averages it away.}:
\be\label{eq:asympt:gamma-smallq}
\gamma(q)\sim a\,q^2-b\,q^{3-n_s}\,,\qquad a,b>0\,,
\ee
the minus sign because $j_2(u)<u^2/15$ for every $u>0$: the analytic bulk term, built from the toe extended to all $u$, overshoots $\gamma$, and the peak term corrects it down. We get the same exact result for $-(\zeta(q)-\zeta(0))$ but with different coefficients $a$ and $b$. 

The exponents in the two terms in eq.~(\ref{eq:asympt:gamma-smallq}) are nearly equal ($2$ versus $3-n_s\approx 2.035$) and the terms are of opposite sign, so at large but finite $q$, they partly cancel and the local exponents $\beta_\gamma$ and $\beta_\zeta$ (eq.~\ref{eq:asympt:scaling-inputs}) drift. Measuring those exponents numerically over $q$ in the range $q=(10^{-4}$ to 1)$\mathrm{Mpc}/h$, we find $\beta$ is monotonically increasing and is within
\be\label{eq:asympt:beta-numeric}
2\;>\;\beta\;>\;1\,,
\ee
where $\beta $ refers to both $\beta_\gamma$ and $\beta_\zeta$ (though the two are slightly different). Only as $q\to0$ does the lower power $q^2$ wins in eq.~(\ref{eq:asympt:gamma-smallq}) and $\beta\to2$ for both exponents. 

\subsection{Small-$\tilde k$ regime ($\sigma_v\tilde k\lesssim 1$, $D\gg1$)\label{sec:asympt:smallk}}

For $|\tilde k_i\tilde k_j\,\psi_{ij}(\bm q)|\le\tilde k^2\sigma_v^2\ll 1$, eq.~(\ref{eq:asympt:PZ-factored}) Taylor-expands as
\be\label{eq:asympt:expand-smallk}
e^{\tilde k_i\tilde k_j\,\psi_{ij}(\bm q)} - 1 =\tilde k_i\tilde k_j\,\psi_{ij}(\bm q) + O\!\big((\tilde k\sigma_v)^4\big)\,,
\ee
and the leading $q$-integral is the Fourier transform of $\psi_{ij}$ at $\bm k=\tilde{\bm k}/D$, as can be seen from eq.~(\ref{psi}):
\be\label{eq:asympt:FT-psi}
\int\! d^3q\, e^{-i\bm q\cdot\bm k}\,\psi_{ij}(\bm q)=\frac{k_i k_j}{k^4}\,P_L(k)\,.
\ee
Contracting against the prefactor $\tilde k_i\tilde k_j$ gives $\tilde k_i\tilde k_jk_ik_j/ k^4=D^2$, which allows us to write
\be\label{eq:asympt:PZ-smallk}
\;\;P_Z\!\left(\frac{\tilde{\bm k}}{D}, D\right)\approx e^{-\tilde k^2\sigma_v^2}\,D^2\, P_L\!\left(\tilde k/D\right)\,.
\ee

At $\tilde k\sigma_v\to 0$ the Gaussian prefactor goes to $1$ and $P_Z$ recovers the $D^2 P_L(k)=P_L(k,D)$ growth of linear theory (quick reminder: we define $P_L(k)\equiv P_L(k,D=1)$); as $\tilde k\sigma_v\to 1$ the deviation of $P_Z$ from $D^2 P_L$ becomes the Gaussian itself --- the behavior visible in Fig.~\ref{fig:pz_asympt_late_D_low_k} at $D=100$ and $30{,}000$, where the ratio $P_Z/P_L$ collapses onto the green curve representing the Gaussian smoothing kernel $\exp(-\sigma^2_v\tilde k^2)$ which is due to free-streaming. 

An important refinement of eq.~(\ref{eq:asympt:PZ-smallk}) can help us understand the $D=1$ curve. Only displacements at linear scales smaller than $1/k=D/\tilde k$ smooth the equal-time density at wavevector $k$ --- larger-scale modes shift the density field coherently \cite{Tassev_Zaldarriaga_2012} and therefore cannot result in smoothing. The relevant rms\footnote{The reason eq.~(\ref{eq:asympt:PZ-smallk}) has $\sigma_v$ instead of $\sigma_v(>k)$ is that we did not consistently expand to linear order in $P_L$ -- only the exponent containing $\psi_{ij}(\bm q)$ was expanded, but not the one with $\sigma_v$. Consistent bookkeeping must preserve translation invariance, and restore $\sigma_v(>k)$. For our analysis at large $D$, this difference is unimportant.} is thus $\sigma_v^2(>k)\equiv\int_k^\infty dk'\,P_L(k')/(6\pi^2)$, and
\be\label{pz-approx-smalltidlek}
P_Z(\tilde k/D, D)\approx e^{-\sigma_v^2(>\tilde k/D)\,\tilde k^2}\,D^2\,P_L(\tilde k/D)\,.
\ee
At $D\gg 1$ with $\tilde k$ finite, $\tilde k/D\to 0$ and $\sigma_v^2(>\tilde k/D)\to\sigma_v^2$, recovering eq.~(\ref{eq:asympt:PZ-smallk}). At $D=1$ that approximation fails: $\sigma_v^2(>\tilde k)<\sigma_v^2$, so the effective smoothing scale sits at much larger $\tilde k$ than $1/\sigma_v$, explaining the deviation of $P_Z/P_L$ at $D=1$ from the Gaussian in Fig.~\ref{fig:pz_asympt_late_D_low_k}.

For $D\gg 1$ the argument $\tilde k/D$ lies below the turnover, $\tilde k/D\ll k_{\rm eq}$, and therefore $P_L(\tilde k/D)\propto(\tilde k/D)^{n_s}$ (its IR slope); then eq.~(\ref{eq:asympt:PZ-smallk}) becomes
\be\label{eq:asympt:PZ-smallk-power}
P_Z\!\left(\frac{\tilde{\bm k}}{D}, D\right)\propto D^{2-n_s}\,\tilde k^{n_s}\,e^{-\sigma_v^2\tilde k^2}\,,\qquad \tilde k\sigma_v\lesssim 1\,,\ \ \tilde k\ll D\,k_{\rm eq}\,.
\ee

In the linear regime at large linear scales, we can see the same result $P_Z(\tilde k/D,D)\approx D^2 P_L(\tilde k/D)\propto D^{2-n_s}\tilde k^{n_s}$ by following the same type of analysis done in Section~\ref{sec:asympt:alphabeta}. The angular average over $\mu=\bm{\hat q}\cdot\bm{\hat k}$ in eq.~(\ref{eq:asympt:FT-psi}) produces Spherical Bessel function kernels over $qk$. At the peak of those kernels, we have $q\sim 1/k=D/\tilde k\gg D\sigma_v\gg 1/k_{\rm eq}$ where to write the last inequality we used $D\gg 1$ and the fact that $\sigma_v\sim 10/k_{\rm eq}$. Thus, the peak of the kernels samples the large $q$ regime, where $\psi \propto q^{-\alpha}=q^{-1-n_s}$. Performing the integral on the left hand side of eq.~\ref{eq:asympt:FT-psi} for that power law, gives a converging answer scaling as $k^{n_s-2}$.  Multiplying that result by two powers of $\tilde k$ in eq.~(\ref{eq:asympt:expand-smallk}) gives $D^2 k^{n_s}$, which recovers the linear regime answer for $P_Z(k,D)$. The contributions from the opposite end of the scales (small $q$) lie at low $q\,k=q\,\tilde k/D$ at the toe of the Bessel kernels, and are therefore producing an analytic contribution to $P_Z$, scaling as $\tilde k^2$. At low $\tilde k$, the kernel peak contributions scaling as $\tilde k^{n_s}$  are dominant, and therefore we see no contributions from the bulk\footnote{We use \textit{bulk} here in a similar way we referred to the contributions from the bulk of $P_L$ in Section~\ref{sec:asympt:alphabeta}.} of $\psi$.

\subsection{Large-$\tilde k$ regime ($\sigma_v\tilde k\gg 1$, $D\gg1$)\label{sec:asympt:largek}}

For $\tilde k\sigma_v\gg 1$ we start with the exact eq.~(\ref{eq:asympt:PZ-Deltapsi}). The integrand is peaked near $\bm q=\bm 0$ where $\Delta\psi_{ij}\to 0$, and is exponentially suppressed beyond some scale $q_\ast$ set by $\tilde k_i\tilde k_j\Delta\psi_{ij}(\bm q_\ast)\sim 1$. The small-$q$ form (eq.~\ref{eq:asympt:scaling-inputs}) reads
\be\label{eq:asympt:Deltapsi-smallq}
\Delta\psi_{ij}(\bm q)\sim \sigma_v^{2-\beta}\,q^\beta\,\tilde A_{ij}(\hat{\bm q})\,,\qquad q\to 0\,,
\ee
where $\beta$ is an effective exponent standing in for the two (slightly different) exponents $\beta_\gamma\approx \beta_\zeta$, which both  eventually converge to a common $\beta\to 2$. Here we used the fact that within an order of magnitude $\sigma_v$ sets the scale of $\psi_{ij}$. We also define $\tilde A_{ij}(\hat{\bm q})$ as a dimensionless tensor collecting the angular structure of $\psi_{ij}(\bm q)$ at leading $q^\beta$.

Applying our heuristic estimate $\tilde k_i\tilde k_j\Delta\psi_{ij}(\bm q_\ast)\sim 1$ combined with eq.~(\ref{eq:asympt:Deltapsi-smallq}) results in
\be\label{eq:asympt:qstar}
q_\ast\sim\sigma_v(\tilde k\,\sigma_v)^{-2/\beta}\,,
\ee
which implies $q_\ast\ll\sigma_v$ for the limit $\tilde k\sigma_v\gg 1$ of this section. That serves as a confirmation of the small-$q$ form we used above for $\psi$. Indeed, at large $q$, $\tilde k_i\tilde k_j\Delta \psi_{ij}(q)\to\sigma_v^2\tilde k^2\gg1$, which exponentially suppresses the term in square brackets in eq.~(\ref{eq:asympt:PZ-Deltapsi}). 

Next, we turn to the Fourier phase at $q_\ast$, which to an order of magnitude is given by
\be\label{eq:asympt:phasescaling}
\big|\bm q_\ast\cdot\tilde{\bm k}/D\big|\sim D^{-1}\,(\tilde k\sigma_v)^{1-2/\beta}\,.
\ee
For $\beta<2$, $1-2/\beta<0$, so the phase tends to $0$ in either limit: $D\to\infty$ at fixed $\tilde k$, or $\tilde k\to\infty$ at fixed $D$. In either case, we can set $e^{-i\bm q\cdot\tilde{\bm k}/D}\to 1$ in the dominant region of the integrand in eq.~(\ref{eq:asympt:PZ-Deltapsi}). Substituting $\bm q=\sigma_v(\tilde k\sigma_v)^{-2/\beta}\bm y$ in eq.~(\ref{eq:asympt:PZ-Deltapsi}), we then find:
\be\label{eq:asympt:PZ-largek-1}
P_Z\!\left(\frac{\tilde{\bm k}}{D},D\right)\sim
\sigma_v^3\,(\tilde k\sigma_v)^{-6/\beta}\!\int\!d^3y\,\exp\!\big(-y^\beta\,\tilde A_{ij}(\hat{\bm y})\hat{\tilde k}_i\hat{\tilde k}_j\big)\,,\quad \sigma_v\tilde k\gg1, \ D\gg 1\,.
\ee
The $\bm y$-integral is a finite dimensionless $C(\beta)$, so
\be\label{eq:asympt:PZ-largek-final}
P_Z\!\left(\frac{\tilde{\bm k}}{D},D\right)\sim C(\beta)\,\sigma_v^3\,(\tilde k\sigma_v)^{-6/\beta}\,,\quad \sigma_v\tilde k\gg1, \ D\gg 1\,,
\ee
which depends on $\tilde k$ but \textit{not} separately on $D$.

Throughout the analysis above, we implicitly assumed that $\beta$ is a constant. We justify that approximation from the numerics: we find that $\beta_\gamma\approx \beta_\zeta\approx 1.63$ at $q=10^{-4}\mathrm{Mpc}/h$ and that value drops by less than $1\%$ at $q=10^{-3}\mathrm{Mpc}/h$, eventually asymptoting in the other direction to $2$ as $q\ll \sigma_v$. Allowing for this running implies that $\beta$ in eq.~(\ref{eq:asympt:PZ-largek-final}) is an effective exponent that arises from the integration over $\bm q$, and could have some slight $D$ dependence, separate from the dependence on $\tilde k$. However, numerically we do not see such dependence of $P_Z(\tilde k/D,D)$ for $D\gtrsim\mathcal{O}(10^2)$ and $\tilde k \gtrsim 1h/\mathrm{Mpc}$ (Fig.~\ref{fig:pz_asympt_late_D_high_k}). Indeed, at that $\tilde k=1h/\mathrm{Mpc}$, numerically we find $P_Z(\tilde k/60{,}000,D=60{,}000)/P_Z(\tilde k/50,D=50)-1=6\times10^{-4}$ and falling fast to zero with increasing $\tilde k$. Having said that, we do expect the high-$\tilde k$ regime to start at later $\tilde k$ as the Gaussian decay at $\sigma_v\tilde k\sim 1$ bites into it (see below), but that would be for much larger $D$'s than the ones shown in the Figure.

The above analysis assumed $\beta<2$. For $\beta=2$, eq.~(\ref{eq:asympt:PZ-Deltapsi}) turns into a straightforward Gaussian integral, which in the $D\gg1$ limit reduces to $P_Z(\tilde k/D,D)\propto \tilde k^{-3}$ --- the universal $-3$ pancake exponent of \cite{Schneider_Bartelmann_1995,Konrad_2021,Konrad_2022}. 

For $\beta\in(1,2)$ (eq.~\ref{eq:asympt:beta-numeric}) the $P_Z$ slope sits in $-6/\beta\in(-6,-3)$, matching what we numerically find for the large $\tilde k$'s of Fig.~\ref{fig:pz_asympt_late_D_high_k}. Using the physical wavevector $k$, we can write
\be\label{eq:asympt:PZ-largek-k}
P_Z(k,D)\propto (Dk\sigma_v)^{-6/\beta}\,,\qquad Dk\sigma_v\gg 1\,,\ D\gg 1\,,
\ee
so at fixed $k$ the small-scale tail of the Zel'dovich power spectrum decays as $D^{-6/\beta}$ at late times --- opposite to the $D^2$ growth of the linear theory result at large scales.

\paragraph{Onset of the large-$\tilde k$ regime.}
For $\sigma_v\tilde k\gtrsim 1$, $P_Z$ leaves the linear power spectrum because of the  Gaussian smoothing kernel due to free-streaming (following eq.~\ref{eq:asympt:PZ-smallk-power}), and eventually ``lands'' on the high-$\tilde k$ prediction (eq.~\ref{eq:asympt:PZ-largek-final}) where the two intersect at $\tilde k_{\rm caustic}$. Equating them at large $D$, one gets
\be\label{eq:asympt:caustic-onset}
\tilde k_{\rm caustic}\,\sigma_v\simeq\sqrt{(2-n_s)\ln D+\mathrm{const.}+\mathcal{O}(\ln(\tilde k_{\rm caustic}\,\sigma_v))}\,,
\ee
so the onset creeps to higher $\tilde k$ only sub-logarithmically in $D$: near $\tilde k\sigma_v\approx 3-3.5$ at $D=100$, and still only $\tilde k\sigma_v\approx 4$ by $D\sim10^4$--$10^5$, consistent with the displayed curves.

\subsection{Summary and explaining Fig.~\ref{fig:zeldovich_check} \label{sec:asympt:summary}}

Combining eqs.~(\ref{eq:asympt:PZ-smallk}) and (\ref{eq:asympt:PZ-largek-final}), $P_Z(\tilde k/D,D)$ at large $D$ traces three regimes separated by $\sigma_v$:
\begin{enumerate}
	\item \textbf{Linear} ($\tilde k\sigma_v\ll 1$): $P_Z(\tilde k/D,D)\approx D^2 P_L(\tilde k/D)$, scaling as $D^{2-n_s}\tilde k^{n_s}$ when $\tilde k/D$ is smaller than the matter--radiation--equality turnover $k_{\rm eq}$.
	\item \textbf{Free-streaming smoothing} ($\tilde k\sigma_v\sim 1$): the prefactor $e^{-\tilde k^2\sigma_v^2}$ in  eqs.~(\ref{eq:asympt:PZ-smallk}) damps the linear contribution for scales smaller than the rms displacement scale.
	\item \textbf{Asymptotically Caustic} ($\tilde k\sigma_v\gg 1$): $P_Z(\tilde k/D,D)\approx C(\beta)\,\sigma_v^3\,(\tilde k\sigma_v)^{-6/\beta}$, $D$-independent, with $\beta$ the effective small-$q$ exponent of $\Delta\psi_{ij}$. The rescaled wavevector $\tilde k=Dk$ is the natural scaling variable because the power depends on the physical scale $1/k$ only through its ratio to the displacement smoothing scale $D\sigma_v$, namely $1/(kD\sigma_v)=1/(\tilde k\sigma_v)$. In the strict $\tilde k\sigma_v\to\infty$ limit, the analytic small-$q$ behavior $\Delta\psi\propto q^2$ dominates  (first term in eq.~\ref{eq:asympt:gamma-smallq}), reflecting the local differentiability of the Lagrangian-to-Eulerian map and thereby yielding the universal fold-caustic asymptote $P_Z\propto \tilde k^{-3}$ \cite{Schneider_Bartelmann_1995}. At lower $\tilde k$ within the nonlinear regime, the nonanalytic correction to $\Delta\psi$ (i.e. the UV roughness of the displacement field) inherited from the UV structure of $P_L$ at $k=\tilde k/D$ (second term in eq.~\ref{eq:asympt:gamma-smallq}) remains important. By reducing the effective $\beta$ below $2$ it makes the spectrum decay more steeply than $\tilde k^{-3}$. The regime $\tilde k\sigma_v\gg 1$ is therefore only \emph{asymptotically} caustic: the observed slope approaches the universal fold caustic result only gradually as $\tilde k\sigma_v\to\infty$ \cite{Konrad_2022}.
\end{enumerate}
 It is worth reiterating that at a fixed physical $k$, $P_Z(k,D)$ grows as $D^2$ in the linear regime, and decays as $D^{-6/\beta}$ in the asymptotically caustic regime\footnote{Applying the above analysis to $\mathcal{P}(\bm k,\bm l, \bm m)$ (eq.~\ref{Pcurly})   to study its asymptotics for generic arguments at high $p$ is tempting. However, the sign  of the exponent $(-l_im_j\psi_{ij}(\bm q))$ entering the integrand of $\mathcal{P}$ need no longer be positive as it was for $P_Z$, which means the $\psi(\bm q)$-dependent  exponential factor in  $\mathcal{P}$ can be exponentially increasing with $q$, instead of being exponentially suppressed (which allowed us to write eq.~\ref{eq:asympt:qstar} for $P_Z$). That implies that the oscillations of $e^{-i\bm{q}\cdot\bm k}$ in $\mathcal{P}$ can be important for certain arguments and we cannot replace  $e^{-i\bm{q}\cdot\bm k}$ with 1 as we did right after eq.~(\ref{eq:asympt:phasescaling}). Thus, the $q$'s that contribute, especially for $\mu_{lm}>0$, can be intermediate or even large and the small-$q$ approximation we did for $P_Z$ need not apply. Thus, the asymptotics at  high $p$ of $\mathcal{P}$ need not be caustic driven for all arguments and depend strongly on $\mu_{lm}$. That is not to say that further study of the asymptotics of $\mathcal{P}$ is not warranted. Dividing $\mathcal{P}$ by its asymptotic behavior at large $p$ and then expanding the resulting ratio in a manner similar to the path we took to get eq.~(\ref{Pexp_XZ}), may speed up the convergence of the numerical implementation of $\mathcal{P}$.}\,\footnote{It would be interesting to see the same asymptotic analysis performed on the full non-linear density power spectrum using the non-linear displacement obtained from simulations (e.g. as done in \cite{Tassev_2014}). In that case higher $n$-point functions of the displacement need to be taken into account. While such a study is beyond the scope of this paper, we note that the BBGKY hierarchy would automatically generate those non-linear higher $n$-point corrections. How robust those BBGKY results would be against modifications to the ZA closure remains to be seen.}.

Going back to Fig.~\ref{fig:zeldovich_check}, the results above explain what we see in it. Focusing on panel (a) of that Figure, the quantities plotted by the rainbow colored curves are $\mathcal{P}(\bm k,\sqrt{p}\hat{\bm k},-\sqrt{p}\hat{\bm k})\exp(-\sigma_v^2p)=P_Z(k,D=\tilde k/k)$ as seen from eq.~(\ref{PZ_check}) with $\sqrt p=Dk=\tilde k$. 

Thus, for low $\tilde k$ (i.e. low $\sqrt{p}$, or curves in yellow and towards the blue hues), the curves are proportional to $P_Z(k,D=\tilde k/k)\propto \tilde k^2\, k^{n_s-2}=p\,k^{n_s-2}$ when $k\ll Dk=\tilde k$. Thus, they fall roughly as $1/k$, and each successive curve is shifted to larger values in proportion to $p$.

At large $\tilde k$, for $k\ll Dk=\tilde k$, we have curves that are proportional to $P_Z(k,D=\tilde k/k)\propto \tilde k^{-6/\beta}$. These curves are constant in $k$, which explains the horizontal asymptotes at low $k$ for the curves towards the red hues of the figure (high $\sqrt{p}=\tilde k$). Those horizontal  curves shift to lower values in proportion to  $\tilde k^{-6/\beta}=p^{-3/\beta}$.

\section{Numerical implementation pipeline for $\mathcal{P}$\label{app:numerics}}

\paragraph{Angular part.} The $\mathcal{C}_{LM}$ coefficients are assembled with FLINT~\cite{flint} in three steps, all carried out at $1000$-bit working precision and once for every $(L,M)$ in the range fixed by the angular cutoff (see \emph{Sum truncation and convergence} below).
\begin{enumerate}
\item The finite inner sums in eq.~(\ref{CLM_FINAL}) (over $U$, $V$ and the dependent integer labels $u,v,w,x,y,z,d$) are evaluated at each $(L,M)$ and collected into the rational coefficients ($\mathrm{coeff}$) of the $(t,s,\mu_{lm})$ monomials, yielding $$\mathcal{C}_{LM}(t,s,\mu_{lm})=\sum_{C_x,C_y,C_z}\mathrm{coeff}^{(L,M)}_{C_x C_y C_z}\,t^{C_x}\,s^{C_y}\,\mu_{lm}^{C_z}$$ as a finite polynomial in $(t,s,\mu_{lm})$ for each $(L,M)$.
\item $\mu_{lm}$ is sampled on a linear grid of $101$ values; at each grid point, the $\mu_{lm}^{C_z}$ factors are summed into the rational coefficients. This leaves a polynomial in $(t,s)$ at fixed $\mu_{lm}$ for every $(L,M)$.
\item Eq.~(\ref{Psub_def}) is applied term-by-term to that polynomial and the result is re-expanded by binomial multiplication, producing the coefficient table $\mathcal{C}_{LM,\,ab}(\mu_{lm})$ of eq.~(\ref{CLM_ab_def}).
\end{enumerate}
The numerical values $\mathcal{C}_{LM,\,ab}(\mu_{lm,i})$ are written to disk per $\mu_{lm}$-grid value as an array indexed by $(L,M,a,b)$, together with the integer exponent labels $(a,b)$. These tables depend only on $(L,M,\mu_{lm})$, are independent of both cosmology and $k$, and are computed once and reused across runs.

\paragraph{Radial part.} The $q$-integrals in eqs.~(\ref{Glmn},\ref{Elmn}) are evaluated with Fast Hankel Transforms using FFTLog~\cite{Hamilton_2000,Simonovic_2018}.\footnote{As part of the numerical cross-checks we used both a C++ Fast Hankel Transform (FHT) implementation and SciPy's FHT code. In that process a bug in the SciPy implementation was identified, reported, and subsequently fixed; see \href{https://github.com/scipy/scipy/issues/21661}{scipy/scipy\#21661}.} The product $p=lm$ is sampled on a logarithmic grid with the functions $i_M(\gamma(q)\,p)$ and $i_N(\zeta(q)\,p)$ tabulated in advance. To suppress ringing at the endpoints of the $q$-interval, the integrand is multiplied by a Planck-taper window. To control aliasing, we run two FFTLog transforms at biases $+0.9$ and $-0.9$ and stitch them on the $k$-grid as follows: within the overlap region the two biased results are matched at the $k$-value where their summands and the first and second derivatives thereof agree to within prescribed tolerances, giving a smooth splicing. Appendix~\ref{app:interp} cross-checks this splicing against two independent direct-integration schemes for the same $q$-integrals.

\paragraph{Contraction stage.} The final assembly contracts the precomputed $\mathcal{C}_{LM}$ coefficients against the radial kernels. This stage uses Neumaier-compensated long-double accumulation, which is sufficient to track the contraction cancellations below the tolerance we require without falling back to arbitrary precision for every evaluation.

\paragraph{Sum truncation and convergence.} The angular truncation levels $(L_{\max},M_{\max})$ needed for convergence vary with $p=lm$: small-$p$ configurations converge at much lower order than large-$p$ ones, so a uniform cutoff over the full $p$-range would be wasteful at small $p$ and insufficient at large $p$. We partition the $p$ grid into three regimes and run each at its own truncation: $(L_{\max},M_{\max})=(10,50)$ for the low-$p$ portion ($p<4\times10^{-4}\,(h/\mathrm{Mpc})^2$), $(30,100)$ for an intermediate stretch ($4\times10^{-4}\,(h/\mathrm{Mpc})^2<p<0.04\,(h/\mathrm{Mpc})^2$), and $(50,120)$ for the high-$p$ tail ($0.04\,(h/\mathrm{Mpc})^2<p<2\,(h/\mathrm{Mpc})^2$). The boundaries between these tiers were set empirically, by inspection of where the cheaper truncations stopped meeting the convergence criterion. Convergence is confirmed per-$(k,\mu_{lm},p)$ by two conditions, both at $2\%$ relative tolerance: the last four $L$ terms and the last four $M$ terms of the final sum must each contribute below the tolerance relative to the full sum. Points that fail the criterion are flagged and excluded from the converged range.

\section{Bessel integrals over interpolated functions}\label{app:interp}

Evaluating eq.~(\ref{Glmn}) and eq.~(\ref{Elmn}) with FFTLog \cite{Hamilton_2000,Simonovic_2018} can produce ringing and aliasing artifacts.
Those come from the fact that FFTLog represents the integrand as a sum of power laws together with periodic copies. The FFTLog bias parameter can change the location of the artifacts: a positive bias improves accuracy at large $kq$ but degrades it at small $kq$, while a negative bias does the reverse. Our main calculation therefore uses two FFTLog runs at biases $\pm 0.9$ and stitches them on the $k$ grid (see the procedure stated in Section~\ref{sec:Pnumerics}). The purpose of this appendix is to verify that this $\pm 0.9$ splicing is faithful, by comparing it against two independent direct-integration schemes that do not rely on FFTLog at all. 

The two schemes both perform the radial integrals in equations~(\ref{Glmn}, \ref{Elmn}) after replacing the non-Bessel part of the integrand --- $\mathcal{H}_M(q,p,\mu_{lm})$ for eq.~(\ref{Glmn}) or $\mathcal{F}_{MN}(q,p)$ for eq.~(\ref{Elmn}) --- by a piecewise interpolant. The first scheme uses a piecewise-linear interpolant in $q$; the second uses a cubic spline. Numerical results are collected in Figure~\ref{fig:appendix_convergence}. Before turning to either scheme, we record the closed-form evaluation of the Bessel-weighted moment that both schemes share.

For integer $n\ge 0$ and integer $L\ge 0$, define
\be\label{anL}
a^{(n,L)}(k,r)\equiv\frac{1}{r^{(n+1)}}\int\limits_0^{r} dr' \left(r'\right)^nJ_{2 L+\frac{1}{2}}(k r') =&&\nonumber\\ 
&&\hspace{-18em}= 
\frac{1}{2}  \left(\frac{k r}{2}\right)^{2L+\frac{1}{2}} \Gamma \left(L+\frac{n}{2}+\frac{3}{4}\right) \, _1\bm{F}_2\left(L+\frac{n}{2}+\frac{3}{4};2
L+\frac{3}{2},L+\frac{n}{2}+\frac{7}{4};-\frac{(kr)^2}{4}\right)\nonumber
\\
&&\hspace{-18em}= \left(\frac{2}{k r}\right)^{n+1}   \left(L-\frac{n}{2}+\frac{3}{4}\right)^{\overline{n}} \textrm{Fr}_{(n \bmod 2)}\left(\sqrt{\frac{2}{\pi }} \sqrt{k r}\right)+\nonumber\\
&&\hspace{-17em}
+2^{1-n}\sqrt{\frac{2}{\pi}}\left(\frac{ \text{PC}_{n,L}\left(\frac{1}{k^2 r^2}\right) \cos (k r)}{(kr)^{3/2}}+\frac{ \text{PS}_{n,L}\left(\frac{1}{k^2 r^2}\right) \sin (k r)}{ (k r)^{5/2}}\right)\ ,
\ee
where $x^{\overline{n}}=x (x+1) \dots (x+n-1)$ denotes the rising factorial; $\mathrm{Fr}_0$ equals the Fresnel $S$ function and $\mathrm{Fr}_1$ equals the Fresnel $C$ function with the normalization of \cite{abramowitz+stegun}; and $\mathrm{PC}_{n,L}(x)$ and $\mathrm{PS}_{n,L}(x)$ are finite-degree polynomials. In Table~\ref{table:PSA} we show some of those polynomials for reference.

\begin{table}[h]
\centering
\begin{tabular}{|c|c|c|c|c|}
\hline
& L=0 & L=1 & L=2 & L=3 \\
\hline\hline
$\mathrm{PC}_{0,L}$ & $0$ & $1$ & $15x$ & $1 - 90x + 945x^2$ \\
\hline
$\mathrm{PS}_{0,L}$ & $0$ & $-1$ & $5 - 15x$ & $-10 + 405x - 945x^2$ \\
\hline\hline
$\mathrm{PC}_{1,L}$ & $-1$ & $1$ & $-1 + 42x$ & $1 - 196x + 2310x^2$ \\
\hline
$\mathrm{PS}_{1,L}$ & $0$ & $-6$ & $6 - 42x$ & $-28 + 966x - 2310x^2$ \\
\hline\hline
$\mathrm{PC}_{2,L}$ & $-2$ & $2$ & $-2 + 140x$ & $2 - 360x + 5940x^2$ \\
\hline
$\mathrm{PS}_{2,L}$ & $3$ & $-9$ & $23 - 140x$ & $-45 + 2340x - 5940x^2$ \\
\hline\hline
$\mathrm{PC}_{3,L}$ & $-4 + 15x$ & $4 - 45x$ & $-4 + 255x$ & $4 - 981x + 16632x^2$ \\
\hline
$\mathrm{PS}_{3,L}$ & $10$ & $-22$ & $50 - 840x$ & $-94 + 4536x - 16632x^2$ \\
\hline
\end{tabular}
\caption{The polynomials $\mathrm{PC}_{n,L}(x)$ and $\mathrm{PS}_{n,L}(x)$ for $L=0,1,2,3$ and $n=0,1,2,3$. Included here for reference.}
\label{table:PSA}
\end{table}

\subsection{Linear interpolant}

Suppose $f(r)$ is sampled at ordered radii $r_i$, with values $f_i\equiv f(r_i)$. Approximating $f(r)$ by its piecewise-linear interpolant on each interval $r_i<r<r_{i+1}$ gives
\be\label{linearInt}
\int\limits_{r_0}^{r_{N-1}} dr J_{2L+\frac{1}{2}}(k r) f(r)\approx
\sum\limits_{i=0}^{N-2} \left(f_i c^{(L)}_{i,i+1}-f_{i+1} c^{(L)}_{i+1,i}\right)\nonumber\\
\textrm{with\ \ } c^{(L)}_{i,j}\equiv \frac{r_j \left(a^{(0,L)}_i r_i-a^{(0,L)}_j r_j\right)-a^{(1,L)}_i r_i^2+a^{(1,L)}_j r_j^2}{r_i-r_j}\ ,
\ee
where $a^{(n,L)}_i\equiv a^{(n,L)}(k,r_i)$. For fixed $L$, these coefficients depend only on the sampling radii and on $k$, so they can be precomputed once the grid is chosen. To achieve machine precision for the ranges of $k$, $r$, and $L$ required by our code, we evaluate the Fresnel-function representation of $a^{(n,L)}$ in eq.~(\ref{anL}) using FLINT with 3000 bits of intermediate precision. 

We use eq.~(\ref{linearInt}) to evaluate eq.~(\ref{Glmn},\ref{Elmn}) by using linear interpolation for the non-$J$ part of the integrands. Using those numerical results, we calculate the contributions to $\mathcal{P}$ for various fixed values of $L$ and show those in the upper right panel of Figure~\ref{fig:appendix_convergence}. At low $k$ the result clearly matches the FHT calculation with FFTLog bias$=-0.9$ (lower right panel of Fig.~\ref{fig:appendix_convergence}) for low $L$ (for higher $L$, the FHT result is drowned in artifact noise). At high $k$ we see ringing artifacts in the linear interpolant graph (especially when compared to the FHT result with bias$=0.9$, see lower left panel of Fig.~\ref{fig:appendix_convergence}) as the interpolated function is not smooth. Thus, we proceed to evaluate the integrals using cubic spline interpolation next.

\subsection{Cubic spline interpolant}

Similar to linear interpolation, we also implement cubic-spline interpolation to check for convergence. Linear interpolation is about 30 times slower than the FHT calculation, and the integration based on cubic spline interpolation is slower than that based on linear interpolation by another factor of about 5. To perform the cubic spline interpolation integration we again use eq.~(\ref{anL})  but instead of using linear order $f(r)$ as in eq.~(\ref{linearInt}), we use a cubic approximation to $f(r)$ with coefficients set by the cubic spline method. The piecewise-cubic version of eq.~(\ref{linearInt}) involves the same closed-form $a^{(n,L)}$, but with $n$ now running up to $3$ instead of $1$; that is why Table~\ref{table:PSA} lists the polynomials $\mathrm{PC}_{n,L}$, $\mathrm{PS}_{n,L}$ for $n=0,1,2,3$.

\begin{figure}[h!]
\centering
\includegraphics[width=\textwidth]{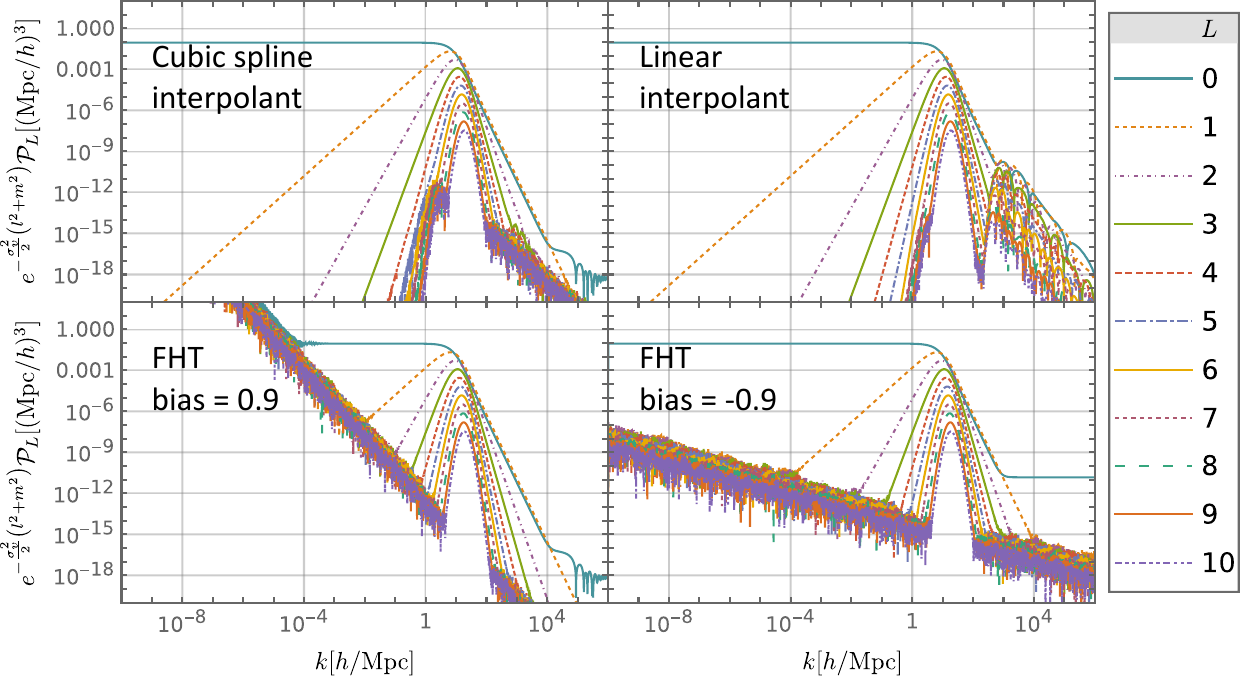}
\caption{For this figure, we write $\mathcal{P}_L$ for the fixed-$L$ contribution to $\mathcal{P}$, so that $\mathcal{P}=\sum_L\mathcal{P}_L$ in the notation of Section~\ref{sec:corrZA} (eqs.~\ref{FINALP_exp},~\ref{FINALP}). The figure shows   $\mathcal{P}_L$ (for the values of $L$ shown in the legend) for a demanding configuration, $l=m=4\,h/\mathrm{Mpc}$ and $\mu_{kl}=-\mu_{lm}=1$, chosen to push the radial integrand into the regime where FFTLog artifacts are readily visible. \emph{Top right:} piecewise-linear interpolant scheme; agrees with the FFTLog result at bias$=-0.9$ (bottom right panel) for low $k$ and low $L$, and develops ringing at high $k$ because the interpolant derivative is not continuous. \emph{Top left:} cubic-spline interpolant scheme; reproduces the linear-interpolant result at low $k$ and the FFTLog result at bias $+0.9$ (bottom left panel) at high $k$ and low $L$. The agreement of the spline scheme with the bias$=0.9$ FFTLog result on the high-$k$ side, combined with the linear/spline-scheme agreement with bias$=-0.9$ on the low-$k$ side, is what reassures us that the $\pm 0.9$ FFTLog splicing used in Section~\ref{sec:Pnumerics} gives sensible results.
}
\label{fig:appendix_convergence}
\end{figure}

An example result is shown in the upper left panel of Figure~\ref{fig:appendix_convergence}. At low $k$ we reproduce the linear interpolation results, and at high $k$ and low $L$ we reproduce the FHT result with bias$=0.9$, which gives us reassurance that the splicing of two FHT's with different biases (such as $\pm 0.9$) can give us a faithful representation for the integrals in eq.~(\ref{Glmn},\ref{Elmn}).

Thus, for our main numerical calculations for $\mathcal{P}$, we use two FHTs, one at bias $+0.9$ and one at bias $-0.9$, and stitch them together separately for each $(L,M)$ in  eq.~(\ref{Glmn}): within the overlap region the two results are matched at the $k$-value where their summands and the first and second derivatives thereof agree up to a prescribed threshold, so that we can obtain a smooth splicing.

\acknowledgments
This work was done for the pure pleasure of finding things out, and was funded by no one.
	Gratitude in supporting this project goes first and foremost to S., who is always kind,  supportive, and encouraging of my work. Enormous thank you to my kiddos, who bring joy and meaning to everything I do. The privilege of having enough hours in the day to be a dad and a partner, and to work on a hobby project such as this, is afforded to me by my union job as a public high-school teacher. Therefore, I would also like to thank the Braintree Education Association for fighting vehemently for working conditions that allow rank-and-file union members like me to have fulfilling lives both inside and outside the classroom. I am especially grateful to my  colleagues, Truong Dinh and Molly FitzGerald, for working relentlessly to strengthen our teachers' union, and thus for making this work possible.

\bibliography{biblio}  

\providecommand{\href}[2]{#2}\begingroup\raggedright\begin{thebibliography}{10}

\bibitem{Tassev_Zaldarriaga_Eisenstein_2013}
S.~Tassev, M.~Zaldarriaga and D.~Eisenstein, \emph{{Solving Large Scale
  Structure in Ten Easy Steps with COLA}},
  \href{https://doi.org/10.1088/1475-7516/2013/06/036}{\emph{JCAP} {\bfseries
  06} (2013) 036} [\href{https://arxiv.org/abs/1301.0322}{{\ttfamily
  1301.0322}}].

\bibitem{Tassev_Eisenstein_Wandelt_Zaldarriaga_2015}
S.~Tassev, D.J.~Eisenstein, B.D.~Wandelt and M.~Zaldarriaga, \emph{{sCOLA: The
  N-body COLA Method Extended to the Spatial Domain}},
  \href{https://arxiv.org/abs/1502.07751}{{\ttfamily 1502.07751}}.

\bibitem{Valageas_2007}
P.~Valageas, \emph{Using the zeldovich dynamics to test expansion schemes},
  \href{https://doi.org/10.1051/0004-6361:20078065}{\emph{Astronomy \&
  Astrophysics} {\bfseries 476} (2007) 31–58}.

\bibitem{Bernardeau_2002}
F.~Bernardeau, S.~Colombi, E.~Gaztanaga and R.~Scoccimarro, \emph{{Large scale
  structure of the universe and cosmological perturbation theory}},
  \href{https://doi.org/10.1016/S0370-1573(02)00135-7}{\emph{Phys. Rept.}
  {\bfseries 367} (2002) 1}
  [\href{https://arxiv.org/abs/astro-ph/0112551}{{\ttfamily
  astro-ph/0112551}}].

\bibitem{Catelan_1995}
P.~Catelan, \emph{{Lagrangian dynamics in nonflat universes and nonlinear
  gravitational evolution}},
  \href{https://doi.org/10.1093/mnras/276.1.115}{\emph{Mon. Not. Roy. Astron.
  Soc.} {\bfseries 276} (1995) 115}
  [\href{https://arxiv.org/abs/astro-ph/9406016}{{\ttfamily
  astro-ph/9406016}}].

\bibitem{Matsubara_2008}
T.~Matsubara, \emph{{Resumming Cosmological Perturbations via the Lagrangian
  Picture: One-loop Results in Real Space and in Redshift Space}},
  \href{https://doi.org/10.1103/PhysRevD.77.063530}{\emph{Phys. Rev. D}
  {\bfseries 77} (2008) 063530}
  [\href{https://arxiv.org/abs/0711.2521}{{\ttfamily 0711.2521}}].

\bibitem{Tassev_Zaldarriaga_2012}
S.~Tassev and M.~Zaldarriaga, \emph{{The Mildly Non-Linear Regime of Structure
  Formation}}, \href{https://doi.org/10.1088/1475-7516/2012/04/013}{\emph{JCAP}
  {\bfseries 04} (2012) 013} [\href{https://arxiv.org/abs/1109.4939}{{\ttfamily
  1109.4939}}].

\bibitem{Tassev_2014}
S.~Tassev, \emph{{Lagrangian or Eulerian; Real or Fourier? Not All Approaches
  to Large-Scale Structure Are Created Equal}},
  \href{https://doi.org/10.1088/1475-7516/2014/06/008}{\emph{JCAP} {\bfseries
  06} (2014) 008} [\href{https://arxiv.org/abs/1311.4884}{{\ttfamily
  1311.4884}}].

\bibitem{Porto_Senatore_Zaldarriaga_2014}
R.A.~Porto, L.~Senatore and M.~Zaldarriaga, \emph{{The Lagrangian-space
  Effective Field Theory of Large Scale Structures}},
  \href{https://doi.org/10.1088/1475-7516/2014/05/022}{\emph{JCAP} {\bfseries
  05} (2014) 022} [\href{https://arxiv.org/abs/1311.2168}{{\ttfamily
  1311.2168}}].

\bibitem{Porto_2014}
R.A.~Porto, L.~Senatore and M.~Zaldarriaga, \emph{The lagrangian-space
  effective field theory of large scale structures},
  \href{https://doi.org/10.1088/1475-7516/2014/05/022}{\emph{Journal of
  Cosmology and Astroparticle Physics} {\bfseries 2014} (2014) 022–022}.

\bibitem{Baumann:2010tm}
D.~Baumann, A.~Nicolis, L.~Senatore and M.~Zaldarriaga, \emph{{Cosmological
  Non-Linearities as an Effective Fluid}},
  \href{https://doi.org/10.1088/1475-7516/2012/07/051}{\emph{JCAP} {\bfseries
  07} (2012) 051} [\href{https://arxiv.org/abs/1004.2488}{{\ttfamily
  1004.2488}}].

\bibitem{Carrasco_Hertzberg_Senatore_2012}
J.J.M.~Carrasco, M.P.~Hertzberg and L.~Senatore, \emph{{The Effective Field
  Theory of Cosmological Large Scale Structures}},
  \href{https://doi.org/10.1007/JHEP09(2012)082}{\emph{JHEP} {\bfseries 09}
  (2012) 082} [\href{https://arxiv.org/abs/1206.2926}{{\ttfamily 1206.2926}}].

\bibitem{Will_2014}
C.M.~Will, \emph{The confrontation between general relativity and experiment},
  \href{https://doi.org/10.12942/lrr-2014-4}{\emph{Living Reviews in
  Relativity} {\bfseries 17} (2014) }.

\bibitem{Carlson_2012}
J.~Carlson, B.~Reid and M.~White, \emph{Convolution lagrangian perturbation
  theory for biased tracers},
  \href{https://doi.org/10.1093/mnras/sts457}{\emph{Monthly Notices of the
  Royal Astronomical Society} {\bfseries 429} (2012) 1674–1685}.

\bibitem{peebles}
P.J.E.~{Peebles}, \emph{{The large-scale structure of the universe}}, Princeton
  University Press (1980).

\bibitem{Tassev_2011}
S.V.~Tassev, \emph{The helmholtz hierarchy: phase space statistics of cold dark
  matter}, \href{https://doi.org/10.1088/1475-7516/2011/10/022}{\emph{Journal
  of Cosmology and Astroparticle Physics} {\bfseries 2011} (2011) 022–022}.

\bibitem{PhRvD91j3507A}
Y.~{Ali-Ha{\"\i}moud}, \emph{{Perturbative interaction approach to cosmological
  structure formation}},
  \href{https://doi.org/10.1103/PhysRevD.91.103507}{\emph{\prd} {\bfseries 91}
  (2015) 103507} [\href{https://arxiv.org/abs/1502.00580}{{\ttfamily
  1502.00580}}].

\bibitem{Nascimento:2024hle}
C.~Nascimento and M.~Loverde, \emph{{Cosmological perturbation theory for large
  scale structure in phase space}},
  \href{https://doi.org/10.1088/1475-7516/2025/06/002}{\emph{JCAP} {\bfseries
  06} (2025) 002} [\href{https://arxiv.org/abs/2410.05389}{{\ttfamily
  2410.05389}}].

\bibitem{Chisari_Zaldarriaga_2011}
N.E.~Chisari and M.~Zaldarriaga, \emph{{Connection between Newtonian
  simulations and general relativity}},
  \href{https://doi.org/10.1103/PhysRevD.84.089901}{\emph{Phys. Rev. D}
  {\bfseries 83} (2011) 123505}
  [\href{https://arxiv.org/abs/1101.3555}{{\ttfamily 1101.3555}}].

\bibitem{Wagner_2015}
C.~Wagner, F.~Schmidt, C.-T.~Chiang and E.~Komatsu, \emph{{Separate Universe
  Simulations}}, \href{https://doi.org/10.1093/mnrasl/slu187}{\emph{Mon. Not.
  Roy. Astron. Soc.} {\bfseries 448} (2015) L11}
  [\href{https://arxiv.org/abs/1409.6294}{{\ttfamily 1409.6294}}].

\bibitem{1964Tell...16....1L}
E.N.~{Lorenz}, \emph{{The problem of deducing the climate from the governing
  equations}},
  \href{https://doi.org/10.1111/j.2153-3490.1964.tb00136.x10.3402/tellusa.v16i1.8893}{\emph{Tellus}
  {\bfseries 16} (1964) 1}.

\bibitem{tao2011introduction}
T.~Tao, \emph{An Introduction to Measure Theory}, Graduate Studies in
  Mathematics, American Mathematical Society (2011).

\bibitem{NIST:DLMF}
``{\it NIST Digital Library of Mathematical Functions}.''
  \url{https://dlmf.nist.gov/}, Release 1.2.1 of 2024-06-15.

\bibitem{zeldovich}
Y.B.~{Zel'dovich}, \emph{{Gravitational instability: An approximate theory for
  large density perturbations.}}, {\emph{Astronomy \& Astrophysics} {\bfseries
  5} (1970) 84}.

\bibitem{Valageas_2004}
P.~Valageas, \emph{A new approach to gravitational clustering: A path-integral
  formalism and large-nexpansions},
  \href{https://doi.org/10.1051/0004-6361:20040125}{\emph{Astronomy \&
  Astrophysics} {\bfseries 421} (2004) 23–40}.

\bibitem{Planck2020}
{Planck Collaboration}, N.~Aghanim, Y.~Akrami, M.~Ashdown, J.~Aumont,
  C.~Baccigalupi et~al., \emph{{Planck} 2018 results. {VI}. {Cosmological}
  parameters},
  \href{https://doi.org/10.1051/0004-6361/201833910}{\emph{Astronomy \&
  Astrophysics} {\bfseries 641} (2020) A6}.

\bibitem{Hamilton_2000}
A.J.S.~Hamilton, \emph{Uncorrelated modes of the non-linear power spectrum},
  \href{https://doi.org/10.1046/j.1365-8711.2000.03071.x}{\emph{Monthly Notices
  of the Royal Astronomical Society} {\bfseries 312} (2000) 257–284}.

\bibitem{Simonovic_2018}
M.~Simonovi{\'c}, T.~Baldauf, M.~Zaldarriaga, J.J.~Carrasco and J.A.~Kollmeier,
  \emph{{Cosmological perturbation theory using the FFTLog: formalism and
  connection to QFT loop integrals}},
  \href{https://doi.org/10.1088/1475-7516/2018/04/030}{\emph{JCAP} {\bfseries
  04} (2018) 030} [\href{https://arxiv.org/abs/1708.08130}{{\ttfamily
  1708.08130}}].

\bibitem{Schneider_Bartelmann_1995}
P.~Schneider and M.~Bartelmann, \emph{The power spectrum of density
  fluctuations in the zel'dovich approximation},
  \href{https://doi.org/10.1093/mnras/273.2.475}{\emph{Monthly Notices of the
  Royal Astronomical Society} {\bfseries 273} (1995) 475}.

\bibitem{Konrad_2021}
S.~Konrad and M.~Bartelmann, \emph{On the asymptotic behaviour of cosmic
  density-fluctuation power spectra},
  \href{https://doi.org/10.1093/mnras/stac1795}{\emph{Monthly Notices of the
  Royal Astronomical Society} {\bfseries 515} (2022) 2578}.

\bibitem{Konrad_2022}
S.~Konrad, Y.B.~Ginat and M.~Bartelmann, \emph{On the asymptotic behaviour of
  cosmic density-fluctuation power spectra of cold dark matter},
  \href{https://doi.org/10.1093/mnras/stac2064}{\emph{Monthly Notices of the
  Royal Astronomical Society} {\bfseries 515} (2022) 5823}.

\bibitem{BBKS_1986}
J.M.~Bardeen, J.R.~Bond, N.~Kaiser and A.S.~Szalay, \emph{{The Statistics of
  Peaks of Gaussian Random Fields}},
  \href{https://doi.org/10.1086/164143}{\emph{Astrophys. J.} {\bfseries 304}
  (1986) 15}.

\bibitem{flint}
.~The FLINT~team, \emph{{FLINT}: {F}ast {L}ibrary for {N}umber {T}heory}, 2023.

\bibitem{abramowitz+stegun}
M.~Abramowitz and I.A.~Stegun, \emph{Handbook of Mathematical Functions with
  Formulas, Graphs, and Mathematical Tables}, Dover, New York (1964).

\end{thebibliography}\endgroup

\end{document}